\documentclass[a4paper,fleqn,usenatbib,useAMS]{mnras}
\usepackage[latin9]{inputenc}
\usepackage{float}
\usepackage{amstext}
\usepackage{amssymb}
\usepackage{graphicx}
\usepackage{epstopdf}
\usepackage{wasysym}
\usepackage[T1]{fontenc}
\usepackage{ae,aecompl}

\makeatletter

\providecommand{\tabularnewline}{\\}

\usepackage{pifont}
\usepackage[usenames,dvipsnames,svgnames,table]{xcolor}
\usepackage{multirow}
\def\jref@jnl#1{{\rm#1}}

\def\actaa{\jref@jnl{Acta Astron.}}      % Acta Astron.
\def\aj{\jref@jnl{AJ}}                   % Astronomical Journal
\def\araa{\jref@jnl{ARA\&A}}             % Annual Review of Astron and Astrophys
\def\apj{\jref@jnl{ApJ}}                 % Astrophysical Journal
\def\apjl{\jref@jnl{ApJ}}                % Astrophysical Journal, Letters
\def\apjs{\jref@jnl{ApJS}}               % Astrophysical Journal, Supplement
\def\ao{\jref@jnl{Appl.~Opt.}}           % Applied Optics
\def\apss{\jref@jnl{Ap\&SS}}             % Astrophysics and Space Science
\def\aap{\jref@jnl{A\&A}}                % Astronomy and Astrophysics
\def\aapr{\jref@jnl{A\&A~Rev.}}          % Astronomy and Astrophysics Reviews
\def\aaps{\jref@jnl{A\&AS}}              % Astronomy and Astrophysics, Supplement
\def\azh{\jref@jnl{AZh}}                 % Astronomicheskii Zhurnal
\def\baas{\jref@jnl{BAAS}}               % Bulletin of the AAS
\def\jrasc{\jref@jnl{JRASC}}             % Journal of the RAS of Canada
\def\jcap{\jref@jnl{JCAP}}               % Journal Cosmology & Astroparticle Physics
\def\memras{\jref@jnl{MmRAS}}            % Memoirs of the RAS
\def\mnras{\jref@jnl{MNRAS}}             % Monthly Notices of the RAS
\def\na{\jref@jnl{New Astronomy}}        % New Astronomy
\def\pra{\jref@jnl{Phys.~Rev.~A}}        % Physical Review A: General Physics
\def\prb{\jref@jnl{Phys.~Rev.~B}}        % Physical Review B: Solid State
\def\prc{\jref@jnl{Phys.~Rev.~C}}        % Physical Review C
\def\prd{\jref@jnl{Phys.~Rev.~D}}        % Physical Review D
\def\pre{\jref@jnl{Phys.~Rev.~E}}        % Physical Review E
\def\prl{\jref@jnl{Phys.~Rev.~Lett.}}    % Physical Review Letters
\def\pasa{\jref@jnl{PASA}}               % Publ Astronomical Soc Aust
\def\pasp{\jref@jnl{PASP}}               % Publications of the ASP
\def\pasj{\jref@jnl{PASJ}}               % Publications of the ASJ
\def\qjras{\jref@jnl{QJRAS}}             % Quarterly Journal of the RAS
\def\rmxaa{\jref@jnl{Rev.Mex.AA}}       % Rev.Mex.AA
\def\skytel{\jref@jnl{S\&T}}             % Sky and Telescope
\def\solphys{\jref@jnl{Sol.~Phys.}}      % Solar Physics
\def\sovast{\jref@jnl{Soviet~Ast.}}      % Soviet Astronomy
\def\ssr{\jref@jnl{Space~Sci.~Rev.}}     % Space Science Reviews
\def\zap{\jref@jnl{ZAp}}                 % Zeitschrift fuer Astrophysik
\def\nar{\jref@jnl{NewAR}}               % New Astronomy Reviews
\def\nat{\jref@jnl{Nature}}              % Nature
\def\iaucirc{\jref@jnl{IAU~Circ.}}       % IAU Cirulars
\def\aplett{\jref@jnl{Astrophys.~Lett.}} % Astrophysics Letters
\def\apspr{\jref@jnl{Astrophys.~Space~Phys.~Res.}}
\def\bain{\jref@jnl{Bull.~Astron.~Inst.~Netherlands}} 
\def\fcp{\jref@jnl{Fund.~Cosmic~Phys.}}  % Fundamental Cosmic Physics
\def\gca{\jref@jnl{Geochim.~Cosmochim.~Acta}}   % Geochimica Cosmochimica Acta
\def\grl{\jref@jnl{Geophys.~Res.~Lett.}} % Geophysics Research Letters
\def\jcap{\jref@jnl{JCAP}}      %
\def\jcp{\jref@jnl{J.~Chem.~Phys.}}      % Journal of Chemical Physics
\def\jgr{\jref@jnl{J.~Geophys.~Res.}}    % Journal of Geophysics Research
\def\jqsrt{\jref@jnl{J.~Quant.~Spec.~Radiat.~Transf.}}
\def\memsai{\jref@jnl{Mem.~Soc.~Astron.~Italiana}}
\def\nphysa{\jref@jnl{Nucl.~Phys.~A}}   % Nuclear Physics A
\def\physrep{\jref@jnl{Phys.~Rep.}}   % Physics Reports
\def\physscr{\jref@jnl{Phys.~Scr}}   % Physica Scripta
\def\planss{\jref@jnl{Planet.~Space~Sci.}}   % Planetary Space Science
\def\procspie{\jref@jnl{Proc.~SPIE}}   % Proceedings of the SPIE

\def\kms{{\rm km}~{\rm s}^{-1}}

\def\Ha{H$\alpha$}
\def\HeII{{\ion{He}{ii}}}
\def\MgII{{\ion{Mg}{ii}}}
\def\C4{{\ion{C}{iv}}}
\def\Tbb{T_\textsc{bb}}

\def\aabboovvee{\"{}}

\definecolor{Orange1}{rgb}{1.0,0.9,0.7}
\definecolor{Orange2}{rgb}{1.0,0.95,0.85}
\definecolor{Orange3}{rgb}{0.5,0.15,0.00}
\colorlet{Plum0}{Plum!20!white}
\colorlet{Plum1}{Plum!50!black}
\makeatother

\title[Spectral features of TDE-candidates]{%
Spectral features of tidal-disruption candidates
and alternative origins for such transient flares}

\author[Saxton et al.]{Curtis J. Saxton,\thanks{%
\mbox{E-mail: saxton@physics.technion.ac.il~(CJS)};
\mbox{hperets@physics.technion.ac.il~(HBP)};
\mbox{alexei@physics.technion.ac.il~(AB)};
\newline
$^\dagger$  Present address:  Soreq Nuclear Research Center, Yavne 8180000, Israel
}
Hagai B. Perets \& Alexei Baskin$^\dagger$
\\ 
Physics Department, Technion - Israel institute of Technology, Haifa, Israel 3200002}

\date{Accepted 2017 November 08. Received 2017 November 08 ; in original form 2016 December 21}

\begin{document}

\pagerange{\pageref{firstpage}--\pageref{lastpage}} \pubyear{2017}

\maketitle

\label{firstpage}

\begin{abstract}
UV and optically selected candidates for stellar tidal disruption
events (TDE) often exhibit broad spectral features
(\HeII\, emission, H$\alpha$ emission, or absorption lines)
on a blackbody-like continuum ($10^{4}\mathrm{K}\la T\la10^{5}\mathrm{K}$).
The lines presumably emit from TDE debris
   or circumnuclear clouds photoionized by the flare.
Line velocities however are much lower than expected
   from a stellar disruption by supermassive black hole (SMBH),
   and are somewhat faster than expected
   for the broad line region (BLR) clouds
   of a persistently active galactic nucleus (AGN).
The distinctive spectral states
   are not strongly related to observed luminosity and velocity,
   nor to SMBH mass estimates.
We use exhaustive photoionization modelling
   to map the domain of fluxes and cloud properties that yield
   (e.g.) a He-overbright state
   where a large He II(4686\AA\,)/H$\alpha$ line-ratio
   creates an illusion of helium enrichment.
Although observed line ratios occur in a plausible minority of cases,
   AGN-like illumination can not reproduce the observed equivalent widths.
We therefore propose to explain these properties
   by a light-echo photoionization model:
   the initial flash of a hot blackbody (detonation) excites BLR clouds,
   which are then seen superimposed on continuum from a later, expanded,
   cooled stage of the luminous source.
The implied cloud mass is substellar,
   which may be inconsistent with a TDE.
Given these and other inconsistencies with TDE models
   (e.g. host-galaxies distribution)
   we suggest to also consider alternative origins for these nuclear flares,
   which we briefly discuss
   (e.g. nuclear supernovae and starved/subluminous AGNs).  
\end{abstract}
\begin{keywords}
accretion, accretion discs
\textemdash{}
black hole physics
\textemdash{}
galaxies: active
\textemdash{}
galaxies: nuclei
\textemdash{}
quasars: emission lines
\end{keywords}

\section{INTRODUCTION}

Early theorists considered that active galactic nuclei (AGN) and quasars
   might consist of a supermassive black hole (SMBH, of mass $m_{\bullet}$)
   accreting gas derived from the tidal disruption of stars in the nuclear
   cluster
	\citep{hills1975,frank1976,young1977,ozernoi1978,kato1978}.
A star of mass $m_{\bigstar}$ and radius $R_{\bigstar}$ disrupts
   if it passes within the `tidal radius,'
   $R_{\mathrm{t}}=R_{\bigstar}(m_{\bullet}/m_{\bigstar})^{1/3}$.
The consequences would be similar if the nucleus hosts
   any other sufficiently dense and relativistic
   type of dark supermassive object
        \citep[e.g.][]{goel2015,meliani2015,saxton2016a}.
Larger SMBH ($m_{\bullet}\ga\mathrm{few}\times10^{8}m_{\odot}$)
   can swallow main sequence stars whole,
   without a tidal disruption event (TDE).
The TDE rate needed to power AGN turned out to be
   difficult to reconcile with stellar densities,
   at least in the nuclei of modern galaxies
   (and gas accretion now seems more important to AGN).
Nonetheless it was recognized that a TDE could produce a distinctive flare,
   peaking in much less than a year
   and potentially remaining observable during a few years of fading
   \citep{lidskii1979,gurzadian1981,lacy1982,luminet1985,rees1988}.
TDEs by lower mass SMBHs ($m_{\bullet}\la3\times10^{7}m_{\odot}$)
   are more likely to accrete at rates exceeding the
   \citet{eddington1918b}
   radiation pressure limit \citep{ulmer1999}.

More than ten transient events detected in ultraviolet and optical (UV/O),
   X-ray and $\gamma$-ray monitoring and serendipitous observations
   are currently classified as TDE candidates \citep{komossa2015b}.
These were selected for their central locations in the host galaxies.
Candidates are usually excluded
   if the host showed conventional signatures of an AGN
   (in spectroscopy, variability, and radio counterparts)
   before and long after the transient\footnote{%
	Realistically, TDE could just as well occur around active nuclei,
	but observers prefer to avoid false-positive identifications.}.
The relation between TDE candidates 
   observed as $\gamma$-ray/X-ray flares
   and those detected in the UV/O is not yet clear.
We note that much smaller,
   stellar black holes, may also tidally disrupt stars.
These would lead to $\mu$TDEs
   which might resemble ultra-long $\gamma$-ray bursts/X-ray flares
   and jetted-TDE candidates
   \citep{perets2016_06},
   but not the UV/O TDE-candidates on which we focus here. 

\citet{arcavi2014}
   classified the UV/O candidates into a sequence
   by their broad line spectral states:
   He-overbright cases;
   some with \HeII\, and H$\alpha$;
   and others with H$\alpha$ dominating.
PS1-11af had broad absorption lines
   (beyond the scope of our emission modelling)
   but it is notable that the velocity widths were
   comparable to velocity widths of emission-line TDE candidates
	\citep{chornock2014}.
We note that \citet{drake2011}
   classified another transient
   (in a Seyfert host with prior radio detection)
   as an exotic supernova rather than a TDE,
   partly
   because the temperature was below predictions of early TDE theories.
We disagree with this exclusion, since the properties were actually
   consistent with more recent UV/O TDE candidates,
   and suggest to include this as an additional TDE candidate
   (for our table of currently identified UV/O TDE-candidates,
   see the Appendix~\ref{s.candidates}).
Although spectoscopic observations have been sparse in time,
   a few UV/O events have evolved from one emission line state to another.
%   the \HeII-bright state
%   or the \Ha-bright state
%   into the intermediate state with comparable \HeII and \Ha emission.
Our paper will study the spectroscopic features of TDE candidates,
   and analyse them using detailed photoionization models.

Previous studies by
   \citet{gaskell2014}
   and
   \citet{guillochon2014}
   tried to explain the spectral features of the UV/O TDE candidates.
However, as we show in the following,
   none of the simple models suggested in the literature,
   typically involving AGN-like illumination,
   can actually explain the spectral properties of the candidates.
In particular,
   though the line-ratios can be explained by models as suggested by
	\citet{guillochon2014} and \citet{gaskell2014},
   they fail to explain the \emph{observed} equivalent width.
We suggest a model which may overcome these difficulties,
   and provide novel predictions for the spectral temporal evolution.
We consider how states similar to
   the observed line ratios and equivalent widths
   can arise in generic models of TDE, AGN outbursts, and other explosive events.
Using models of photoionized spectra,
   we distinguish among luminous source spectra
   that can reproduce the observed lines.

Besides the failure of the simple picture
   to explain the observed spectral features,
   our current interpretation of candidate TDEs as
   bona-fide `real' TDEs
   encounters several other difficulties.
We complement our photoionization modeling
   with an analysis of these additional
   challenges and their implications,
   and briefly discuss other possible origins for the observed TDE candidates.

Our paper is organized as follows.
We begin by describing our modeling
   of the spectral features of TDEs in Section~\ref{s.model},
   and show that none of the simple models
   can explain both the line ratios and equivalent widths seen in observations.
We propose a novel model which could
   potentially resolve these difficulties (Section~\ref{s.echo}),
   and then discuss
   our results as well the possible origins of such nuclear flares
   (Section~3),
   and in particular possible alternatives to the TDE origin,
   including subluminous AGN flaring and nuclear supernovae
   (Section~\ref{s.supernovae}).

\section{MODELING}
\label{s.model}

The principal luminous media in a TDE include
   the dense tidal stream of bound debris,
   which orbits round to shock its own tail
   and form an accretion disc;
   unbound debris escaping the system ballistically;
   and a radiation-driven wind.
If the TDE continuum emission photoionizes circumnuclear gas clouds,
   like those of AGN,
   their line emissions may resemble
   an AGN's broad emission line regions (BLRs)
   except evolving on shorter time-scales.
For the moment,
   we are agnostic about whether the luminous source
   originated as a TDE or some other type of transient event.
(It is also unimportant whether the clouds originated from the flare,
   or already orbited dormantly in the inner galaxy.)
In the following,
   we model the expected spectral features,
   and compare them with the observed features.
We find that current models are inconsistent with observed features,
   and we discuss a novel echo model that
   could be reconciled with observations.

\subsection{Emission line ratios and equivalent widths}

To interpret observations, we require predictions of both the
   \HeII(4686\AA) and H$\alpha$(6563\AA) intensities,
   as well as the line ratios,
   the underlying continuum, and other lines that might appear.
These synthetic observables are generic and time-independent:
   they should apply equally to slowly evolving structure
   under variable illumination
   (e.g. BLR during an AGN flare)
   or dynamically evolving clouds under
   a steady or variable radiation source (e.g. TDE debris).
In order to do so,
   our calculations (below) follow photoionization models employing
   the \textsc{Cloudy} code \citep{ferland2013}.
The standard approach \citep{baldwin1995,korista2004}
   is to compute emission line output of clouds across the
   `flux-density plane', ($n_{\textsc{h}},\Phi_{\textsc{h}}$),
   where $n_{\textsc{h}}$ is the cloud density,
   and $\Phi_{\textsc{h}}$ is ionizing flux from an isotropic source
   with a specified spectral energy distrbution (SED), given by
\begin{equation}
	\Phi_\textsc{h}={{{L}}\over{{4\upi r^{2}\bar{\epsilon}}}}\ ,
\label{eq.Phi}
\end{equation}
   where $\bar{\epsilon}$
   is an effective mean ionizing photon energy for the given SED. 

As calculated by \citet{guillochon2014} and \citet{gaskell2014},
   emission line ratios consistent with the \HeII-overbright
   state are attainable in specific cloud conditions,
   though they considered an AGN SED model of
   \citealt{mathews1987}
   (hereafter `MF1987');
   moreover, see \citealp{strubbe2015} for different conclusions).
However,
   the observed ratios are a necessary but not \emph{sufficient} constraint;
   any successful model should also provide
   the correct \emph{equivalent widths} of the lines.
As we show in the following, \emph{all} the models,
   and in particular current SED models
   (constructed to be consistent with observations),
   on which we focus thus far fail to produce both
   the line ratios and the equivalent widths
   (but this is also true for the MF1987 model,
   considered here only for comparison with previous works).

If the line intensity emerging from a cloud is $I_\mathrm{line}$
   and the incident continuum is $F_{\lambda}$,
   the equivalent width (`EW') is
   $W_{\lambda}\equiv I_\mathrm{line}/F_{\lambda}$.
The integrated EW of a composite system depends on
   the covering fraction of the clouds,
   $\Omega$.
Observationally, emission lines with $W_{\lambda}<1$\AA\,
   might be undetectable.
\citet{gezari2012}, observing PS1-10jh,
   found an \HeII\, line luminosity of
   $(9\pm1)\times10^{40}\mathrm{erg}\,\mathrm{s}^{-1}$,
   which corresponds to $W_{4686}=66.4\pm13.9$\AA\,;
   while \citet{vanvelzen2011b}
   report a H$\alpha$ value $W_{6563}=87\pm5$\AA\, for TDE2.
The \HeII-overbright early state of ASASSN-15oi
   had an equivalent width $W_{4686}=99\pm13$\AA\,
   \citep{holoien2016a}.
By visual inspection,
   other UV/O candidates' broad lines typically
   have equivalent widths of similar order
   \citep{wang2011,drake2011,holoien2014,arcavi2014}.
These facts require models of the continuum and clouds that provide
$W_{\lambda}$ of at least a few \AA,
   for some covering factor.
Therefore, models predicting smaller $W_{\lambda}$ values for a full coverage
   are not viable matches for known UV/O transients,
   even if the line-ratios match.

Previous calculations
   \citep{gaskell2014,guillochon2014}
   allowed a spatially extensive line-emitting medium,
   analogous to the BLR of an AGN,
   and employed the AGN-like MF1987 SED model.
This SED model is inconsistent with observed AGN SEDs
	\citep[][and citations thereafter]{laor1997}.
\cite{gaskell2014} and \citet{guillochon2014}
   find that the He-overbright state
   (e.g. with the line ratio
   $w\equiv I_{4686}/I_{6563}\approx4$
   observed in PS1$-$10jh)
   occurs naturally in specific photoionization conditions,
   without requiring helium-enriched gas.
\citet{guillochon2014} obtained time-dependent functions of
   $(n_{\textsc{h}},\Phi_{\textsc{h}})$
   from hydrodynamic simulations of TDE streams,
   and showed that these tracks cross
   a He-overbright region of the flux-density plane
   (at least when the column $N_{\textsc{h}}=10^{23}\,\mathrm{cm}^{-2}$
   exactly).
%The observed equivalent widths were not confronted directly.
%Here we follow a similar approach,
%   but we also consider the EWs.

\cite{roth2016}
   performed radiative transfer simulations
   of a TDE's opaque quasi-spherical outflow,
   with a prescribed density profile,
   resulting in line emission (at an assumed velocity width)
   escaping from layers near the continuum source's photosphere.
\HeII-overbright states with EW $\ga100$\AA\,
   (by inspection of their figures)
   appeared naturally in some configurations
   with outer radii $\le2\times10^{15}$cm.
Are these conditions special, or commonplace,
   and do they persist as the source evolves?
In their model,
   the line emission depends on the properties of the entire surface
   ($\Omega=1$)
   and layers beneath.
We might wonder whether the photosphere's compactness
   would lead to significant intra-day spectral variability.

In our paper,
   we consider line-emission from discrete external BLR-like clouds
   well outside the central photosphere
   and tidal radius.
In an approach that is initially similar to \cite{gaskell2014},
   we focus on a variety of more modern SED models:
   an AGN spectrum with a spectral slope $1.2$
   \citep{baskin2014a};
   and another AGN case with $2.0$
   (these correspond to the \emph{observed} range of slopes);
   a blackbody peaking at 4\,Ryd;
   another at $10^6$K;
   and a hotter $10^{7}$K blackbody
   (harder than the observed continuum).
For comparison with the previous works we
   also run models using the MF1987 SED.

For each SED, we run several cubes of calculations:
   exploring a rectangle of $(n_{\textsc{h}},\Phi_{\textsc{h}})$
   in steps of $0.1$\,dex,
   and saving depth profiles along the $N_{\textsc{h}}$ or spatial axis.
%For greater resolution, and to check consistency,
We generate cubes where the maximum column was set to
   $N_{\textsc{h}}=10^{23},10^{24},10^{25}~\mathrm{cm}^{-2}$,
   and then another cube of results where $N_{\textsc{h}}$ was adjusted
   for each $(n_\textsc{h},\Phi_\textsc{h})$
   so that electron scattering was almost optically thick
   ($\tau_\mathrm{es}=0.5$).
{\sc Cloudy}'s approximate representation of scattering
   becomes less accurate at greater optical depths.
We refer the interested reader to
   Appendix~\ref{appendix.cubes}
   for the technical details.% of the {\sc Cloudy} calculations.

Using these calculations,
   we densely map
   the range of possible values of H$\alpha$ to \HeII\, emission line ratio
   and \HeII\, equivalent widths
   obtainable when the luminous nuclear source
   is emitting steadily with a specified SED.
Composition is solar.
%   and photoionizing a distant cloud (of solar composition).
We record intensities from the cloud's illuminated face,
   as well as the total including the shadowed face too.
These computational surveys proceed in full generality,
   but we apply the results later (in post-processing)
   in the contexts of specific astronomical events.
Table~\ref{table.peaks}
   presents the maximum attainable \HeII/\Ha\, line ratio ($w_\dag$)
   from the illuminated side,
   for each SED model,
   the cloud parameters where this occurs,
   and the equivalent widths of five emission lines in these conditions
   (assuming covering factor, $\Omega=1$).

\begin{figure*}
\begin{centering}
\begin{tabular}{ccc}
\includegraphics[width=150mm]{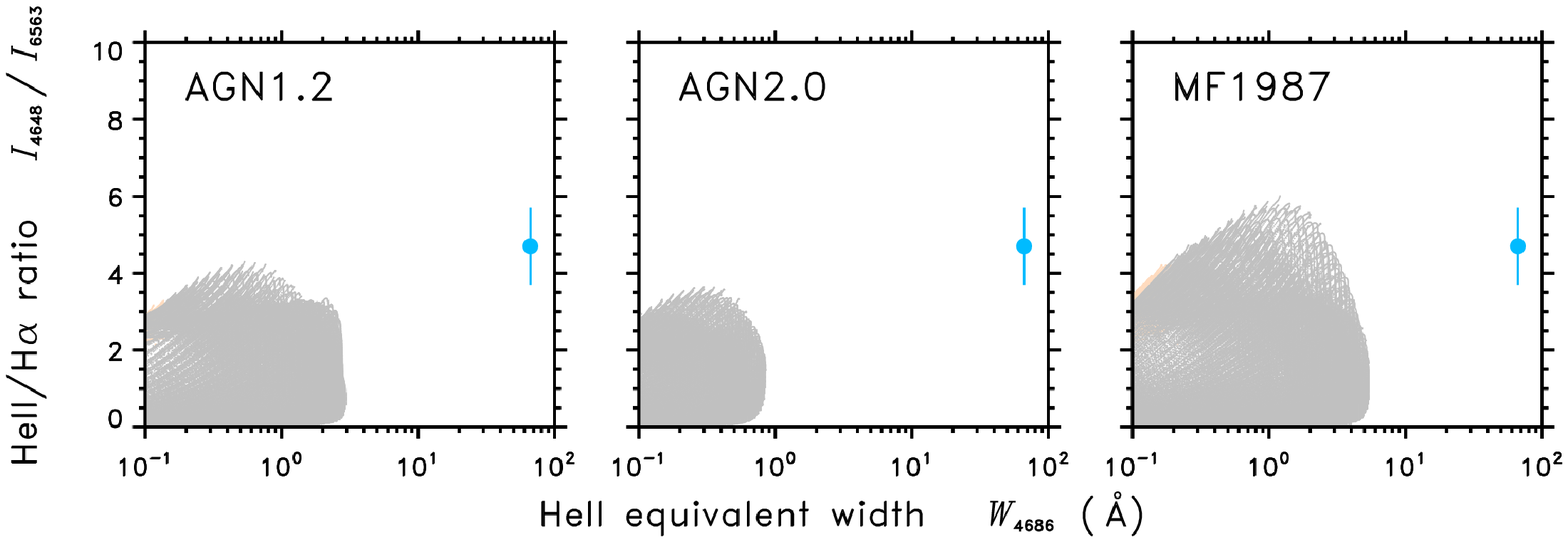}  &  & \tabularnewline
\includegraphics[width=150mm]{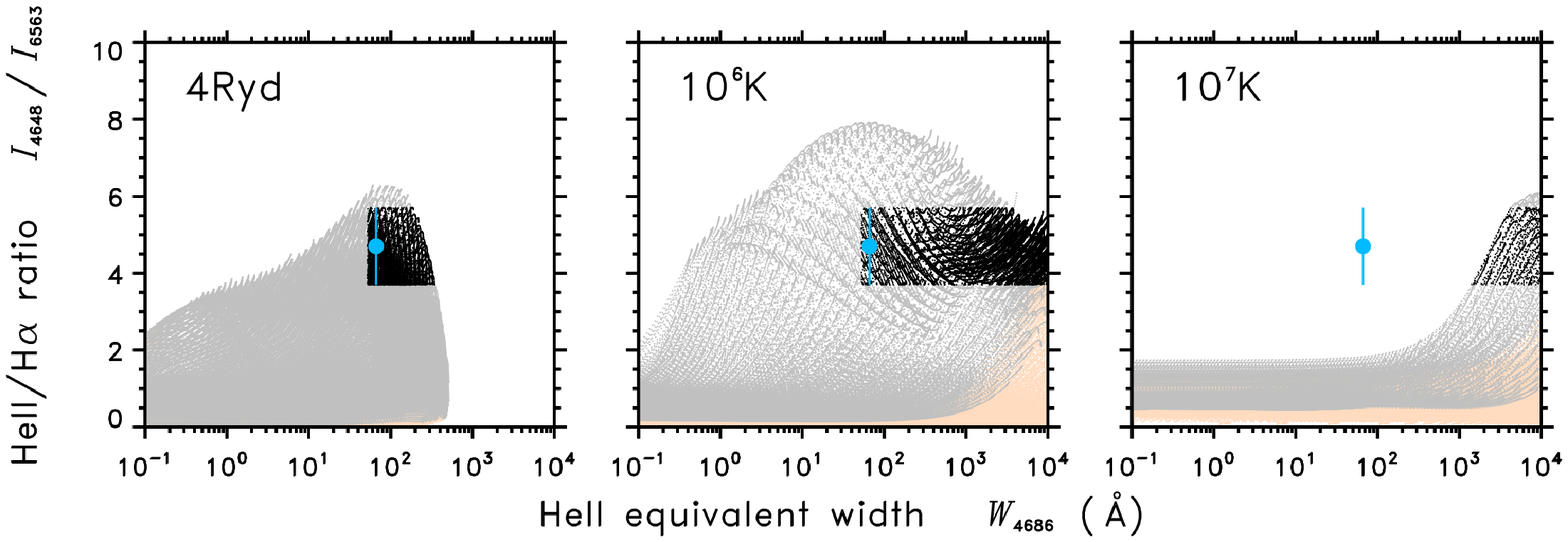}  &  & \tabularnewline
\end{tabular}
\par\end{centering}
\caption{The distribution of possibilities for
   the $I_{4686}/I_{6563}$ line ratio
   vs \HeII\, equivalent width,
   assuming a nominal covering factor $\Omega=0.1$.
The top panels assume an AGN-like source;
   the bottom panels assume a steady blackbody source.
The cyan datum shows the observed line ratio
   and \HeII\, equivalent width for PS1$-$10jh.
Black areas are compatible with observations,
   for the fiducial $\Omega$ or smaller.
Grey areas fail to match.
Orange areas are physically rejected
   since they would reside inside the photosphere.
The source luminosity is assumed to be time-independent,
   and was normalized to match the observed continuum
   around wavelength 4686\AA.
Note that a smaller choice of $\Omega$ corresponds
   to a proportional translation of dots
   leftwards parallel to the horizontal axis.
%Consistent models in this context would require
%    $\Omega>5.3\times10^{-2}$ for the 4Ryd model;
%    $\Omega>6/9\times10^{-4}$ for $10^{6}$K;
%    and $\Omega>5.5\times10^{-5}$ for $10^{7}$K models.
}
\label{fig.ew.ratio}
\end{figure*}

Fig.~\ref{fig.ew.ratio}
   shows the distribution of {\sc Cloudy} results
   for radiation from the illuminated cloud face when $\tau_\mathrm{es}\le0.5$,
   in regions of parameter-space near the peak \HeII/\Ha.
%, since {\sc Cloudy}'s approximate treatment of scattering
%   becomes imprecise at higher optical depths).
A dot represents every result pair
   $(I_{4686}/I_{6563},W_{4686})$
   obtained from each input condition
   $(N_{\textsc{h}},n_{\textsc{h}},\Phi_{\textsc{h}})$
   in the photoionization parameter cube.
Numerically,
   setting aside precautions against Compton-thick conditions,
   higher $N_\textsc{h}$
   enables slightly higher $w_\dagger$,
   with much the same range of equivalent widths
   (which is unsurprising because the bright side of an opaque body
   should dominate radiation from the shadoy parts).
For the sake of illustration,
   we assume a covering factor $\Omega=0.1$,
   but the results for another chosen $\Omega$
   are easily obtained by moving the points appropriately
   (e.g left, if $\Omega$ is reduced).
For a specific application,
   the observation of PS1-10jh is overplotted in cyan, where,
   using the results and plots of \citet{gezari2015},
   the \HeII\, line luminosity is
   $L_{4686}=(9\pm1)\times10^{40}\,\mathrm{erg}~\mathrm{s}^{-1}$;
   the blackbody flux of the observed continuum under the line was
   $F_{\mathrm{o}}\approx(1.7\pm0.3)\times10^{-17}\mathrm{erg}~\mathrm{cm}^{-2}~\mathrm{s}^{-1}~$\AA~$^{-1}$.
The fitted temperature ($T_{\mathrm{o}}\approx3\times10^{4}$K)
   implies a photospheric luminosity
   $L_{\mathrm{o}}\approx8.7\times10^{43}\,\mathrm{erg}~\mathrm{s}^{-1}$.
The \HeII\, line's equivalent width is
   $W_{4686}=66.4\pm13.9${\AA\,}.
The set of solutions that would be consistent with observations
   for smaller covering factors ($\Omega<0.1$ a free parameter)
   are plotted in black.
Grey dots cannot fit PS1-10jh.
Orange dots are unphysical cases that would hide {\em inside} the source
   (if it were approximately a spherical, opaque {\em photosphere} surface).

AGN-like spectra (top row of the Figure)
   produce line ratios compatible with PS1$-$10jh
   under a wide range of conditions,
   however,
   the \HeII\, equivalent widths are \emph{inconsistent} with \emph{any}
   AGN-like models,
   even for covering factor much larger than would be realistically expected.
(MF1987 might work in unrealistic extreme cloud models,
   with numerically unsafe $\tau_\mathrm{es}>0.5$,
   positions within 3~light~days of the source,
   and huge covering factors, $\Omega\ga0.8$ at 67\% confidence.)

Given the failure of these AGN models,
   we now consider other types of sources,
   in order to overcome the EW problem.
In particular,
   we consider a set of simple blackbody SED models.
For the 4Ryd, $10^{6}$K and $10^{7}$K blackbody models
   (bottom row of the Figure)
   there is an extensive range of plausible solutions
   that satisfy observational constraints on \emph{both} the line ratio and EW,
   either for the standard covering factor
   and even for smaller values ($\Omega\le0.1$).
The potential disadvantage of these hot sources is that they would be
   difficult to reconcile with the observed
   $T_{\mathrm{o}}$$\approx$$3\times10^{4}$K
   continuum
   (which remains steady at late times).
Nevertheless, in the following,
   we suggest a generic scenario
   which can accommodate these constraints
   when the luminous source has a varying SED and a light echo occurs.

%%%%%%%%%%%%%%%%%%%%%%%%%%%%%%%%%%%%%%%%%%%%%%%%%%%%%%%%%%%%%%%%%%%%%%%%
\begin{table*}
\caption{%
\textsc{Cloudy} model conditions for a given source SED that give
   the peak ratio of \HeII(4686\AA\,) to H$\alpha$(6563\AA\,)
   emission line intensities ($w_{\dag}\equiv I_{4686}/I_{6563}$)
   from the illuminated face of the cloud
   (and limited to cases with low electron scattering optical depths $\tau_\mathrm{es}<0.5$).
At the peak,
   the hydrogen column density,
   number density, and ionising flux
   are labelled
$(N_{\textsc{h}},n_{\mathrm{h}},\Phi_{\textsc{h}})=(N_{\dag},n_{\dag},\Phi_{\dag})$.
The last five columns give the equivalent widths of emission lines,
   calculated for full coverage ($\Omega$=$1$)
   ($W_{\lambda}$) for \HeII, H$\alpha$,
   H$\beta$,
   H$\gamma$,
   \MgII\, and \C4\,
   emission lines
   (and `---' denotes an absent line).
By rows, the illuminating SED was
   Mathews \& Ferland~(1987);
   AGN with slope $1.2$;
   AGN with slope $2.0$;
   4Ryd blackbody;
   $10^{6}$K blackbody;
   $10^{7}$K blackbody.
}
\label{table.peaks}{\footnotesize{} \centering{}$\begin{array}{lr@{.}lr@{.}lr@{.}lr@{.}lr@{.}lr@{.}lr@{.}lr@{.}lr@{.}lr@{.}lr@{.}l}
\hline \\
\mbox{SED}
& \multicolumn{2}{c}{\log{\bar{\epsilon}}}
& \multicolumn{2}{c}{\log N_{\dag}}
& \multicolumn{2}{c}{\log n_{\dag}}
& \multicolumn{2}{c}{\log\Phi_{\dag}}
& \multicolumn{2}{c}{~w_{\dag}~~~}
%& \multicolumn{2}{c}{\log\frac{{W_{_{4686}}}}{{\Omega}}}
%& \multicolumn{2}{c}{\log\frac{{W_{_{6563}}}}{{\Omega}}}
%& \multicolumn{2}{c}{\log\frac{{W_{_{4340}}}}{{\Omega}}}
%& \multicolumn{2}{c}{\log\frac{{W_{_{1549}}}}{{\Omega}}}
%& \multicolumn{2}{c}{\log\frac{{W_{_{2798}}}}{{\Omega}}}\\
& \multicolumn{2}{c}{\log{W_{_{4686}}}}
& \multicolumn{2}{c}{\log{W_{_{6563}}}}
& \multicolumn{2}{c}{\log{W_{_{4861}}}}
& \multicolumn{2}{c}{\log{W_{_{4340}}}}
& \multicolumn{2}{c}{\log{W_{_{2798}}}}
& \multicolumn{2}{c}{\log{W_{_{1549}}}}
\\
\\
%&\multicolumn{2}{c}{(\mathrm{erg})}
&\multicolumn{2}{c}{(\mathrm{eV})}
&\multicolumn{2}{c}{(\mathrm{cm}^{-2})}
&\multicolumn{2}{c}{(\mathrm{cm}^{-3})}
&\multicolumn{2}{c}{(\mathrm{cm}^{-2}\,\mathrm{s}^{-1})}
&\multicolumn{2}{c}{~}
&\multicolumn{2}{c}{(\mbox{\AA})}
&\multicolumn{2}{c}{(\mbox{\AA})}
&\multicolumn{2}{c}{(\mbox{\AA})}
&\multicolumn{2}{c}{(\mbox{\AA})}
&\multicolumn{2}{c}{(\mbox{\AA})}
&\multicolumn{2}{c}{(\mbox{\AA})}\\
\\
\hline \\
\mbox{MF1987}
%& -9 & 914%-9.91440
& 1&881%-9.91440
& 23&8
& 11&7
&22&7
&5&82
&1&08
&0&456
&0&127
&0&0221
&-0&0742
&1&46
\\
\\
\mbox{AGN1.2}
%& -9 & 677%-9.67660
&2&118
&23&8
&11&8
&22&7
&4&26
&0&732
&0&306
&-0&0175
&-0&164
&-0&403
&1&20
\\
\\
\mbox{AGN2.0}
%& -9 & 730%-9.72950
&2&065
&23&8
&11&3
&22&5
&3&62
&0&473
&0&116
&-0&151
&-0&268
&-0&444
&0&962
\\
\\
\mbox{4Ryd}
%& -10 & 223%-10.2228
&1&572
&23&8
&12&0
&23&0
&6&21
&2&79 
&2&56
&1&81
&1&52 
&0&957
&1&51
\\
\\
10^{6}\mathrm{K}
%& -9 & 426%-9.42620
&2&369
&23&5
&13&0
&25&2
&7&90
&2&89
&2&56
&1&95
&1&71
&\multicolumn{2}{c}{\mbox{---}}
&-7&13
\\
\\
10^{7}\mathrm{K}
%& -8 & 428%-8.42830
&3&367
&23&4
&13&6
&24&9
&5&96
&5&28
&5&02
&4&28
&3&90
&\multicolumn{2}{c}{\mbox{---}}
&-2&20
\\
\\
\hline
%%%%%%%%%%%%%%%%%%
%%%%%%%%%%%%%%%%%%
\end{array}$ }
\end{table*}

%%%%%%%%%%%%%%%%%%%%%%%%%%%%%%%%%%%%%%%%%%%%%%%%%%%%%%%%%%%%%%%%%%%%%%%%

%%%%%%%%%%%%%%%%%%%%%%%%%%%%%%%%%%%%%%%%%%%%%%%%%%%%%%%%%%%%%%%%%%%%%%%%

\subsection{A novel scenario: Light-echo photoionization of nuclear clouds}
\label{s.echo}

A hard ionizing continuum that peaks at $h\nu\sim4$\,Ryd
   and a high density gas $n_\textsc{h}\sim10^{12}$~${\rm cm}^{-3}$
   is required to produce
   both a strong \HeII\,$\lambda$4686 emission-line
   and a large \HeII/\Ha\ line ratio.
Such densities are observed to occur in the circumnuclear media of AGN
   \citep[e.g.][]{netzer2013}.
The hard continuum induces high gas temperatures,
   which makes the \textsc{${\rm He}^{++}$}
   recombination coefficient larger relative to ${\rm H}{}^{+}$;
   and the high gas density makes the gas
   optically thick to H$\alpha$ (see \citealt{strubbe2015}).
However,
   observations in the optical of \HeII-bright TDE candidates
   are consistent with a black-body continuum source
   with a temperature $\Tbb$\ of a few $10^{4}$\,K only
   (e.g.\ \citealt{gezari2012}).
Two scenarios can alleviate this discrepancy.
In the first scenario,
   two continuum sources exist concurrently:
   a `cold' source which is observed in the optical;
   and another `hot' source (e.g.\ a black body with $k\Tbb\sim4$~Ryd)
   which ionizes the line-emitting gas.
As the `hot' source cools,
   the ionizing continuum becomes softer,
   and the object transforms from a \HeII- to \Ha-bright.
In the second scenario,
   there is a single continuum source only.
The source cools from a very high $T$ (e.g.\ $\Tbb\sim10^{9}$\,K)
   to the observed $\Tbb\sim10^{4}$\,K.
During a later phase,
   the source temperature might settle at this level or vary slowly,
   according to observations.
%   in agreement with observations
%   and more detailed outflow models,
%	\citep[e.g.][]{metzger2016}.
The line-emitting gas resides
   at a distance of the order of several to a few tens of light-days
   from the continuum source
   and from our line-of-sight to the source.
The observed line emission is delayed w.r.t.\ the observed continuum emission.
Thus, in \HeII-bright objects, we measure line emission
   which corresponds to a continuum source with $k\Tbb\sim4$~Ryd (or more),
   when the observed continuum source itself has already cooled to $\Tbb\sim10^{4}$\,K.
The fading system transforms from \HeII-bright
   to \HeII+\Ha,
   and then to \Ha-bright
   when the delayed
   line emission corresponds to
   an ionizing continuum with $\Tbb\la10^{5}$\,K.
This light-echo scenario has a few consequential implications,
   which are described below.

To observe the ionizing source during its brief, hot, initial phase,
   TDE candidates must be found in X-rays {\em before}
   they are optically detectible.
By the time an emission-line light echo appears,
   the central source has already cooled to
   $\Tbb$$\sim$$3\times10^4$K.
%\footnote{%
%In our fiducial model outlined below,
%   e.g. eqn~(\ref{eq:Ropt}),
%   if the early ionizing source was near $10^6$K for 10\,days,
%   it was previously around $10^8$K for only 0.1\,day.
%}
Hotter source conditions were briefer and earlier.
From this view,
   it is unremarkable that
   many UV/O-selected candidates lack X-ray dectections
   (e.g. limits in Table~S3 of \citealt{vanvelzen2016a}).
A few candidates do possess a relatively constant X-ray counterpart
	\citep{holoien2016a,holoien2016b}.
These X-rays may be produced by a corona,
   by a jet,
   by reprocessing,
   or by shocked ISM gas.
Our models omit this minor X-ray component,
   as its contribution to the flux of ionizing photons is negligible.

\subsubsection{The continuum source}
\label{subsec:echo-model}

The continuum during the broad line phases of the UV/O candidates
   indicates temperatures of
   $10^{4}\,\mathrm{K}\la T\la10^{5}\,\mathrm{K}$.
In time-independent modeling,
   a luminous source at such temperatures seems
   unable to photoionize clouds to produce \HeII\, and H$\alpha$
   equivalent widths in the observed range.
Hotter sources could induce emission lines
   with the observed equivalent widths,
   but require covering factors orders of magnitude smaller.
The lax constraints on $\Omega$ are favourable,
   but the high-temperature continuum is inconsistent with observation.
This leads us to the possibility of a time-dependent source,
   which might provide the best features
   of both the low- and high-temperature SEDs.

Firstly suppose that the source was a sudden energetic detonation
   in a (possibly inhomogenous) gaseous medium,
   at a super-Eddington rate.
It might have been a tidal disruption event,
   a large-amplitude AGN flare,
   a peculiar kind of superluminous supernova,
   or any unrecognised new class of explosion
   with sufficient energy and brevity.
Regardless of what the specific injection mechanism might have been,
   the affected gas will expand as an opaque,
   radiation-dominated wind or blast-wave.
If the injection was abrupt,
   then the photosphere expands adiabatically
   with a temperature vs radius relation of $T\propto1/R$.
If there was continuous energy injection,
   due to the power of a hiddenl accreting source,
   then different relations govern the observable light curve.

Next, in order to set spatial and temporal scales,
   suppose that the hot initial `flash' phase
   lasted less than a few days.
The opaque photosphere,
   dominated internally by radiation pressure, expands and cools.
As the initial flash of hard radiation propagates outwards,
   it photoionizes circumnuclear clouds or dense debris
   (perhaps within several light-days of the center).
Depending on distance and the evolving luminosity,
   some clouds emit lines in a He-over-bright state,
   including the strong-EW mode in our $10^{7}$K calculations.
Light from the photosphere reaches telescopes at Earth directly,
   but line emission from the clouds is delayed
   by the light-travel time to the cloud
   ($R_\mathrm{c}/c$, where $c$ is the speed of light).
Thus the He-overbright clouds' emission
   is superimposed as a light-echo on
   a later, expanded and cooled view of the nucleus photosphere.
Obtaining the observed equivalent widths ($W$$\sim$60\AA)
   and underlying continuum temperature ($10^{4}$K to $10^{5}$K)
   implies constraints on
   the photospheric temperature and luminosity evolution,
   the covering factor and location of the cloud.
The total mass of clouds is derivable
   from the column density ($N_{\textsc{h}}$) assumed in each
   photoionization model.
%   the total mass of clouds is derivable.

In the light-echo model,
   we suppose that
   the continuum source is an optically thick
   photosphere of the flare or wind
   expanding according to power-laws in time: 
\begin{equation}
	T\propto t^{a}
\label{eq.expansion.T}
\end{equation}
\begin{equation}
	L\propto t^{b}\propto T^{b/a}\ .
\label{eq.expansion.L}
\end{equation}
For a blackbody emitting photosphere, the radius 
\begin{equation}
	R%\propto\sqrt{L/T^{4}}
	\propto T^{x}
\label{eq.expansion.R}
\end{equation}
   has index $x=[(b/a)-4]/2$.
For an explosion expanding adiabatically at constant velocity,
    $a=-1$ and $b=-2$.
For a relativistic event driven by a power source
   ($P\propto t^{q}$),
   we might have
   $a=(q-2)/4(q+2)$,
   $b=(2+3q)/(2+q)$
	\citep{blandford1976,cohen1999}.
For some models of the rising phase of a TDE,
   the indices $a=-7/36$, $b=11/9$
	\citep{strubbe2009,lodato2009}.
In the recent \cite{metzger2016} model of
   a TDE's accretion-powered massive outflow,
   the temperature initially declines over tens of days,
   with indices $a=-11/12$ and $b=-5/3$,
   which approximates adiabatic behaviour well enough.
After $\sim$30\,days of the dimming and cooling,
   their model's photospheric temperature steadies (or rises slowly)
   consistently with late-time $\Tbb$ in observed UV/O flares.
Our calculations omit this stage,
   because we're mostly interested in the photoionization echo of the initial flash.
The simpler adiabatic model suffices for now.
%This paper assumes the simpler adiabatic model.

\subsubsection{The line emitting source}

\begin{figure*}
\begin{centering}
\begin{tabular}{ccc}
\includegraphics[width=150mm]{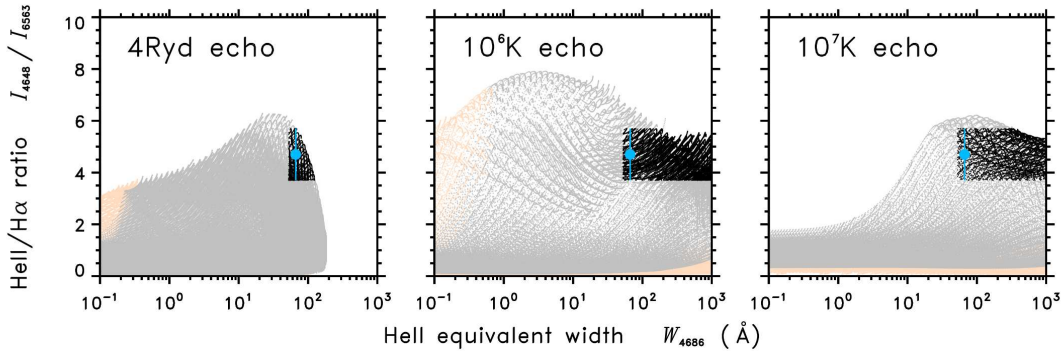}  &  & \tabularnewline
\end{tabular}
\par\end{centering}
\caption{%
The distribution of line ratios vs 
   %\HeII\, line luminosity (top row)
   %and 
   \HeII\, equivalent widths
  %(bottom row)
   in with continuum evolution in blackbody light-echo models.
The assumed photosphere expands adiabatically at constant velocity,
   illuminating a cloud with covering factor $\Omega=0.1$.
As in Fig.~\ref{fig.ew.ratio},
   black dots are solutions consistent
   with observations (when $\Omega\le0.1$);
   the PS1$-$10jh data are the cyan dot;
   unphysical solutions are orange;
   observationally incompatible points are grey.
%Note that a smaller choice of $\Omega$ corresponds
%   to a proportional translation of dots
%   leftwards parallel to the horizontal axis.
%, i.e. consistent models in this
%context would require $\Omega>0.027$ for the 4Ryd model; $\Omega>7.6\times10^{-4}$
%for $10^{6}$ K; and $\Omega>8.6\times10^{5}$ for $10^{7}$K.
}
\label{fig.lum.echo} 
\end{figure*}

We explore a broad realm of possibilities across the
   $(N_{\textsc{h}},n_{\textsc{h}},\Phi_{\textsc{h}})$
   space,
   while correcting the model continuum
   for the time-lagged expansion of the photosphere.
The \HeII-overbright state is a limited occurrence
   (e.g. Fig.~\ref{fig.ratio.echo} in Appendix~\ref{appendix.cubes}).
It occurs throughout a sausage of the parameter cube,
   spaning a considerable range in $\Phi_\textsc{h}$,
   implying that \HeII-brightness can continue a while as the flare fades.
Cloud density may be more constraining.
%(The diversity of temporal light-curves is best left for future work.)
Here however, we illustrate the instantaneous possibilities
   in terms of predicted observables.
Fig.~\ref{fig.lum.echo}
   %(top row)
   shows the \HeII/H$\alpha$ ratio vs \HeII,
   in a light-echo model with PS1-10jh parameters,
   and assuming a constant-velocity adiabatic expansion of the central source.
%Bottom panels of Fig.~\ref{fig.ew.ratio}
%   are corresponding plots where the horizontal axis
%   is the \HeII\, equivalent width.
Black dots are compatible with observations,
   for our fiducial covering factor or a smaller value
   ($\Omega\le0.1$).
In addition to rejecting models where the cloud
   would reside underneath the photosphere,
   we also omit cases where
   $R_{\mathrm{c}}$ would exceed 60~light~days,
   as this would be beyond the characteristic scale implied by
   the 22~day rise time of the observed light-curve.
The light-echo model reduces the equivalent widths
   for the hotter blackbody SED scenarios.
When the covering factor is a free parameter,
   there are still many acceptable fits that have small $\Omega$.
%When $\Omega$ is treated as a free parameter,
The limits from PS1-10jh
   would require $\Omega>0.053$ for the 4Ryd model;
   $\Omega>6.9\times10^{-4}$ for $10^{6}$\,K;
   and $\Omega>5.5\times10^{-5}$ for $10^{7}$\,K.

%It is worth noting that
Some observed TDE candidates
   changed from \HeII- or \Ha-bright states
   to the intermediate case with both \HeII and \Ha\,
   (Appendix~\ref{s.candidates}).
None has yet been caught transitioning fully
   from \HeII\, to \Ha\, (or vice versa).
Theoretically, we do not expect {\em every TDE}
   to reach the \HeII-bright condition,
   since it occurs in only part of the
   $(N_{\textsc{h}},n_{\textsc{h}},\Phi_{\textsc{h}})$
   space:
   the location and gas density prerequisites are non-trivial
   (see Fig.~\ref{fig.ratio.echo} in the appendices).
Beyond this zone,
   most flares would begin as \Ha-bright and end as \Ha-bright
   (or lose their broad lines).
Catching a complete transformation
   requires a monitoring campaign with $\sim$1\,day cadence,
   spanning many weeks.
It is also possible that a seemingly intermediate \HeII+\Ha\, state
   occurs through superposition,
   when there are clouds at different locations and densities,
   which can be \HeII-bright or \Ha-bright individually.
(ASASSN-14ae's change from \Ha-bright to intermediate state
   might be such a case.)

The emission line broadening results from gas motions.
% which are transverse to the impinging ionizing radiation.
\citet{strubbe2015}
   pointed out that a velocity gradient
   which is parallel to the radiation reduces the gas opacity.
Transparency increases the \Ha\, output,
   and thus reduces the \HeII/\Ha\, line ratio.
We carried out photoionization
   calculations with `turbulent' velocity
   which corresponds to the observed line width in PS1-10jh
   ($\sigma_\mathrm{turb}=5400$~km~s$^{-1}$).
These calculations indeed result in \HeII/\Ha$<1$.
Subsection~\ref{s.puzzles}
   continues to discuss the velocity-opacity problem
   that could suppress the \HeII-overbright state.

\begin{figure*}
\begin{centering}
\begin{tabular}{ccc}
\includegraphics[width=150mm]{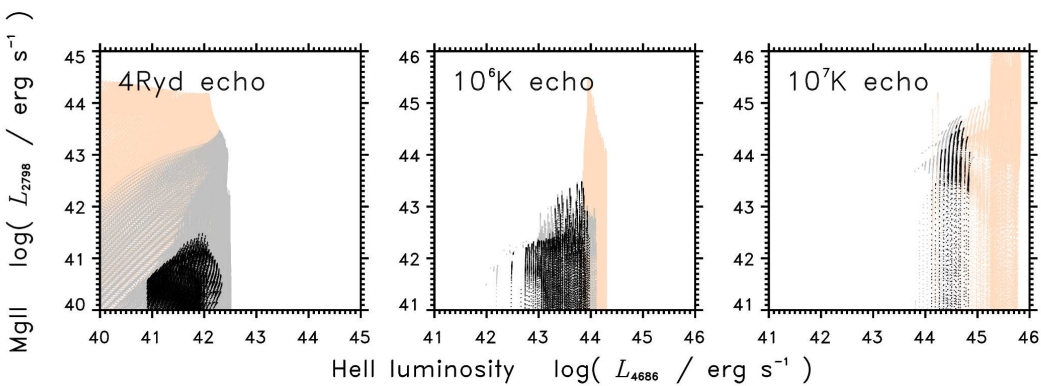}  &  & \tabularnewline
\end{tabular}
\par\end{centering}
\caption{%
The distributions of predicted
   \MgII\, line luminosity (vertical axes)
%   \MgII\,, \C4\, and H$\beta$ line luminosities (vertical axes)
   vs \HeII\, (horizontal axes)
   from blackbody light-echo models.
Solutions are selectively coloured
   (as in Fig.\ref{fig.ew.ratio})
   for their consistency with H$\alpha$ and \HeII\, observed from PS1$-$10jh.
%The shown \C4 and \MgII\, properties
%   are unforced, independent predictions.
}
\label{fig.ew.lines} 
\end{figure*}

Fig.~\ref{fig.ew.lines}
   shows the predicted luminosities of
   \MgII\, emission lines
%   other emission lines
   from the illuminated face of the cloud
   in comparison to the \HeII\, luminosity.
   %\C4 and \MgII\, 
%Specific results, coinciding with the \HeII/\Ha\, peak,
%   appear in Table~\ref{table.peaks}.
%Each of these EW values is calculated relative to
%   the continuum seen directly under the respective line,
   (in the echo model fiducially assuming $\Omega=0.1$).
The selection of credible solutions for PS1$-$10jh
   (marked black)
   relies only on the \Ha\, and \HeII\, emission properties,
   as in Fig.~\ref{fig.lum.echo}.
Among the set of solutions that are compatible with PS1-10jh,
   the \MgII\, luminosities and 
   can in some cases be comparable to the \HeII\, value
  (up to a few times brighter),
   but other times undetectable.
Under the same model selections,
   \C4 luminosities are distributed similarly:
   undetectable in many cases,
   but sometimes exceeding the \HeII\, luminosity
   (by up to $\sim$1\,dex).
%H$\beta$ correlates with the \HeII\, luminosities.

\begin{figure*}
\begin{centering}
\begin{tabular}{ccc}
\includegraphics[width=142mm]{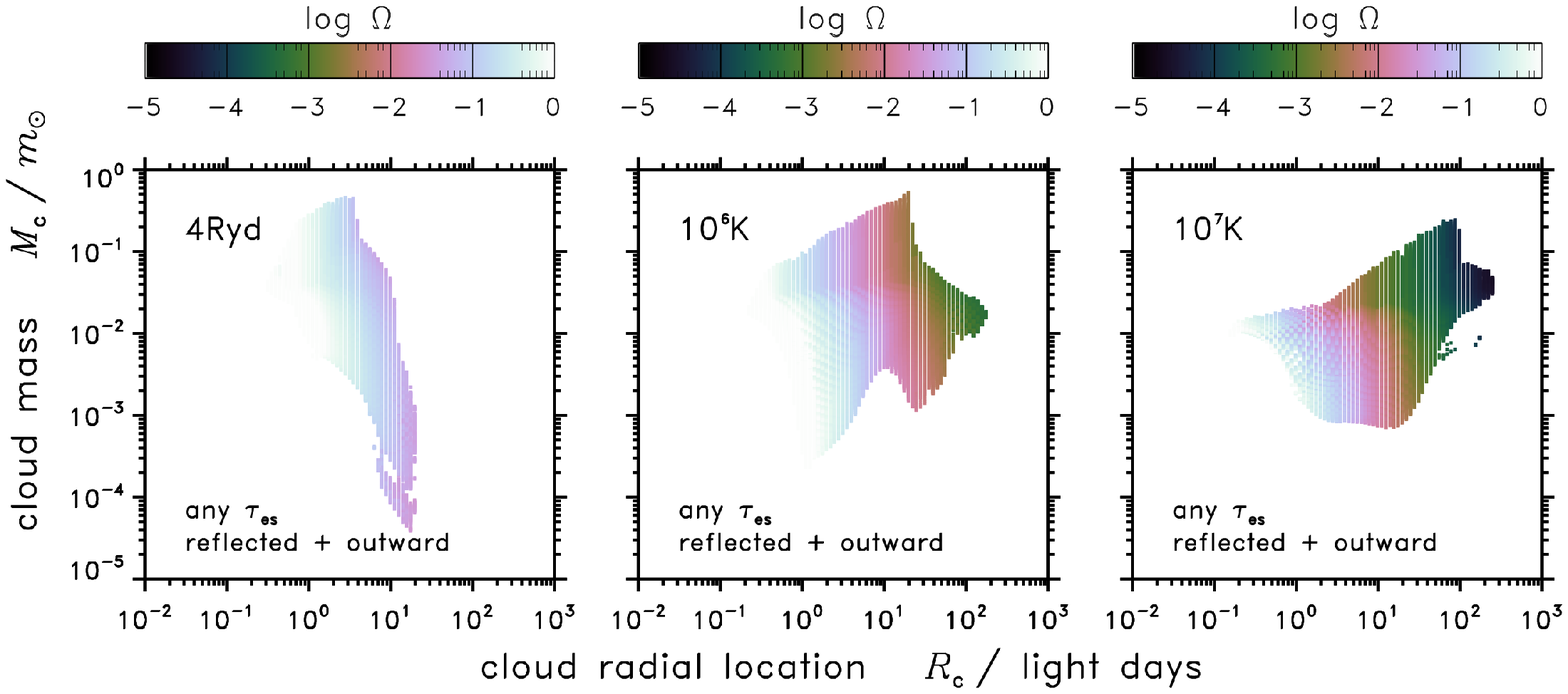}  &  & \tabularnewline
\includegraphics[width=142mm]{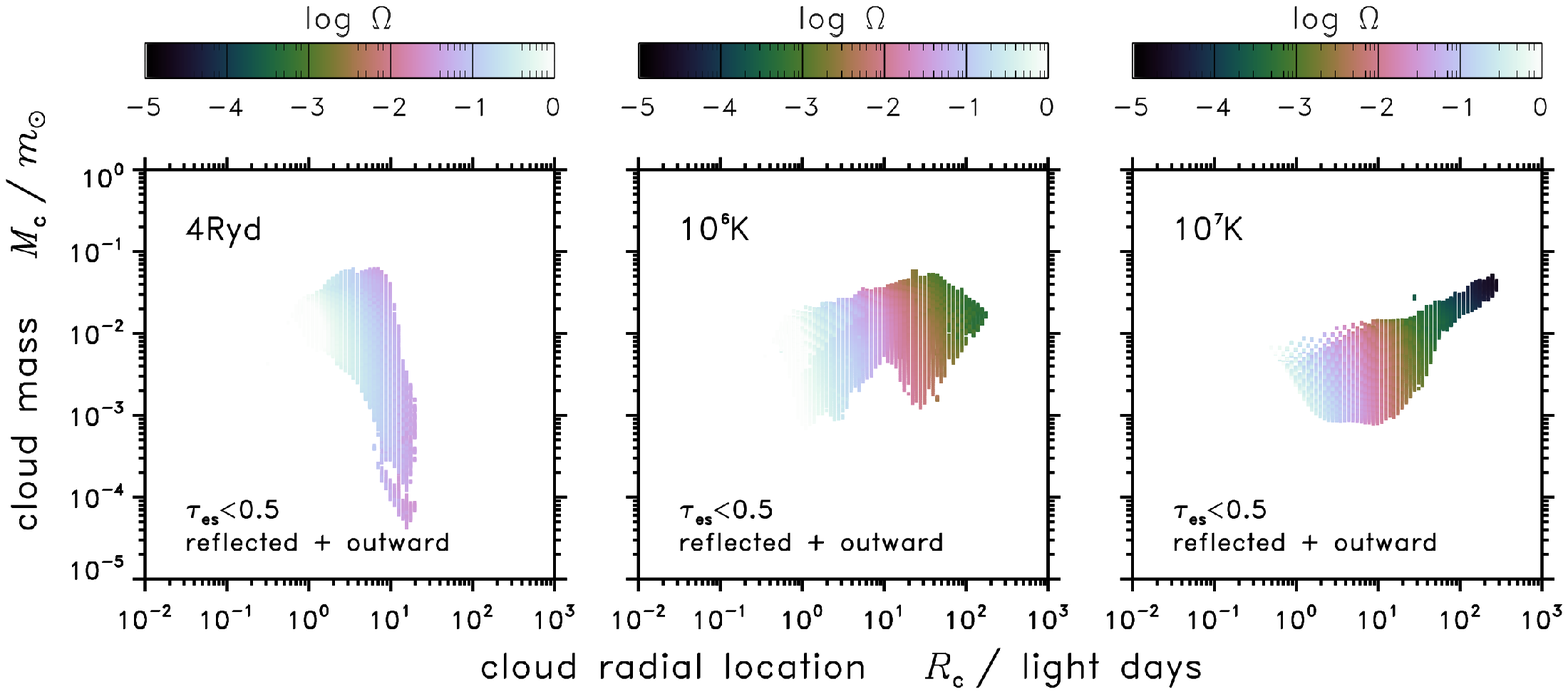}  &  & \tabularnewline
\includegraphics[width=142mm]{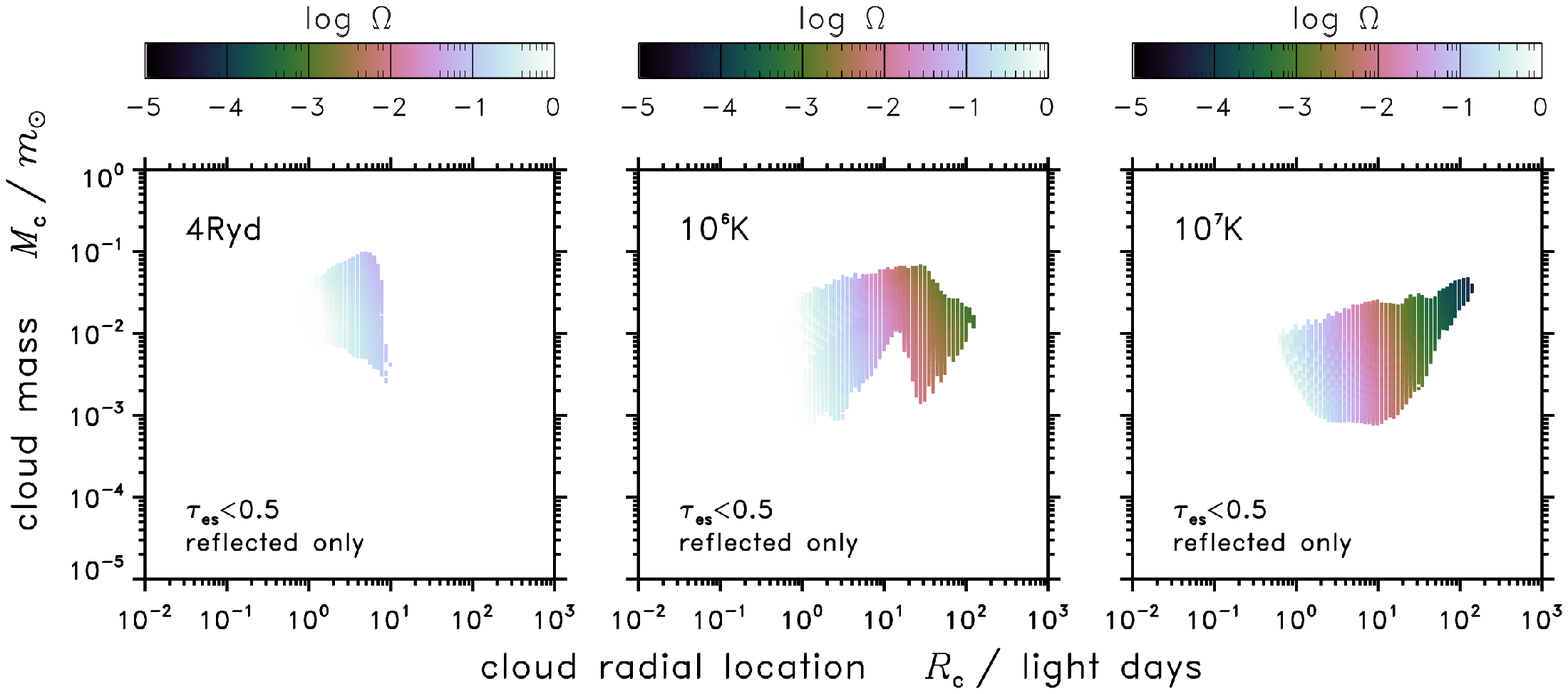}  &  & \tabularnewline
\end{tabular}
\par\end{centering}
\caption{%  
Distributions of possible cloud masses ($M_{\mathrm{c}}$)
   and locations ($R_{\mathrm{c}}$)
   that are compatible with PS1$-$10jh observations,
   for some covering factor ($\Omega$, now a fitted parameter).
Columns show three blackbody light-echo models.
Colours indicate the cloud covering factor.
Only the observationally consistent solutions
   (black in Fig.~\ref{fig.ew.ratio}) are shown.
The implied cloud mass never exceeds a subsolar value.
}
\label{fig.clouds} 
\end{figure*}

Fig.~\ref{fig.clouds} plots the distribution of 
   cloud mass and position
   $(M_{\mathrm{c}},R_{\mathrm{c}})$
   for the realistic light-echo model fits to PS1-10jh.
Each dot corresponds to one of the successful solutions that were
   marked black in Fig.~\ref{fig.lum.echo}.
Now the colour scale indicates $\Omega$
   (as a fitted parameter).
Many likely locations are at light-days or tens of light-days
   from the source,
   which agrees with the light-echo expectations.
A few \HeII-bright solutions exist for clouds
   located at $\ga10^2$ light days out.
The three rows of this Figure show different constraints on the photoionization models.
The top row includes the total line emission from both faces of the cloud,
   and considers any column density up to $N_\textsc{h}=10^{25}~\mathrm{cm}^{-2}$.
The middle row shows the total line emission,
   but we omit the opaque clouds with $\tau_\mathrm{es}>0.5$,
  (where {\sc Cloudy}'s approximate treatment of scattering is less realistic).
This excludes the higher-$M_{\mathrm{c}}$ solutions.
In the bottom row,
   we show only the radiation from the illuminated face of the cloud,
   which best describes the light-echo scenario.
Some of the lowest-$M_{\mathrm{c}}$ solutions vanish
   when the source spectrum is 4Ryd;
   but the limits for $10^6$K and $10^7$K SEDs aren't changed significantly.
In all cases investigated, we find the following: 
\begin{itemize}
%\item
%Although the steady-spectrum AGN-like models
%   can provide the observed line ratio in many cases,
%   they are unsuccessful regarding the \HeII\, equivalent width.
%The few exceptions, using the MF1987 SED,
%   require extreme covering factors $\Omega\ga0.82$.
\item
The median cloud masses are subsolar.
Taken without bias across the whole parameter space,
   the masses tend to be planetary in scale.
Masses as high as $\sim$$0.1m_{\odot}$ are possible,
   but require fine-tuning. 
\item
Cloud positions at $\sim$10~light days occur naturally,
   without premeditation or fine-tuning.
The implied covering factors are mostly moderate
   ($10^{-3}\le\Omega\le10^{-1}$),
   which retrospectively justifies our fiducial assumptions,
   and supports the light echo scenario. 
\item
At the favoured cloud locations of $\sim$10 light days,
   the orbital velocities would be low enough
   to solve the velocity problem of the UV/O optical candidates
   \citep[e.g. discussed by][]{arcavi2014}.
The emission line widths are unrelated to the inner nucleus
   where the tidal disruption or luminous flare occurs
   (see Appendix~\ref{s.scaling}).
They may simply characterize the motions of
   pre-existing clouds in the impoverished BLR
   of a previously dormant, declining or erratic weak AGN. 
\end{itemize}

Interestingly,
   several recent papers have reported infra-red light-echo effects
   from dusty media around TDE candidates.
\cite{jiang2016} and \cite{vanvelzen2016b}
   infer clouds at $\sim$$0.1$pc from the nucleus.
\cite{dou2016}
   find IR light echo effects over longer timescales,
   implying more clouds a few light years farther out.
Since circumnuclear material was present over this range of radii,
   it is not unlikely that there is also gas
   at even smaller scales,
   i.e. a few 10\,ld
   (consistent with clouds in our echo photoionization models).
These inner clouds would be IR-invisible
   due to sublimation of their entrained dust.
   %which is consistent with the cloud locations in our photoionization models.
The apparent persistence of some broad emission lines
   for $\ga10^2$\,days
	\citep[e.g. ASASSN-14ae][]{holoien2014}
%   \HeII\, emission at 250\,days  in PS1-10jh
%	\citep{gezari2015}
   might also be an echo effect,
   in reflections off dust at large distances,
   and this echo could be prolonged by the convex spatial distribution of reflectors
   \citep[as in][]{dou2017}.
The $9000\kms$ widths of these later delayed emission lines
   exceeds keplerian velocities of the $0.1$pc zone,
   but might be the echo of conditions near $\sim$10\,ld.

%%%%%%%%%%%%%%%%%%%%%%%%%%%%%%%%%%%%%%%%%%%%%%%%%%%%%%%%%%%%%%%%%%%%%%%%%%%%%%

%%%%%%%%%%%%%%%%%%%%%%%%%%%%%%%%%%%%%%%%%%%%%%%%%%%%%%%%%%%%%%%%%%%%%%%%

%%%%%%%%%%%%%%%%%%%%%%%%%%%%%%%%%%%%%%%%%%%%%%%%%%%%%%%%%%%%%%%%%%%%%%%%

%%%%%%%%%%%%%%%%%%%%%%%%%%%%%%%%%%%%%%%%%%%%%%%%%%%%%%%%%%%%%%%%%%%%%%%%%%%%%%

\section{DISCUSSION}

As we discussed above,
   any acceptable explanation for UV/O TDE-candidates should be able to accommodate
   both the observed line ratio \emph{and} the EW constraints.
%\emph{None}
%None of the previously suggested,
%   steadily illuminated AGN-like models stands to the test.
Previously suggested,
   steadily illuminated AGN-like models struggle with this test.
\cite{roth2016}
   propose an internal, radiative transfer model
   in which plausible line emission
   can in some cases
   emerge from the luminous object's own photospheric structure.
We reassert a modified external model,
   in which the line emission comes from
   the photoionization light-echo affecting remote BLR-like clouds,
   tens of light days away.
Independent of the origin of the TDE candidates,
   our echo scenario could reproduce the observations,
   for various assumed types of transients.
Nevertheless,
   it can constrain details of the explosive origin
   (within a given theory).
   %it can have implications for the explosive origin.
   %and might relate it to some types of supernovae.
In the following subsections,
   we first discuss direct implications of the light-echo,
   then overview some of the difficulties with the TDE scenario
   for the observed UV/O candidates.
%for these transient events.
We consider possible alternatives to TDE origins,
   including nuclear SNe
   %(in the light-echo model),
   or intermittency in subluminous/starved AGNs.

\subsection{Implications of the light-echo model}

\subsubsection{Implications for the continuum source}

To check several key qualitative properties of the system
   (to order of magnitude)
   let us assume a single continuum source
%Assuming a single continuum source
   that cools from $\Tbb\sim10^{9}$ to $10^{4}$\,K
   and photoionizes a distant, physically passive gas cloud.
%   enables evaluation of several key properties.
%Our goal here is to provide qualitative order-of-magnitude estimates,
%   rather than quantitative calculations,
%   since the latter require extensive numerical modeling
%   which is beyond the scope of this paper.
We make two additional simplifying assumptions.
Firstly, the continuum source is a spherical black body,
   which cools through adiabatic radial expansion
   with an adiabatic index of $\gamma=4/3$
   (i.e.\ dominated by radiation pressure).
Secondly, all relevant time scales are of the order of
   the light travel time from the continuum source to the line emitting gas,
   i.e.\ $\tau_\mathrm{c}\sim R_\mathrm{c}/c$,
   where $R_\mathrm{c}$ is the distance between the source and the cloud,
   and $c$ is the speed of light.
The first approximation is a rough one,
   as gas with $T\leq10^{6}$\,K can cool through line emission,
   which breaks the adiabatic approximation
   and slows the expansion.

The flux of H-ionizing photons ($h\nu\geq1$~Ryd)
   emitted by a black-body with a temperature $T_\textsc{bb}$
   is approximately
\begin{equation}
\Phi\simeq
\frac{4\upi k^{3}T_\textsc{bb}^{3}}{h^{3}c^{2}}\zeta(3)
	=1.5\times10^{29}\,T_{\textsc{bb},6}^{3}\mbox{~cm\ensuremath{^{-2}}~s\ensuremath{^{-1}}}
\ ,
\label{eq:phi_for_TBB}
\end{equation}
   where $k$ is the Boltzmann constant,
   $h$ is Planck's constant,
   $\zeta$ is the Riemann zeta function {[}$\zeta(3)\simeq1.2${]}
   and $T_\textsc{bb}=10^{6}\,T_{6}$\,K.\footnote{%
	Although the discussion focuses on
	line emission of \HeII\ which requires photons with $h\nu\geq4$~Ryd,
	we include photons already from $h\nu\geq1$~Ryd in $\Phi$,
	as this is the standard definition
	of $\Phi$ in photoionization codes (e.g.\ {\sc Cloudy}).}
The expression is correct within 2\% and 30\%
   for $T_{6}\geq0.5$ and 0.1, respectively.
There exists an optimal source $T_\textsc{bb}$ ($T_{\mathrm{opt}}$)
   and corresponding optimal $\Phi$
   at the illuminated face of the distant cloud ($\Phi_{\mathrm{opt}}$)
   which produce maximal \HeII\ and minimal \Ha\ emission.
These two parameters are somewhat degenerate
   and depend on the gas column.
%They can be roughly approximated by 
Good fiducial values are
   $T_{\mathrm{opt}}\sim10^{6}$\,K
   and $\Phi_{\mathrm{opt}}\sim10^{23}$~cm$^{-2}$~s$^{-1}$.
The radius of a black body with $T=T_{\mathrm{opt}}$
   that provides $\Phi_{\mathrm{opt}}$
   at $r=R_\mathrm{c}$ is 
\begin{equation}
	R_{{\rm opt}}\approx\sqrt{\frac{\Phi_{{\rm opt}}}{\Phi}}R_\mathrm{c}
	=2.1\times10^{13}\,T_{{\rm opt,6}}^{-1.5}\Phi_{23}^{0.5}r_{{\rm 10ld}}\mathrm{~cm},
\label{eq:Ropt}
\end{equation}
   where $\Phi_{{\rm opt}}=10^{23}\Phi_{23}$~cm$^{-2}$~s$^{-1}$
   and $R_\mathrm{c}=10r_{{\rm 10ld}}\mbox{~ld}=2.6\times10^{16}\,r_{{\rm 10ld}}$~cm
   %which corresponds to the observable time-scale of spectral variations
	\citep[compatible with the observable time-scale of spectral variations;][]{gezari2012,holoien2016b}.
The assumption of adiabatic expansion
   implies the continuum source's initial radius, 
\begin{equation}
	R_{{\rm i}}
	%=\left(\frac{T_{{\rm opt}}}{T_{{\rm i}}}\right)^{\frac{1}{3(\gamma-1)}}R_{{\rm opt}}
	=\left(\frac{T_{{\rm opt}}}{T_{{\rm i}}}\right)^{4/3\gamma}
	=2.1\times10^{10}\,T_{{\rm opt,6}}^{-0.5}T_{{\rm i,9}}^{-1}\Phi_{23}^{0.5}r_{{\rm 10ld}}\mathrm{~cm},
\label{eq:r_in}
\end{equation}
   where $T_{{\rm i}}=10^{9}\,T_{{\rm i,9}}$\,K
   is the initial temperature.
This value of $R_\mathrm{i}$
   is smaller than the expected \cite{schwarzschild1916} radius
   ($R_\bullet\sim10^{12}$\,cm for $m_\bullet=10^7m_\odot$).
However, the continuum source in our model
   is not limited to a TDE
   (see above, and later in subsections~\ref{s.supernovae}, \ref{s.agn}).
For stellar explosions in the parsec-scale nuclear star cluster,
   $R_\bullet$ is inapplicable.
In an AGN jet context,
   it is conceivable for transient luminous features to vary on scales smaller than $R_\bullet$
   (e.g. \citealp{aleksic2014}).
Thus $R_\mathrm{i}>R_\bullet$ is not required.
It should be noted that if $T_\mathrm{i}\sim10^6$\,K,
   rather than $10^9$\,K, then $R_\mathrm{i}\sim R_\bullet$ and
   is consistent with a TDE origin of the continuum source
   (see also below).
Mutatis mutandis,
   the adiabatic phase ends at a final radius
\begin{equation}
R_{{\rm f}}\sim2.1\times10^{14}\,T_{{\rm opt,6}}^{-0.5}T_{{\rm f,5}}^{-1}\Phi_{23}^{0.5}r_{{\rm 10ld}}\mathrm{~cm},
\end{equation}
   where $T_\mathrm{f}=10^{5}T_{{\rm f,5}}$\,K.
We use $T_\mathrm{f}\sim10^{5}$\,K
   %as the final $T_\textsc{bb}$,
   rather than $10^{4}$\,K,
   since below $T\sim10^{5}$\,K
   radiative cooling dominates and
   expansion slows significantly.
   %the source likely no longer expands significantly,
   %and most of the cooling is attributed to radiative losses (see above).
Until then,
  the photospheric expansion velocity is approximately
\begin{equation}
	v_{\mathrm{exp}}
	\sim\frac{R_{{\rm f}}}{R_\mathrm{c}/c}
	=2400\,T_{\mathrm{opt,6}}^{-0.5}T_{\mathrm{f,5}}^{-1}\Phi_{23}^{0.5}\mbox{~km~s\ensuremath{^{-1}}}.
\label{eq:v_exp}
\end{equation}
Note that $v_{{\rm exp}}$ is independent of the assumed $R_\mathrm{c}$.
The \HeII-bright state is produced from $\Tbb=10^{6}$\,K down to $10^{5}$\,K,
   i.e.\ 1~dex in $\Tbb$.
Since $\Tbb\propto R^{-1}$ (eq.~\ref{eq:r_in}),
   the cloud remains in the \HeII-bright state for a time of at least
\begin{equation}
\Delta t\approx10R_{{\rm opt}}/v_{{\rm exp}}=10\,T_{{\rm opt,6}}^{-1}T_{{\rm f,5}}r_{{\rm 10ld}}\mathrm{~d},
\label{eq.hot.time}
\end{equation}
   where we used eq.~(\ref{eq:Ropt}).
The {\em apparent} broad-line phase could be prolonged
   by contributions from multiple clouds in different orbits,
   or monochromatic reflections off distant dust.
By eq.~(\ref{eq.hot.time}),
   the hard X-ray beginning
   was brief enough ($\la0.1$\,day)
   to go undetected before the UV/O transient became conspicuous.
%   and happened days before.
%   (therefore easily missable).
%The persistent long-term X-ray luminosity of some transients
%   may be a secondary component
%   (perhaps a corona around the photosphere,
%   or weak jet activity)
%   but the X-ray dimness of other events
%   \citep{vanvelzen2016a}
%   is natural to our model.

Assuming isentropic expansion and using eq.~(\ref{eq:v_exp})
   we can infer the total mass $M$ of the continuum source.
The velocity of isentropic expansion is approximately
\begin{equation}
	v_{{\rm ise}}
	\approx\frac{2}{\gamma-1}\sqrt{\gamma\frac{aT_{i}^{4}/3}{M(4\upi R_{i}^{3}/3)^{-1}}},
\label{eq.v.ise}
\end{equation}
   where $a=4\sigma/c$,
   and $\sigma$ is the Stefan-Boltzmann constant
	\citep[e.g.][]{zeldovich1967}.
Taking $v_{{\rm ise}}=v_{{\rm exp}}$
   and inserting eq.~(\ref{eq:r_in}) yields
$M\approx4\times10^{4}\,T_{i,9}T_{{\rm opt,6}}^{-0.5}T_{{\rm 5,f}}^{2}\Phi_{23}^{0.5}r_{{\rm 10ld}}^{3}m_{\odot}$.
%   which is an unreasonable value for an explosion of stellar progenitor.
The total bolometric luminosity
   across the photosphere is
%   of the continuum source is 
\begin{eqnarray}
L_{{\rm bol}}
	&=&4\upi R^{2}\sigma T_\textsc{bb}^{4}
	=4\upi R_{{\rm opt}}^{2}\sigma T_{{\rm opt}}^{2}T_\textsc{bb}^{2}
	\nonumber\\
	&=&3.2\times10^{45}\,T_{5}^{2}T_{{\rm opt,6}}^{-1}\Phi_{23}r_{{\rm 10ld}}^{2}\mbox{~erg~s\ensuremath{^{-1}}},
\end{eqnarray}
   where $\Tbb=10^{5}T_{5}$\,K.
This estimated $L_{{\rm bol}}$ is broadly consistent with observations
   \citep{vanvelzen2011b,cenko2012a,chornock2014,cenko2016,holoien2016b}.
Thus, an adiabatic expansion predicts that for $v_{{\rm exp}}t\gg R_{{\rm in}}$,
   $L_{{\rm bol}}\propto t^{-2}$.
The total energy emitted during the adiabatic-expansion phase is 
\begin{eqnarray}
	E
	&\sim&
	%\int_{0}^{R_\mathrm{c}/c}L_{{\rm bol}}\ddt
	\int_{0}^{R_\mathrm{c}/c}L_{{\rm bol}}\mathrm{d}t
	\simeq\frac{4\upi\sigma R_{{\rm opt}}^{4}T_{{\rm opt}}^{4}}{R_{{\rm i}}R_{{\rm f}}}
	{{R_\mathrm{c}}\over{c}}
	\nonumber\\
	&=&
	3\times10^{55}\,T_{{\rm opt,6}}^{-1}T_{{\rm i,9}}T_{{\rm f,5}}\Phi_{23}r_{{\rm 10ld}}^{3}\mbox{~erg}
	\ .
\label{eq:E_tot}
\end{eqnarray}

This energy scale (\ref{eq:E_tot})
   and the mass derived from equation (\ref{eq.v.ise})
   imply some interesting parameter constraints,
   depending on what type of explosive event
   erupted into the continuum source.
Taking fiducial input values,
   the estimate (\ref{eq:E_tot})
   exceeds the energies of super-luminous SNe
   by a factor of $\sim10^{3}$
   \citep{galyam2012,inserra2013,vreeswijk2014,dong2016}.
The corresponding mass $M\sim2\times10^4m_\odot$
   would also be extraordinary for a stellar progenitor.
The energy implied by eq.~(\ref{eq:E_tot})
   would also require extraordinary mass accretion or annihilation
   ($E/\eta c^2$, with $\eta<0.1$)
   besides inefficiencies such as neutrino production,
   and would exceed the observational fluence estimates.
What are the alternatives?
More detailed tests show that
   the energy and progenitor mass
   are sensitive to details of different $\gamma$
   and models of the photosphere's expansion.
A softer equation of state with $\gamma=5/4$
   yields $M\sim10m_{\odot}$
   (and a corresponding $R_{{\rm i}}\sim10^{9}$~cm).
The smaller value of $\gamma$ is relevant for the highest $T$ and small $R$,
   where the source conditions may be favourable for pair production,
   reducing $\gamma$ below $\gamma=4/3$.
(Plausibly, 
   some entrainment of upswept circumnuclear gas might also alter $\gamma$.)
Alternatively,
   a lower initial tempreature ($T_{{\rm i,9}}$) could be an explanation.
Replacing $R_{{\rm i}}$ with $R_{{\rm opt}}$ in eq.~(\ref{eq:E_tot})
   (i.e., starting the integration at $\Tbb=10^{6}$\,K rather than $10^{9}$\,K)
   yields $E\sim3\times10^{52}$~erg,
   consistent with observations,
   and more compatible with a stellar event such as a supernova
   (subsection~\ref{s.supernovae}).
Thus the continuum source energy problem
   is both flexible and sensitive to subtle details of the flare's expansion and cooling evolution
   (at this back-of-the-envelope level).
The properties of the external photoionized cloud are more robustly constrained,
   and we discuss those firmer findings next.

\subsubsection{Implications for the line-emitting gas}
\label{sec:gas}

To produce a \HeII-bright state,
   the line-emitting gas must react to variations in the ionizing continuum
   quicker than the variability time-scale of the continuum source.
Two time-scales affect the gas:
   the ionization and recombination time-scales.
The gas reacts to changes in the continuum
   on the longer of these two scales.
We continue to use
   $\tau_{{\rm var}}\equiv R_\mathrm{c}/c=8.6\times10^{5}r_{{\rm 10ld}}$~s
   as the variability time-scale of the continuum.
The ionization time-scale $\tau_{{\rm ion}}$ can be estimated as follows.
The photoionization cross-section of He$^{+}$
   that is averaged over a black-body continuum ($\bar{\sigma}$)
   is approximately
\begin{equation}
	\bar{\sigma}\approx\left\{ \begin{array}{ll}
	1.41\times10^{-18}\,T_{6}\mbox{~cm\ensuremath{^{2}}},
	&\text{for }0.1\leq T_{6}<0.3\\
	1.10\times10^{-19}\,T_{6}^{-1.7}\mbox{~cm\ensuremath{^{2}}},
	&\text{for }T_{6}\geq0.3
\end{array}\right.,
\end{equation}
   where we use the expression for $\sigma(\nu)$ from \citet{verner1996}.
The approximation is correct within a factor of 3.
The implied $\tau_{{\rm ion}}\equiv1/\bar{\sigma}\Phi$ is 
\begin{equation}
	\tau_{{\rm ion}}\approx\left\{ \begin{array}{ll}
	7\times10^{-6}\,\Phi_{23}^{-1}T_{6}^{-1}\mbox{~s},&\text{for }0.1\leq T_{6}<0.3\\
	9\times10^{-5}\,\Phi_{23}^{-1}T_{6}^{1.7}\mbox{~s},&\text{for }T_{6}\geq0.3
	\end{array}\right..
\label{eq:tau_ion}
\end{equation}
Thus, $\tau_{{\rm ion}}\ll1$~s for $\Phi_{23}\sim1$ and $T_{{\rm opt,6}}\sim1$,
   and is negligible compared to $\tau_{{\rm var}}$.
The assumption of adiabatic expansion yields
   $\tau_{{\rm ion}}\ll\tau_{{\rm var}}$
   also for $T_{6}\sim1000$,
   i.e.\ when the continuum source is at maximum $\Tbb$,
   since $\Phi_{2}=\Phi_{1}T_{2}/T_{1}$
   (eqs.~\ref{eq:phi_for_TBB} and \ref{eq:r_in}).
The constraint on the recombination time scale,
   $\tau_{{\rm rec}}\equiv1/\alpha n_{{\rm gas}}<\tau_{{\rm var}}$,
   implies a lower limit on the gas number density $n_{{\rm gas}}$,
   where $\alpha$ is the recombination coefficient.
We approximate the dependence of $\alpha$ of He$^{++}$ on $T_{{\rm gas}}$
   by $\alpha\approx10^{-7}T_{{\rm gas}}^{-1.07}$~cm$^{3}$~s$^{-1}$
   \citep{verner1996},
   which is accurate within a factor of $2.5$
   in the $10^{4}\leq T_{{\rm gas}}\leq10^{9}$\,K
range. The implied constraint on the gas density is 
\begin{equation}
n_{{\rm gas}}
	>\frac{1}{\alpha}\frac{1}{R_\mathrm{c}/c}\approx3\times10^{6}\,T_{{\rm gas,5}}^{1.07}r_{{\rm 10ld}}^{-1}\mbox{~cm\ensuremath{^{-3}}},
\label{eq:n_gas}
\end{equation}
   where $T_{{\rm gas}}=10^{5}\,T_{{\rm gas,5}}$\,K.\footnote{%
We use the scaling of $T_{{\rm gas}}\sim10^{5}$\,K in eq.~(\ref{eq:n_gas}),
   as it is roughly the thermodynamic equilibrium $T$ of gas that is
   located at a distance $R_\mathrm{c}$ from a continuum source with
   $T_\textsc{bb}=T_{i}\sim10^{9}$\,K,
   i.e.\ $T_{{\rm gas}}\approx T_{{\rm i}}\sqrt{R_{i}/R_\mathrm{c}}$
   (see eq.~\ref{eq:r_in}).}
This value of $n_{{\rm gas}}$ is well below $n_{{\rm gas}}\sim10^{12}$~cm$^{-3}$
   which is required by photoionization calculations
   to produce the observed high \HeII/\Ha\ line ratio.

\subsection{Puzzles and difficulties for the TDE model}
\label{s.puzzles}

Several puzzles and problems emerge
   from observational studies of TDE candidates
   (see \citealt{komossa2015b}),
   and may suggest that these events
   might not arise from tidal disruptions by MBHs,
   but rather from other sources.
Here we only focus on issues relating to our results,
   and later discuss possible alternative origins. 
 
1. In some systems, the inferred mass accretion is subsolar
   (see also discussion in section \ref{subsec:echo-model}),
   and indeed much less than any likely donor star.
For UV/O candidates,
   the most credible estimates appear subsolar
	\citep{renzini1995,gezari2012,chornock2014,holoien2014}.
ASASSN-14ae was discovered after its peak,
   but the accretion was only $\ga0.001m_{\odot}$
   \citep{holoien2014}.
A similar limit from
   iPTF16axa gives $3.1\times10^{-3}m_\odot$ \citep{hung2017}.
More clearly subsolar events include
   iPTF16fnl with $<$$10^{-4}m_{\odot}$ \citep{blagorodnova2017};
   PS1-11af with $0.002m_{\odot}$ \citep{chornock2014};
   ASASSN-14li with $0.004m_{\odot}$ \citep{holoien2016a};
   ASASSN-15oi with $0.003m_{\odot}$ \citep{holoien2016a};
   PS1-10jh with $0.012m_{\odot}$ \citep{gezari2012};
   and $0.004m_{\odot}$, $0.11m_{\odot}$ for two transients of \citet{gezari2008}.
Allowing host galaxy extinction
   may be compatible with accretion
   $\la0.1m_\odot$ in PTF09ge
	\citep{vanvelzen2016b}
   and $0.06m_\odot$ in D23H-1
	\citep{gezari2009}.
Bolometric corrections might explain these two events as stellar TDEs,
   but the transients with lower accertion estimates
   are more challenging to explain.

\citet{guillochon2014}
   matched PS1-10jh chromatic light-curves to simulations of
   a $4m_\odot$ or $0.5m_\odot$ star
   in a sub-Eddington TDE
   plus translucent outflow
   %without a strong wind
   (but without detailed fitting of line luminosities or EW).
%It is suggested that some of these were partial disruption events.
If a normal star was only partially disrupted
	\citep[e.g.][]{guillochon2013}
   this could explain the transient's low mass budget,
   but flaring should eventually recur
   on timescales comparable to the orbital period of the remnant stellar core.
Since no secondary flare has been reported,
   the partial TDE scenario would require orbital periods
   longer than several years (in every known TDE candidate).
Though earlier models find that half of TDE debris remains bound,
   \cite{metzger2016}
   argue that a TDE with sufficient Eddington ratio
   ejects most of the gas
   into an opaque, radiation dominated outflow;
   and only a minority accretes.
Alternatively,
   it might be that the accretion process
   was radiatively inefficient for some unknown reason; 
   perhaps because the tidal stream's self-intersection shock
   has been delayed by orbital precession effects
   or other complications
	\citep{guillochon2015,piran2015,svirski2015}.
Unless one of more of these remedies applies to all UV/O candidates,
   the observed low mass indicators
   reveal something unexpected about the accretion source.
%In contrast,
%   our light-echo model of the \HeII-overbright spectral state
%   {\em predicts and requires} subsolar masses for the external line-emitting clouds,
%   which need not originate from the central explosive / disruptive event.

2. For the UV/O TDE candidates,
   the emission line widths tend to be
   less than the orbital velocities expected at $R_{\mathrm{t}}$,
   or else the emission region
   is radially much larger than expected from stellar disruption
	\citep{gezari2009,gezari2012,arcavi2014,chornock2014}.
A potential solution could invoke
   debris circularization occuring far from the tidal-disruption radius,
   \citep{piran2015,svirski2015}
   but whether this is indeed the culprit
   is still an open question.
\cite{roth2016} and \cite{metzger2016}
   attribute the line widths to
   random transverse motions (eddies?)
   on the central source's photosphere,
   as it expands at $\sim$$10^4\kms$.
(Detailed hydrodynamic models,
   including temporal variability,
   aren't yet available.)
We ran {\sc Cloudy} tests of $>$$10^3\kms$ turbulent photoionized slabs,
   and always find that the kinematically reduced \Ha\, opacity
   prevents \Ha\, suppression, which prevents the \HeII-overbright state.
An orderly outflow with radial velocity gradients
   would also suffer the problem of \Ha-transparency.
We suggest that \HeII-overbrightness
   arises more easily
   from a population of well separated clouds,
   each with low internal velocities,
   but high velocity dispersions among them as an {\em ensemble}.
We further discuss the line velocity scaling
   and present the relevant data in the Appendix~\ref{s.scaling}.

3. Reviewing a set of UV/O candidates,
	\citet{arcavi2014}
   noted that most TDE host galaxies seem weak in star formation,
   and several belong to the rare E+A `post-starburst' type.
Exceptionally,
	\citet{holoien2014}
   report a candidate in a spiral galaxy.
E+A `post-starburst' galaxies show stellar absorption lines
   indicating a stellar population truncated at a few Gyr age,
   and ongoing star formation is weak or regionally limited
	\citep{dressler1983,yamauchi2005,brown2009,pracy2014}.
They often exhibit disturbed morphologies
   or close tidally interacting companion galaxies
	\citep{yang2004,kewley2006,goto2008,pracy2009}
   suggesting the possibility of relatively recent (Gyr) galaxy mergers.
If a galaxy nucleus contains a binary SMBH
   then 3-body scattering will gradually harden their orbit
   to a pc-scale separation,
   while ejecting many stars
   and incurring tidal disruptions
   more often than a single supermassive object would
   \citep[e.g.][]{quinlan1996,ivanov2005,chen2009,chen2011,wegg2011}.
A significantly non-spherical stellar background potential
   could also increase the TDE rate \citep[e.g.][]{vasiliev2014a,hamers2017}.
Since SMBH binaries and asymmetric potentials
   are both expected outcomes of galaxy mergers,
   it may be that the overabundance of TDE-candidates in these galaxies
   is related to their relatively recent galaxy merger history.
However, the expected increased TDE rates 
   are typically at most a few times larger,
   while the inferred flaring rates in these hosts
   seems at least an order of magnitude (if not more) larger than expected
%   while the inferred rates in such galaxies
%   appear to be at least an order of magnitude (if not more) larger than expected
   given the rarity of E+A galaxies
   \citep[only as few as $\la10^{-2}$ of all galaxies;][]{zabludoff1996,goto2003,quintero2004,vergani2010}.

Circumnuclear power-law cusps of stellar density
   should raise TDE rates,
   in theory
	\citep{magorrian1999,wang2004}.
If the apparently steep young cusp of one E+A galaxy
   (NGC\,3156, $\rho_\bigstar\propto r^{-2.2}$ at $\sim$$4$\,pc resolution)
   continues to radii $\ll0.1$\,pc,
   %and if such cusps are normal for this galaxy class,
   then this might explain the TDE productivity of this class
	\citep{stone2016a,stone2016b}.
It is unknown how long steep stellar cusps survive dynamically,
   and whether they are special to the E+A phase of galaxy evolution.

Some E+A galaxies seem to feed low-luminosity AGN \citep{yang2006},
   and the evidence of an intermediate (Gyr ago) star-formation epoch
   suggests the existence of a gas reservoir,
   which is lower than star-forming galaxies,
   but can be sufficient for a spasmodic, low rate feeding of an AGN,
   potentially producing starved/subluminous AGNs.
If the TDE-candidate flares are not bona-fide TDEs
   but are related to bursts from subluminous AGNs
   (a possibility discussed in the next section),
   then such E+A galaxies
   might indeed provide favourable conditions,
   %to produce the observed rates,
   and explain why such nuclear flares
   preferentially reside in E+A hosts.
   %the preference of such nuclear flares
   %to reside in E+A host galaxies.

\subsection{Alternative (non TDE) origins for TDE-candidates}

\subsubsection{A supernova among circumnuclear clouds}
\label{s.supernovae}

The nuclear regions of many galaxies may be
   favourable locations for concentrated episodes of supernova activity.
The quiescent Milky Way's inner $0.1$pc
   produced massive stars $\sim$1Myr ago
	\citep[e.g][]{bartko2010},
   which will yield a local SN rate
   of a few $10^{-5}\,\mathrm{yr}^{-1}$:
   within reach of predicted TDE rates.
Evidence that the central object Sgr~A* fluctuates over $\sim$500yr
	\cite[e.g][]{ryu2013}
   also implies a variable gas supply to a starved AGN
   (some of which might be diverted to star formation).
Stronger circumnuclear starbursts
   in highly disturbed galaxies
   might conceivably yield enough SN
   to confuse cosmic TDE monitoring.

While the continuum source of TDE candidates
   may be explained by a (super-luminous) SN explosion,
   the required density of line-emitting gas
   in \HeII-bright cases
   ($n_{{\rm gas}}\sim10^{12}$~cm$^{-3}$)
   is large compared to the densities expected
   for the ISM of quiescent galaxies.
Such a large $n_{{\rm gas}}$
   is however typical in nuclear regions of AGN
   \citep[e.g.][]{netzer2013}.
\cite{french2016}
   report that TDE detections are preferentially found
   (40\%--75\%)
   in Balmer-strong galaxies,
   i.e.\ galaxies
   that recently had a strong starburst
   (roughly within the last Gyr).
These galaxies are only 0.2\% of the local galaxy population.
A strong starburst requires a large supply of gas,
   a fraction of which may reach the host nucleus and feed an AGN.
This activity quenches star formation in the nuclear region;
   the only exception is the
    parsec-scale
   dusty torus where starbursts
   may occur continuously throughout the life of AGN
   \citep{collin1999b,collin1999a,sirko2003}.
Thus, a candidate TDE
   may be the SN explosion of a progenitor star
  that formed locally within the torus.
(This region is circumnuclear but far outside the event horizon
   and hot inner accretion disc,
   and therefore not directly related to the horizon size.)
The explosion occurs sometime after the AGN falls dormant
   ($\la10^{3}$~yr),
   before the dense clumps of gas
   disperse (or collapse).
An AGN has a life-time of $\sim10^{8}$~yr.
A star of a few $10m_{\odot}$ lives for a few $10^{7}$~yr.
Thus, the expected fraction of TDE candidates
   in quiescent Balmer-strong galaxies is
   $\sim10^{7}/10^{8}=10$\%,
   consistent with observations
   \citep{french2016}.
SN explosions occurring after the nucleus has lost its denser clumps
   %may appear as flares that 
   miss the \HeII-bright phase.
%TDE candidates that miss the \HeII-bright phase,
%   may be SN explosions in a nucleus that no longer has dense enough clumps.
%A model of gas with $\tau_{{\rm rec}}>\tau_{{\rm var}}$
%   requires non-equilibrium photoionization calculations,
%   and is beyond the scope of this study.
This scenario naturally explains
   the observed emission-line width of a few 1000~km~s$^{-1}$
%   in TDE candidates,
   which is typical of the BLR region of AGN.

The SN-explosion scenario for TDE imposters may also explain
   the peculiar SNe of type II-L and II-P,
   which exhibit spectral behaviour similar to \HeII-bright TDE candidates
    \citep{smith2015}.
These are likely SN explosions of massive progenitors
   that detonate near a reservoir of dense gas
   (e.g.\ protostellar clouds).
A possible connection between TDE candidates
   and extreme SNe was previously
   pointed out by others 
	\citep{komossa2009,drake2011}.
Here, we establish this connection from photoionization considerations.
\citet{gezari2009}
   disfavour the connection,
   pointing out that the UV continuum emission
   (i.e.\ $T_\textsc{bb}\ga10^{4}$\,K)
   fades significantly faster in SNe compared to TDE candidates
   (weeks versus months).
This discrepancy may be explained by
   the denser ISM than expected in a galactic nucleus.
Ambient gas confinement  could
%is likely to
   alter the SN shock physics and radiative efficiency,
% slow the cooling of SN ejecta,
   producing an abnormally shallow light-curve.
%   and produce a shallower light-curve for SNe in the nucleus.
Finally,
   we note that similar links between nuclear SNe and AGN activity
   has been suggested before
   \citep[e.g. ][]{terlevich1992,aretxaga1994,ulrich1997,aretxaga1999b}.
%   but extending this possibility
%   in the context of singular TDE-candidate events,
%   is more attractive than the previously suggested
%   connection to recurrent AGN activity and flaring. 

\subsubsection{Subluminous AGN flaring scenario}
\label{s.agn}

Another alternative possibility is that the UV/O TDE candidates
   could have been a type of unrecognized flaring from accretion onto a MBH,
   possibly in a subluminous/starved AGN
   that lacks strong emission line regions during a long semi-quiescent state.
Circumnuclear clouds may be present,
   but normally not irradiated strongly enough for an observable BLR (or NLR).
Some hosts may be senescent or starving AGN,
   which have already accreted most of the available interstellar gas.
These could be related to AGNs that change their BLR type and/or missing them
   \citep[e.g.][]{tran2001,bianchi2005,zhang2006b,kaastra2014},
   which can occur even on short timescales
   \citep{storchi1993,aretxaga1999b,shappee2014,oknyanky2016,mcelroy2016}.
Some peculiar types of AGNs could play a similar role, e.g.  XBONG sources
   (\citealp{elvis1981,griffiths1995,comastri2002,fiore2003,severgnini2003};
   a few of which are E+A galaxies \citealp{kim2006}).
In particular,
   variability in blazars and BL~Lac objects
   can exceed several optical magnitudes
   \citep{eggen1973,xie2002,dolcini2005,danforth2013,sandrinelli2014}.

Indeed, there are a few possible examples.
\citet{campana2015}
   proposed a reverse argument
   --- explaining recurrent AGN flaring as TDE in the weak AGN in IC\,3599.
\cite{alloin1986}
   observe recurrent outbursts in a low-powered AGN,
   each of which behaved like an intermediate phenomenon between
   modern TDE candidates and the normal activity of a steady AGN,
   while \citet{r_saxton2015}
   suggested to relate a transient TDE-like event
   to a disc instability around a weak AGN in NGC 3599.

\citet{laor2003}
   proposed that shear destroys any BLR clouds within a limiting orbital radius,
   but the number of clouds irradiated enough to emit the broad lines
   depends on the nucleus' luminosity.
Consequently,
   an AGN weaker than some threshold will not have any observable BLR.
It is therefore conceivable that a nuclear {\em brightening}
   in a previously {\em unrecognised} AGN
   can activate previously unilluminated BLR clouds,
   giving the false appearance of a TDE.
This depends on the weakness of ordinary AGN signatures in the initial state,
   and the availability of at least one non-luminous circumnuclear cloud
   initially outside the \cite{laor2003} limiting radius.
Flaring might be fed by an episodic surge of accretion:
   a disc instability,
   an abrupt change in accretion state,
   or passage of a massive object through the disc.
Accretion might increase if a new gas source arrives,
   derived from a cloud that has scattered into a diving orbit,
   much like the disruption of a star.
Whatever the trigger and the fuel,
   flaring then excites one or more previously dormant circumnuclear clouds;
   if they cover enough of the nucleus then the result is a transient BLR.
One then still needs to explain
   the lack of hard radiation from the inner parts of the disc
   (perhaps due to an inner disc cavity).
%This may occur if a cavity exists in the inner region.

Among conceivable flaring processes around MBHs,
   the disc-instability model of \citet{kim2010}
   might be extended from microquasars to instability in a disc around a MBH.
In this case,
   an instability in the inner accretion disc
   can lead to draining of the inner region,
   producing a cavity,
   while the accreted material powers a flare.
The flaring can later
   be echoed by more distant material,
   (as described in the echo model).
Stochastic recurrence timescales might be longer for larger $m_\bullet$,
   with testable consequences in population statistics
	\citep{vanvelzen2017}.
Given the observational selection bias against AGN hosts,
   such events are likely to be identified
   only in cases where a gaseous disc exists around a MBH,
   but the accretion is insuffcient to power an observable AGN,
   e.g. in cases of starved or subluminous AGNs,
   where circumnuclear gas exist, but is scarce.
These might possibly correspond to E+A galaxies,
   in which star-formation and hence gas supply
   were abundant in the relatively near (but not recent) past,
   while the MBH now feeds only on the (now dwindling) gas inflow
   following the past merger event.
Even drier galaxies would not have sufficient supply,
   and the MBH would not sustain a disc in which instabilities could occur.

Another model by \citet{tanaka2013}
   involving binary MBH may also be relevant.
The binary clears a central cavity in its accretion disc,
   which appears dimmer and softer than an equivalent single-AGN disc.
Depending on the binary period,
   intermittent gas streams of perhaps $0.1m_{\odot}$
   dive towards the SMBHs from the cavity's lip.
Each stream shocks near the recipient SMBH,
   causing a flare resembling a TDE.
If the recurrence period is long
   compared to the era of transient monitoring,
   then a single detection could be misidentified as a TDE.
The recurrence period, debris velocities and energy yields in a cavity flare
   depend on the binary separation,
   and need not correspond to predicted tidal radii of stellar disruptions.
Timescales of cavity flares need not show any clear relation to SMBH masses
   \citep[c.f.][]{vanvelzen2017}.
Subtler details might distinguish binary cavity flares:
   the steepness of their fading,
   or occurrence around a SMBH that is too massive for a TDE.
(For example,
	\citealp{leloudas2017}
   recently infer $m_\bullet>10^8m_\odot$
   in the weak-lined TDE candidate ASASSN-15lh.)
If SMBH binaries are a common product of the mergers forming E+A galaxies,
   then the cavity flare mechanism might explain the peculiar
   incidence of UV/O nuclear transients in these hosts.

Let us consider generically what happens as the amplitude of an AGN flare
   is increased (whether it involves jet internal shocks,
   or an accretion surge from disc instability).
For a small-amplitude event,
   the rising and the fading involve similar physics,
   and phenomena are almost reversible in time.
A more luminous flash exerts an impulsive radiation pressure
   on opaque circumnuclear clouds.
At greater intensities,
   the peak luminosity exceeds the local Eddington limit,
   driving a radiation-dominated outflow,
   which could in turn drive external shocks into the ISM,
   resulting in a prolonged, fading afterglow like those of gamma-ray bursts.
The emergent luminosity is limited to a few times $L_{\mathrm{E}}$,
   but a radiation-driven wind could carry more kinetic power.
Thus an asymmetry between rising and fading rates
    is naturally expected for higher-amplitude events.

If the energy injection is abrupt and intense enough
   to propel an \emph{opaque} radiation-dominated
   explosion then the responsible process is concealed.
In super-Eddington models,
   the wind photosphere hides the accretion disc and accretor
	%\citep[e.g.][]{kasen2010b,metzger2016}.
	%\citep[e.g.][]{ulmer1999,strubbe2009,metzger2016}.
	\citep[e.g.][]{strubbe2009,metzger2016}.
This explains why late-time UV/O candidate temperatures
   are a few times $10^4$K,
   rather than the $10^5$K temperatures expected
   for matter near the tidal radius
	\citep[e.g.][]{jaroszynski1980,loeb1997,ulmer1999}.
This generic self-concealment of super-Eddington events implies that
   a high-amplitude AGN outburst
   may be practically difficult to distinguish from a TDE flare,
   or an unusual supernova detonating in a dense medium \citep{smith2015,moriya2017}.
In this way,
   the continuum source may be the {\em least} informative aspect of the transient,
   and other identifying signatures are needed.
Transient radio, X-ray and $\gamma$-ray components
   could be the result
   of collimated jets erupting through the opaque surface:
   briefly, persistently, or with instabilities and precession effects
   depending on idiosyncrasies of each system
	\citep[e.g.][]{gezari2009,vanvelzen2011a,bloom2011,levan2011,
	vanvelzen2013,saxton2012,cenko2012b,bower2013,
	vanvelzen2016a,k_alexander2016}.
In events where jets ceased after an initial burst,
   %it is conceivable that 
   their emissions might ionize distant clouds,
   like the early hot photosphere's `flash' in our light-echo models.

\section{CONCLUSIONS}

In this study we explored
   the spectral features of tidal disruption event candidates.
Our analysis shows that
%none of the models currently used can
   standard AGN-like irradiation models cannot
   explain both the origin of the line ratios
   as well as the individual equivalent widths seen in observations.
%Others have reported that
%   optical depth effects could produce the line emission
%   near the continuum source's own photosphere
%	\cite{roth2016},
%   and it would be interesting to assess the generality of these conditions.
%As our alternative,
We find that models of blackbody illumination
   of remote BLR-like clouds
   ($n_\mathrm{gas}\sim10^{12}\,\mathrm{cm}^{-3}$)
   could yield realistic line emissions
   across a significant range of conditions,
   but only when considering a hot source ($10^{6}$K),
   while a $10^{4}$\,K blackbody is consistent with
   the continuum observed in UV/O TDE candidates.
We propose a novel light-echo model to resolve these difficulties,
   and this has important implications
   for the early evolution of the explosive event
   and the later temporal evolution of the spectral features.
In the light-echo model,
   an initial hard radiation flash illuminates
   BLR clouds residing at distances on the order of ten light-days.
Their strong line emission
   is then seen delayed and superimposed on
   the softer continuum emitted later from the exploded source.
%   the softened nuclear spectrum emitted at a later time.
Our model can therefore provide a prediction
   that TDE candidates should appear X-ray bright in the continuum
   %concurrently or 
   before the \HeII\, state.

This light-echo photoionization modelling
   is generic to near-super-Eddington outbursts
   in a galaxy nucleus,
   as long as there are some BLR-like external clouds
   at suitable distances.
Diffuser clouds only exhibit
   the \Ha-bright or fully ionized states.
It could be 
   difficult to distinguish between a tidal disruption event,
   an outburst by an impoverished AGN,
   or an exotic superluminous supernova,
   or perhaps other explosive transients not yet recognized.
A weak, starving AGN accreting at a low average rate,
   with a nearly exhausted gas supply,
   might mimic a TDE if a high-amplitude flare photoionizes
   the few clouds of a normally dim and impoverished BLR.
%The He-overbright state can occur in the BLR of a semi-starved AGN.
    %but it helps if the clouds reside predominantly in a narrow radial range.
A superluminous supernova detonating
   among the dense BLR clouds
   could have a similarly evolving outflow,
   %with the opaque photosphere 
   hiding the central engine.
For each single cloud in any of these types of system,
   the occurrence of a \HeII-bright state depends on whether it evolves
   through a limited sausage-shaped zone of
   the $(N_\textsc{h},n_\textsc{h},\Phi_\textsc{h})$ parameter cube.
Densities of $10^{11}<n_{\textsc{h}}<10^{13}\,\mathrm{cm}^{-3}$
   are necessary but not sufficient.
%The collective spectral state depends on
%   the clouds' respective echo time-lags,
%   their internal properties,
%   the fading of the central source,
%   and possibly further reflections off dusty media farther away.
Brightening, fading and temperature changes by the central source control $\Phi_\textsc{h}$,
   while $(N_\textsc{h},n_\textsc{h})$ might change
   if the cloud derives from ballistically evolving TDE debris
	\citep[e.g.][]{guillochon2014}.
Transitions between \HeII-bright, \HeII+\Ha, and \Ha-bright states are conceivable.
Multiple clouds and reflections off remote dust may complicate the collective spectra.
%If such a system develops the He-overbright state at all,
%   then it might happen twice:
%   once while $\Phi_{\textsc{h}}$ is rising and again during the fall.
%If however the transient event involves ballistically
%   or dynamically expanding debris,
%   then the He-overbright state might occur only once.
In future,
   time-dependent emission line models and
   higher-cadence spectroscopic monitoring
   will test these possibilities.
It would also be valuable to explore intermediate cases between
   the theory of a line emission hugging an opaque photosphere
%   the opaque photosphere-hugging emission theory
	\citep[][despite the challenge of velocity and \Ha-transparency]{roth2016}
   and our model of light-echo photoionization of distinct BLR clouds
   (methodologically constrained by $\tau_\mathrm{es}<0.5$).
Temporal variability of the spectra,
   testing the photoionization light echo time-lags,
   may be essential to discriminating among these models
   and diagnosing their spatial structures.

\section*{Acknowledgements}
We thank the referee
   N. Murray
%Norman Murray
   for his constructive criticism;
   and we thank
   B.D. Metzger,
%Brian Metzger,
   N.C. Stone,
%Nick Stone,
   S. van~Velzen,
%Sjoert van~Velzen,
   A. Laor,
%Ari Laor,
   I. Arcavi
%Iair Arcavi
   and
   S. Kaspi
%Shai Kaspi
for helpful discussions.
This work has made use of NASA's Astrophysics Data System.
Calculations were performed with version 13.03 of \textsc{Cloudy},
last described by \citet{ferland2013}.
This publication has made use of code written by
    James R. A. Davenport.\footnote{\texttt{{http://www.astro.washington.edu/users/jrad/idl.html} }}
Specifically,
the figure color scheme\footnote{\texttt{{http://www.mrao.cam.ac.uk/{\textasciitilde{}}dag/CUBEHELIX/} }}
was developed by \citet{green2011}.
CJS and HBP acknowledge support from
   the Israel Science Foundation
   through the astrophysics I-CORE program 1829/12.
CJS thanks
   D.M.Grere,
   R. \& R. Whittaker for their hospitality during this work.

%\bibliographystyle{mnras}
%\bibliographystyle{mn2e}
%\bibliography{blob,jour}

\begin{thebibliography}{147}
\expandafter\ifx\csname natexlab\endcsname\relax\def\natexlab#1{#1}\fi

\bibitem[{{Aleksi{\'c}} {et~al}\mbox{.}(2014){Aleksi{\'c}}, {Ansoldi},
  {Antonelli}, {Antoranz}, {Babic}, {Bangale}, {Barrio}, {Gonz{\'a}lez},
  {Bednarek}, {Bernardini}, {Biasuzzi}, {Biland}, {Blanch}, {Bonnefoy},
  {Bonnoli}, {Borracci}, {Bretz}, {Carmona}, {Carosi}, {Colin}, {Colombo},
  {Contreras}, {Cortina}, {Covino}, {Da Vela}, {Dazzi}, {De Angelis}, {De
  Caneva}, {De Lotto}, {Wilhelmi}, {Mendez}, {Prester}, {Dorner}, {Doro},
  {Einecke}, {Eisenacher}, {Elsaesser}, {Fonseca}, {Font}, {Frantzen}, {Fruck},
  {Galindo}, {L{\'o}pez}, {Garczarczyk}, {Terrats}, {Gaug}, {Godinovi{\'c}},
  {Mu{\~n}oz}, {Gozzini}, {Hadasch}, {Hanabata}, {Hayashida}, {Herrera},
  {Hildebrand}, {Hose}, {Hrupec}, {Idec}, {Kadenius}, {Kellermann}, {Kodani},
  {Konno}, {Krause}, {Kubo}, {Kushida}, {La Barbera}, {Lelas}, {Lewandowska},
  {Lindfors}, {Lombardi}, {Longo}, {L{\'o}pez}, {L{\'o}pez-Coto},
  {L{\'o}pez-Oramas}, {Lorenz}, {Lozano}, {Makariev}, {Mallot}, {Maneva},
  {Mankuzhiyil}, {Mannheim}, {Maraschi}, {Marcote}, {Mariotti},
  {Mart{\'{\i}}nez}, {Mazin}, {Menzel}, {Miranda}, {Mirzoyan}, {Moralejo},
  {Munar-Adrover}, {Nakajima}, {Niedzwiecki}, {Nilsson}, {Nishijima}, {Noda},
  {Orito}, {Overkemping}, {Paiano}, {Palatiello}, {Paneque}, {Paoletti},
  {Paredes}, {Paredes-Fortuny}, {Persic}, {Poutanen}, {Moroni}, {Prandini},
  {Puljak}, {Reinthal}, {Rhode}, {Rib{\'o}}, {Rico}, {Garcia}, {R{\"u}gamer},
  {Saito}, {Saito}, {Satalecka}, {Scalzotto}, {Scapin}, {Schultz}, {Schweizer},
  {Shore}, {Sillanp{\"a}{\"a}}, {Sitarek}, {Snidaric}, {Sobczynska}, {Spanier},
  {Stamatescu}, {Stamerra}, {Steinbring}, {Storz}, {Strzys}, {Takalo},
  {Takami}, {Tavecchio}, {Temnikov}, {Terzi{\'c}}, {Tescaro}, {Teshima},
  {Thaele}, {Tibolla}, {Torres}, {Toyama}, {Treves}, {Uellenbeck}, {Vogler},
  {Zanin}, {Kadler}, {Schulz}, {Ros}, {Bach}, {Krau{\ss}}, \&
  {Wilms}}]{aleksic2014}
{Aleksi{\'c}} J. {et~al.}, 2014, Science, 346, 1080

\bibitem[{{Alexander} {et~al}\mbox{.}(2016){Alexander}, {Berger}, {Guillochon},
  {Zauderer}, \& {Williams}}]{k_alexander2016}
{Alexander} K.~D., {Berger} E., {Guillochon} J., {Zauderer} B.~A., {Williams}
  P.~K.~G., 2016, \apjl, 819, L25

\bibitem[{{Alloin} {et~al}\mbox{.}(1986){Alloin}, {Pelat}, {Phillips},
  {Fosbury}, \& {Freeman}}]{alloin1986}
{Alloin} D., {Pelat} D., {Phillips} M.~M., {Fosbury} R.~A.~E., {Freeman} K.,
  1986, \apj, 308, 23

\bibitem[{{Arcavi} {et~al}\mbox{.}(2014){Arcavi}, {Gal-Yam}, {Sullivan}, {Pan},
  {Cenko}, {Horesh}, {Ofek}, {De Cia}, {Yan}, {Yang}, {Howell}, {Tal},
  {Kulkarni}, {Tendulkar}, {Tang}, {Xu}, {Sternberg}, {Cohen}, {Bloom},
  {Nugent}, {Kasliwal}, {Perley}, {Quimby}, {Miller}, {Theissen}, \&
  {Laher}}]{arcavi2014}
{Arcavi} I. {et~al.}, 2014, \apj, 793, 38

\bibitem[{{Aretxaga}(1999)}]{aretxaga1999b}
{Aretxaga} I., 1999, in IAU Symposium, Vol. 193, Wolf-Rayet Phenomena in
  Massive Stars and Starburst Galaxies, {van der Hucht} K.~A., {Koenigsberger}
  G., {Eenens} P.~R.~J., eds., p. 716

\bibitem[{{Aretxaga} \& {Terlevich}(1994)}]{aretxaga1994}
{Aretxaga} I., {Terlevich} R., 1994, \mnras, 269, 462

\bibitem[{{Baldwin} {et~al}\mbox{.}(1995){Baldwin}, {Ferland}, {Korista}, \&
  {Verner}}]{baldwin1995}
{Baldwin} J., {Ferland} G., {Korista} K., {Verner} D., 1995, \apjl, 455, L119

\bibitem[{{Bartko} {et~al}\mbox{.}(2010){Bartko}, {Martins}, {Trippe}, {Fritz},
  {Genzel}, {Ott}, {Eisenhauer}, {Gillessen}, {Paumard}, {Alexander},
  {Dodds-Eden}, {Gerhard}, {Levin}, {Mascetti}, {Nayakshin}, {Perets},
  {Perrin}, {Pfuhl}, {Reid}, {Rouan}, {Zilka}, \& {Sternberg}}]{bartko2010}
{Bartko} H. {et~al.}, 2010, \apj, 708, 834

\bibitem[{{Baskin}, {Laor} \& {Stern}(2014){Baskin}, {Laor}, \&
  {Stern}}]{baskin2014a}
{Baskin} A., {Laor} A., {Stern} J., 2014, \mnras, 438, 604

\bibitem[{{Bianchi} {et~al}\mbox{.}(2005){Bianchi}, {Guainazzi}, {Matt},
  {Chiaberge}, {Iwasawa}, {Fiore}, \& {Maiolino}}]{bianchi2005}
{Bianchi} S., {Guainazzi} M., {Matt} G., {Chiaberge} M., {Iwasawa} K., {Fiore}
  F., {Maiolino} R., 2005, \aap, 442, 185

\bibitem[{{Blagorodnova} {et~al}\mbox{.}(2017){Blagorodnova}, {Gezari}, {Hung},
  {Kulkarni}, {Cenko}, {Pasham}, {Yan}, {Arcavi}, {Ben-Ami}, {Bue}, {Cantwell},
  {Cao}, {Castro-Tirado}, {Fender}, {Fremling}, {Gal-Yam}, {Ho}, {Horesh},
  {Hosseinzadeh}, {Kasliwal}, {Kong}, {Laher}, {Leloudas}, {Lunnan}, {Masci},
  {Mooley}, {Neill}, {Nugent}, {Powell}, {Valeev}, {Vreeswijk}, {Walters}, \&
  {Wozniak}}]{blagorodnova2017}
{Blagorodnova} N. {et~al.}, 2017, \apj, 844, 46

\bibitem[{{Blandford} \& {McKee}(1976)}]{blandford1976}
{Blandford} R.~D., {McKee} C.~F., 1976, Physics of Fluids, 19, 1130

\bibitem[{{Bloom} {et~al}\mbox{.}(2011){Bloom}, {Giannios}, {Metzger}, {Cenko},
  {Perley}, {Butler}, {Tanvir}, {Levan}, {O'Brien}, {Strubbe}, {De Colle},
  {Ramirez-Ruiz}, {Lee}, {Nayakshin}, {Quataert}, {King}, {Cucchiara},
  {Guillochon}, {Bower}, {Fruchter}, {Morgan}, \& {van der Horst}}]{bloom2011}
{Bloom} J.~S. {et~al.}, 2011, Science, 333, 203

\bibitem[{{Bower} {et~al}\mbox{.}(2013){Bower}, {Metzger}, {Cenko},
  {Silverman}, \& {Bloom}}]{bower2013}
{Bower} G.~C., {Metzger} B.~D., {Cenko} S.~B., {Silverman} J.~M., {Bloom}
  J.~S., 2013, \apj, 763, 84

\bibitem[{{Brown} {et~al}\mbox{.}(2009){Brown}, {Moustakas}, {Caldwell},
  {Palamara}, {Cool}, {Dey}, {Hickox}, {Jannuzi}, {Murray}, \&
  {Zaritsky}}]{brown2009}
{Brown} M.~J.~I. {et~al.}, 2009, \apj, 703, 150

\bibitem[{{Campana} {et~al}\mbox{.}(2015){Campana}, {Mainetti}, {Colpi},
  {Lodato}, {D'Avanzo}, {Evans}, \& {Moretti}}]{campana2015}
{Campana} S., {Mainetti} D., {Colpi} M., {Lodato} G., {D'Avanzo} P., {Evans}
  P.~A., {Moretti} A., 2015, \aap, 581, A17

\bibitem[{{Cenko} {et~al}\mbox{.}(2012{\natexlab{a}}){Cenko}, {Bloom},
  {Kulkarni}, {Strubbe}, {Miller}, {Butler}, {Quimby}, {Gal-Yam}, {Ofek},
  {Quataert}, {Bildsten}, {Poznanski}, {Perley}, {Morgan}, {Filippenko},
  {Frail}, {Arcavi}, {Ben-Ami}, {Cucchiara}, {Fassnacht}, {Green}, {Hook},
  {Howell}, {Lagattuta}, {Law}, {Kasliwal}, {Nugent}, {Silverman}, {Sullivan},
  {Tendulkar}, \& {Yaron}}]{cenko2012a}
{Cenko} S.~B. {et~al.}, 2012{\natexlab{a}}, \mnras, 420, 2684

\bibitem[{{Cenko} {et~al}\mbox{.}(2016){Cenko}, {Cucchiara}, {Roth},
  {Veilleux}, {Prochaska}, {Yan}, {Guillochon}, {Maksym}, {Arcavi}, {Butler},
  {Filippenko}, {Fruchter}, {Gezari}, {Kasen}, {Levan}, {Miller}, {Pasham},
  {Ramirez-Ruiz}, {Strubbe}, {Tanvir}, \& {Tombesi}}]{cenko2016}
{Cenko} S.~B. {et~al.}, 2016, \apjl, 818, L32

\bibitem[{{Cenko} {et~al}\mbox{.}(2012{\natexlab{b}}){Cenko}, {Krimm},
  {Horesh}, {Rau}, {Frail}, {Kennea}, {Levan}, {Holland}, {Butler}, {Quimby},
  {Bloom}, {Filippenko}, {Gal-Yam}, {Greiner}, {Kulkarni}, {Ofek}, {Olivares
  E.}, {Schady}, {Silverman}, {Tanvir}, \& {Xu}}]{cenko2012b}
{Cenko} S.~B. {et~al.}, 2012{\natexlab{b}}, \apj, 753, 77

\bibitem[{{Chen} {et~al}\mbox{.}(2009){Chen}, {Madau}, {Sesana}, \&
  {Liu}}]{chen2009}
{Chen} X., {Madau} P., {Sesana} A., {Liu} F.~K., 2009, \apjl, 697, L149

\bibitem[{{Chen} {et~al}\mbox{.}(2011){Chen}, {Sesana}, {Madau}, \&
  {Liu}}]{chen2011}
{Chen} X., {Sesana} A., {Madau} P., {Liu} F.~K., 2011, \apj, 729, 13

\bibitem[{{Chornock} {et~al}\mbox{.}(2014){Chornock}, {Berger}, {Gezari},
  {Zauderer}, {Rest}, {Chomiuk}, {Kamble}, {Soderberg}, {Czekala}, {Dittmann},
  {Drout}, {Foley}, {Fong}, {Huber}, {Kirshner}, {Lawrence}, {Lunnan},
  {Marion}, {Narayan}, {Riess}, {Roth}, {Sanders}, {Scolnic}, {Smartt},
  {Smith}, {Stubbs}, {Tonry}, {Burgett}, {Chambers}, {Flewelling}, {Hodapp},
  {Kaiser}, {Magnier}, {Martin}, {Neill}, {Price}, \&
  {Wainscoat}}]{chornock2014}
{Chornock} R. {et~al.}, 2014, \apj, 780, 44

\bibitem[{{Cohen} \& {Piran}(1999)}]{cohen1999}
{Cohen} E., {Piran} T., 1999, \apj, 518, 346

\bibitem[{{Collin} \& {Zahn}(1999{\natexlab{a}})}]{collin1999b}
{Collin} S., {Zahn} J.-P., 1999{\natexlab{a}}, \aap, 344, 433

\bibitem[{{Collin} \& {Zahn}(1999{\natexlab{b}})}]{collin1999a}
{Collin} S., {Zahn} J.-P., 1999{\natexlab{b}}, \apss, 265, 501

\bibitem[{{Comastri} {et~al}\mbox{.}(2002){Comastri}, {Mignoli}, {Ciliegi},
  {Severgnini}, {Maiolino}, {Brusa}, {Fiore}, {Baldi}, {Molendi}, {Morganti},
  {Vignali}, {La Franca}, {Matt}, \& {Perola}}]{comastri2002}
{Comastri} A. {et~al.}, 2002, \apj, 571, 771

\bibitem[{{Danforth} {et~al}\mbox{.}(2013){Danforth}, {Nalewajko}, {France}, \&
  {Keeney}}]{danforth2013}
{Danforth} C.~W., {Nalewajko} K., {France} K., {Keeney} B.~A., 2013, \apj, 764,
  57

\bibitem[{{Dolcini} {et~al}\mbox{.}(2005){Dolcini}, {Covino}, {Treves},
  {Palazzi}, {Pian}, {Molinari}, {Chincarini}, {Zerbi}, {Rodon{\'o}}, {Testa},
  {Tosti}, {Vitali}, {Antonelli}, {Conconi}, {Cutispoto}, {Monfardini},
  {Stefanon}, {D'Avanzo}, {Danziger}, {Fernandez-Soto}, \&
  {Meurs}}]{dolcini2005}
{Dolcini} A. {et~al.}, 2005, \aap, 443, L33

\bibitem[{{Dong} {et~al}\mbox{.}(2016){Dong}, {Shappee}, {Prieto}, {Jha},
  {Stanek}, {Holoien}, {Kochanek}, {Thompson}, {Morrell}, {Thompson}, {Basu},
  {Beacom}, {Bersier}, {Brimacombe}, {Brown}, {Bufano}, {Chen}, {Conseil},
  {Danilet}, {Falco}, {Grupe}, {Kiyota}, {Masi}, {Nicholls}, {Olivares E.},
  {Pignata}, {Pojmanski}, {Simonian}, {Szczygiel}, \& {Wo{\'z}niak}}]{dong2016}
{Dong} S. {et~al.}, 2016, Science, 351, 257

\bibitem[{{Dou} {et~al}\mbox{.}(2017){Dou}, {Wang}, {Yan}, {Jiang}, {Yang},
  {Cutri}, {Mainzer}, \& {Peng}}]{dou2017}
{Dou} L., {Wang} T., {Yan} L., {Jiang} N., {Yang} C., {Cutri} R.~M., {Mainzer}
  A., {Peng} B., 2017, \apjl, 841, L8

\bibitem[{{Dou} {et~al}\mbox{.}(2016){Dou}, {Wang}, {Jiang}, {Yang}, {Lyu}, \&
  {Zhou}}]{dou2016}
{Dou} L., {Wang} T.-g., {Jiang} N., {Yang} C., {Lyu} J., {Zhou} H., 2016, \apj,
  832, 188

\bibitem[{{Drake} {et~al}\mbox{.}(2011){Drake}, {Djorgovski}, {Mahabal},
  {Anderson}, {Roy}, {Mohan}, {Ravindranath}, {Frail}, {Gezari}, {Neill}, {Ho},
  {Prieto}, {Thompson}, {Thorstensen}, {Wagner}, {Kowalski}, {Chiang}, {Grove},
  {Schinzel}, {Wood}, {Carrasco}, {Recillas}, {Kewley}, {Archana}, {Basu},
  {Wadadekar}, {Kumar}, {Myers}, {Phinney}, {Williams}, {Graham}, {Catelan},
  {Beshore}, {Larson}, \& {Christensen}}]{drake2011}
{Drake} A.~J. {et~al.}, 2011, \apj, 735, 106

\bibitem[{{Dressler} \& {Gunn}(1983)}]{dressler1983}
{Dressler} A., {Gunn} J.~E., 1983, \apj, 270, 7

\bibitem[{{Eddington}(1918)}]{eddington1918b}
{Eddington} A.~S., 1918, \apj, 48, 205

\bibitem[{{Eggen}(1973)}]{eggen1973}
{Eggen} O.~J., 1973, \apjl, 186, L1

\bibitem[{{Elvis} {et~al}\mbox{.}(1981){Elvis}, {Schreier}, {Tonry}, {Davis},
  \& {Huchra}}]{elvis1981}
{Elvis} M., {Schreier} E.~J., {Tonry} J., {Davis} M., {Huchra} J.~P., 1981,
  \apj, 246, 20

\bibitem[{{Ferland} {et~al}\mbox{.}(2013){Ferland}, {Porter}, {van Hoof},
  {Williams}, {Abel}, {Lykins}, {Shaw}, {Henney}, \& {Stancil}}]{ferland2013}
{Ferland} G.~J. {et~al.}, 2013, \rmxaa, 49, 137

\bibitem[{{Fiore} {et~al}\mbox{.}(2003){Fiore}, {Brusa}, {Cocchia}, {Baldi},
  {Carangelo}, {Ciliegi}, {Comastri}, {La Franca}, {Maiolino}, {Matt},
  {Molendi}, {Mignoli}, {Perola}, {Severgnini}, \& {Vignali}}]{fiore2003}
{Fiore} F. {et~al.}, 2003, \aap, 409, 79

\bibitem[{{Frank} \& {Rees}(1976)}]{frank1976}
{Frank} J., {Rees} M.~J., 1976, \mnras, 176, 633

\bibitem[{{French}, {Arcavi} \& {Zabludoff}(2016){French}, {Arcavi}, \&
  {Zabludoff}}]{french2016}
{French} K.~D., {Arcavi} I., {Zabludoff} A., 2016, \apjl, 818, L21

\bibitem[{{Gal-Yam}(2012)}]{galyam2012}
{Gal-Yam} A., 2012, Science, 337, 927

\bibitem[{{Gaskell} \& {Rojas Lobos}(2014)}]{gaskell2014}
{Gaskell} C.~M., {Rojas Lobos} P.~A., 2014, \mnras, 438, L36

\bibitem[{{Generozov} {et~al}\mbox{.}(2017){Generozov}, {Mimica}, {Metzger},
  {Stone}, {Giannios}, \& {Aloy}}]{generozov2016}
{Generozov} A., {Mimica} P., {Metzger} B.~D., {Stone} N.~C., {Giannios} D.,
  {Aloy} M.~A., 2017, \mnras, 464, 2481

\bibitem[{{Gezari} {et~al}\mbox{.}(2008){Gezari}, {Basa}, {Martin}, {Bazin},
  {Forster}, {Milliard}, {Halpern}, {Friedman}, {Morrissey}, {Neff},
  {Schiminovich}, {Seibert}, {Small}, \& {Wyder}}]{gezari2008}
{Gezari} S. {et~al.}, 2008, \apj, 676, 944

\bibitem[{{Gezari} {et~al}\mbox{.}(2015){Gezari}, {Chornock}, {Lawrence},
  {Rest}, {Jones}, {Berger}, {Challis}, \& {Narayan}}]{gezari2015}
{Gezari} S., {Chornock} R., {Lawrence} A., {Rest} A., {Jones} D.~O., {Berger}
  E., {Challis} P.~M., {Narayan} G., 2015, \apjl, 815, L5

\bibitem[{{Gezari} {et~al}\mbox{.}(2012){Gezari}, {Chornock}, {Rest}, {Huber},
  {Forster}, {Berger}, {Challis}, {Neill}, {Martin}, {Heckman}, {Lawrence},
  {Norman}, {Narayan}, {Foley}, {Marion}, {Scolnic}, {Chomiuk}, {Soderberg},
  {Smith}, {Kirshner}, {Riess}, {Smartt}, {Stubbs}, {Tonry}, {Wood-Vasey},
  {Burgett}, {Chambers}, {Grav}, {Heasley}, {Kaiser}, {Kudritzki}, {Magnier},
  {Morgan}, \& {Price}}]{gezari2012}
{Gezari} S. {et~al.}, 2012, \nat, 485, 217

\bibitem[{{Gezari} {et~al}\mbox{.}(2009){Gezari}, {Heckman}, {Cenko},
  {Eracleous}, {Forster}, {Gon{\c c}alves}, {Martin}, {Morrissey}, {Neff},
  {Seibert}, {Schiminovich}, \& {Wyder}}]{gezari2009}
{Gezari} S. {et~al.}, 2009, \apj, 698, 1367

\bibitem[{{Goel} {et~al}\mbox{.}(2015){Goel}, {Maity}, {Roy}, \&
  {Sarkar}}]{goel2015}
{Goel} A., {Maity} R., {Roy} P., {Sarkar} T., 2015, \prd, 91, 104029

\bibitem[{{Goto} {et~al}\mbox{.}(2008){Goto}, {Kawai}, {Shimono}, {Sugai},
  {Yagi}, \& {Hattori}}]{goto2008}
{Goto} T., {Kawai} A., {Shimono} A., {Sugai} H., {Yagi} M., {Hattori} T., 2008,
  \mnras, 386, 1355

\bibitem[{{Goto} {et~al}\mbox{.}(2003){Goto}, {Nichol}, {Okamura}, {Sekiguchi},
  {Miller}, {Bernardi}, {Hopkins}, {Tremonti}, {Connolly}, {Castander},
  {Brinkmann}, {Fukugita}, {Harvanek}, {Ivezic}, {Kleinman}, {Krzesinski},
  {Long}, {Loveday}, {Neilsen}, {Newman}, {Nitta}, {Snedden}, \&
  {Subbarao}}]{goto2003}
{Goto} T. {et~al.}, 2003, \pasj, 55, 771

\bibitem[{{Green}(2011)}]{green2011}
{Green} D.~A., 2011, Bulletin of the Astronomical Society of India, 39, 289

\bibitem[{{Griffiths} {et~al}\mbox{.}(1995){Griffiths}, {Georgantopoulos},
  {Boyle}, {Stewart}, {Shanks}, \& {della Ceca}}]{griffiths1995}
{Griffiths} R.~E., {Georgantopoulos} I., {Boyle} B.~J., {Stewart} G.~C.,
  {Shanks} T., {della Ceca} R., 1995, \mnras, 275, 77

\bibitem[{{Guillochon}, {Manukian} \& {Ramirez-Ruiz}(2014){Guillochon},
  {Manukian}, \& {Ramirez-Ruiz}}]{guillochon2014}
{Guillochon} J., {Manukian} H., {Ramirez-Ruiz} E., 2014, \apj, 783, 23

\bibitem[{{Guillochon} \& {Ramirez-Ruiz}(2013)}]{guillochon2013}
{Guillochon} J., {Ramirez-Ruiz} E., 2013, \apj, 767, 25

\bibitem[{{Guillochon} \& {Ramirez-Ruiz}(2015)}]{guillochon2015}
{Guillochon} J., {Ramirez-Ruiz} E., 2015, \apj, 809, 166

\bibitem[{{Gurzadian} \& {Ozernoi}(1981)}]{gurzadian1981}
{Gurzadian} V.~G., {Ozernoi} L.~M., 1981, \aap, 95, 39

\bibitem[{{Hamers} \& {Perets}(2017)}]{hamers2017}
{Hamers} A.~S., {Perets} H.~B., 2017, \apj, 846, 123

\bibitem[{{Hills}(1975)}]{hills1975}
{Hills} J.~G., 1975, \nat, 254, 295

\bibitem[{{Holoien} {et~al}\mbox{.}(2016{\natexlab{a}}){Holoien}, {Kochanek},
  {Prieto}, {Grupe}, {Chen}, {Godoy-Rivera}, {Stanek}, {Shappee}, {Dong},
  {Brown}, {Basu}, {Beacom}, {Bersier}, {Brimacombe}, {Carlson}, {Falco},
  {Johnston}, {Madore}, {Pojmanski}, \& {Seibert}}]{holoien2016a}
{Holoien} T.~W.-S. {et~al.}, 2016{\natexlab{a}}, \mnras, 463, 3813

\bibitem[{{Holoien} {et~al}\mbox{.}(2016{\natexlab{b}}){Holoien}, {Kochanek},
  {Prieto}, {Stanek}, {Dong}, {Shappee}, {Grupe}, {Brown}, {Basu}, {Beacom},
  {Bersier}, {Brimacombe}, {Danilet}, {Falco}, {Guo}, {Jose}, {Herczeg},
  {Long}, {Pojmanski}, {Simonian}, {Szczygie{\l}}, {Thompson}, {Thorstensen},
  {Wagner}, \& {Wo{\'z}niak}}]{holoien2016b}
{Holoien} T.~W.-S. {et~al.}, 2016{\natexlab{b}}, \mnras, 455, 2918

\bibitem[{{Holoien} {et~al}\mbox{.}(2014){Holoien}, {Prieto}, {Bersier},
  {Kochanek}, {Stanek}, {Shappee}, {Grupe}, {Basu}, {Beacom}, {Brimacombe},
  {Brown}, {Davis}, {Jencson}, {Pojmanski}, \& {Szczygie{\l}}}]{holoien2014}
{Holoien} T.~W.-S. {et~al.}, 2014, \mnras, 445, 3263

\bibitem[{{Hung} {et~al}\mbox{.}(2017){Hung}, {Gezari}, {Blagorodnova}, {Roth},
  {Cenko}, {Kulkarni}, {Horesh}, {Arcavi}, {McCully}, {Yan}, {Lunnan},
  {Fremling}, {Cao}, {Nugent}, \& {Wozniak}}]{hung2017}
{Hung} T. {et~al.}, 2017, \apj, 842, 29

\bibitem[{{Inserra} {et~al}\mbox{.}(2013){Inserra} {et~al.}}]{inserra2013}
{Inserra} C., {et~al.}, 2013, \apj, 770, 128

\bibitem[{{Ivanov}, {Polnarev} \& {Saha}(2005){Ivanov}, {Polnarev}, \&
  {Saha}}]{ivanov2005}
{Ivanov} P.~B., {Polnarev} A.~G., {Saha} P., 2005, \mnras, 358, 1361

\bibitem[{{Jaroszynski}, {Abramowicz} \& {Paczynski}(1980){Jaroszynski},
  {Abramowicz}, \& {Paczynski}}]{jaroszynski1980}
{Jaroszynski} M., {Abramowicz} M.~A., {Paczynski} B., 1980, \actaa, 30, 1

\bibitem[{{Jiang} {et~al}\mbox{.}(2016){Jiang}, {Dou}, {Wang}, {Yang}, {Lyu},
  \& {Zhou}}]{jiang2016}
{Jiang} N., {Dou} L., {Wang} T., {Yang} C., {Lyu} J., {Zhou} H., 2016, \apjl,
  828, L14

\bibitem[{{Kaastra} {et~al}\mbox{.}(2014){Kaastra}, {Kriss}, {Cappi},
  {Mehdipour}, {Petrucci}, {Steenbrugge}, {Arav}, {Behar}, {Bianchi},
  {Boissay}, {Branduardi-Raymont}, {Chamberlain}, {Costantini}, {Ely},
  {Ebrero}, {Di Gesu}, {Harrison}, {Kaspi}, {Malzac}, {De Marco}, {Matt},
  {Nandra}, {Paltani}, {Person}, {Peterson}, {Pinto}, {Ponti}, {Nu{\~n}ez}, {De
  Rosa}, {Seta}, {Ursini}, {de Vries}, {Walton}, \& {Whewell}}]{kaastra2014}
{Kaastra} J.~S. {et~al.}, 2014, Science, 345, 64

\bibitem[{{Kaspi} {et~al}\mbox{.}(2005){Kaspi}, {Maoz}, {Netzer}, {Peterson},
  {Vestergaard}, \& {Jannuzi}}]{kaspi2005}
{Kaspi} S., {Maoz} D., {Netzer} H., {Peterson} B.~M., {Vestergaard} M.,
  {Jannuzi} B.~T., 2005, \apj, 629, 61

\bibitem[{{Kato} \& {H{\= o}shi}(1978)}]{kato1978}
{Kato} M., {H{\= o}shi} R., 1978, Progress of Theoretical Physics, 60, 1692

\bibitem[{{Kewley}, {Geller} \& {Barton}(2006){Kewley}, {Geller}, \&
  {Barton}}]{kewley2006}
{Kewley} L.~J., {Geller} M.~J., {Barton} E.~J., 2006, \aj, 131, 2004

\bibitem[{{Kim} {et~al}\mbox{.}(2006){Kim}, {Barkhouse}, {Romero-Colmenero},
  {Green}, {Kim}, {Mossman}, {Schlegel}, {Silverman}, {Aldcroft}, {Anderson},
  {Ivezic}, {Kashyap}, {Tananbaum}, \& {Wilkes}}]{kim2006}
{Kim} D.-W. {et~al.}, 2006, \apj, 644, 829

\bibitem[{{Kim}(2010)}]{kim2010}
{Kim} S.-W., 2010, Journal of Korean Physical Society, 57, 673

\bibitem[{{Komossa}(2015)}]{komossa2015b}
{Komossa} S., 2015, Journal of High Energy Astrophysics, 7, 148

\bibitem[{{Komossa} {et~al}\mbox{.}(2009){Komossa}, {Zhou}, {Rau}, {Dopita},
  {Gal-Yam}, {Greiner}, {Zuther}, {Salvato}, {Xu}, {Lu}, {Saxton}, \&
  {Ajello}}]{komossa2009}
{Komossa} S. {et~al.}, 2009, \apj, 701, 105

\bibitem[{{Korista} \& {Goad}(2004)}]{korista2004}
{Korista} K.~T., {Goad} M.~R., 2004, \apj, 606, 749

\bibitem[{{Lacy}, {Townes} \& {Hollenbach}(1982){Lacy}, {Townes}, \&
  {Hollenbach}}]{lacy1982}
{Lacy} J.~H., {Townes} C.~H., {Hollenbach} D.~J., 1982, \apj, 262, 120

\bibitem[{{Laor}(2003)}]{laor2003}
{Laor} A., 2003, \apj, 590, 86

\bibitem[{{Laor} {et~al}\mbox{.}(1997){Laor}, {Fiore}, {Elvis}, {Wilkes}, \&
  {McDowell}}]{laor1997}
{Laor} A., {Fiore} F., {Elvis} M., {Wilkes} B.~J., {McDowell} J.~C., 1997,
  \apj, 477, 93

\bibitem[{{Leloudas} {et~al}\mbox{.}(2016){Leloudas}, {Fraser}, {Stone},
  {van~Velzen}, {Jonker}, {Arcavi}, {Fremling}, {Maund}, {Smartt},
  {Kr{\`i}hler}, {Miller-Jones}, {Vreeswijk}, {Gal-Yam}, {Mazzali}, {De Cia},
  {Howell}, {Inserra}, {Patat}, {de Ugarte Postigo}, {Yaron}, {Ashall}, {Bar},
  {Campbell}, {Chen}, {Childress}, {Elias-Rosa}, {Harmanen}, {Hosseinzadeh},
  {Johansson}, {Kangas}, {Kankare}, {Kim}, {Kuncarayakti}, {Lyman}, {Magee},
  {Maguire}, {Malesani}, {Mattila}, {McCully}, {Nicholl}, {Prentice},
  {Romero-Ca{\~n}izales}, {Schulze}, {Smith}, {Sollerman}, {Sullivan},
  {Tucker}, {Valenti}, {Wheeler}, \& {Young}}]{leloudas2017}
{Leloudas} G. {et~al.}, 2016, Nature Astronomy, 1, 0002

\bibitem[{{Levan} {et~al}\mbox{.}(2011){Levan}, {Tanvir}, {Cenko}, {Perley},
  {Wiersema}, {Bloom}, {Fruchter}, {Postigo}, {O'Brien}, {Butler}, {van der
  Horst}, {Leloudas}, {Morgan}, {Misra}, {Bower}, {Farihi}, {Tunnicliffe},
  {Modjaz}, {Silverman}, {Hjorth}, {Th{\"o}ne}, {Cucchiara}, {Cer{\'o}n},
  {Castro-Tirado}, {Arnold}, {Bremer}, {Brodie}, {Carroll}, {Cooper}, {Curran},
  {Cutri}, {Ehle}, {Forbes}, {Fynbo}, {Gorosabel}, {Graham}, {Hoffman},
  {Guziy}, {Jakobsson}, {Kamble}, {Kerr}, {Kasliwal}, {Kouveliotou},
  {Kocevski}, {Law}, {Nugent}, {Ofek}, {Poznanski}, {Quimby}, {Rol},
  {Romanowsky}, {S{\'a}nchez-Ram{\'{\i}}rez}, {Schulze}, {Singh}, {van
  Spaandonk}, {Starling}, {Strom}, {Tello}, {Vaduvescu}, {Wheatley}, {Wijers},
  {Winters}, \& {Xu}}]{levan2011}
{Levan} A.~J. {et~al.}, 2011, Science, 333, 199

\bibitem[{{Lidskii} \& {Ozernoi}(1979)}]{lidskii1979}
{Lidskii} V.~V., {Ozernoi} L.~M., 1979, Soviet Astronomy Letters, 5, 16

\bibitem[{{Lodato}, {King} \& {Pringle}(2009){Lodato}, {King}, \&
  {Pringle}}]{lodato2009}
{Lodato} G., {King} A.~R., {Pringle} J.~E., 2009, \mnras, 392, 332

\bibitem[{{Loeb} \& {Ulmer}(1997)}]{loeb1997}
{Loeb} A., {Ulmer} A., 1997, \apj, 489, 573

\bibitem[{{Luminet} \& {Marck}(1985)}]{luminet1985}
{Luminet} J.-P., {Marck} J.-A., 1985, \mnras, 212, 57

\bibitem[{{Magorrian} \& {Tremaine}(1999)}]{magorrian1999}
{Magorrian} J., {Tremaine} S., 1999, \mnras, 309, 447

\bibitem[{{Mathews} \& {Ferland}(1987)}]{mathews1987}
{Mathews} W.~G., {Ferland} G.~J., 1987, \apj, 323, 456

\bibitem[{{McElroy} {et~al}\mbox{.}(2016){McElroy}, {Husemann}, {Croom},
  {Davis}, {Bennert}, {Busch}, {Combes}, {Eckart}, {Perez-Torres}, {Powell},
  {Scharw{\"a}chter}, {Tremblay}, \& {Urrutia}}]{mcelroy2016}
{McElroy} R.~E. {et~al.}, 2016, \aap, 593, L8

\bibitem[{{Meliani} {et~al}\mbox{.}(2015){Meliani}, {Vincent},
  {Grandcl{\'e}ment}, {Gourgoulhon}, {Monceau-Baroux}, \&
  {Straub}}]{meliani2015}
{Meliani} Z., {Vincent} F.~H., {Grandcl{\'e}ment} P., {Gourgoulhon} E.,
  {Monceau-Baroux} R., {Straub} O., 2015, Classical and Quantum Gravity, 32,
  235022

\bibitem[{{Metzger} \& {Stone}(2016)}]{metzger2016}
{Metzger} B.~D., {Stone} N.~C., 2016, \mnras, 461, 948

\bibitem[{{Miller} {et~al}\mbox{.}(2015){Miller}, {Kaastra}, {Miller},
  {Reynolds}, {Brown}, {Cenko}, {Drake}, {Gezari}, {Guillochon}, {Gultekin},
  {Irwin}, {Levan}, {Maitra}, {Maksym}, {Mushotzky}, {O'Brien}, {Paerels}, {de
  Plaa}, {Ramirez-Ruiz}, {Strohmayer}, \& {Tanvir}}]{miller2015}
{Miller} J.~M. {et~al.}, 2015, \nat, 526, 542

\bibitem[{{Moriya} {et~al}\mbox{.}(2017){Moriya}, {Tanaka}, {Morokuma}, \&
  {Ohsuga}}]{moriya2017}
{Moriya} T.~J., {Tanaka} M., {Morokuma} T., {Ohsuga} K., 2017, \apjl, 843, L19

\bibitem[{{Netzer}(2013)}]{netzer2013}
{Netzer} H., 2013, {The Physics and Evolution of Active Galactic Nuclei}

\bibitem[{{Oknyansky} {et~al}\mbox{.}(2016){Oknyansky}, {Huseynov},
  {Artamonov}, {Lipunov}, {Gorbovskoy}, {Kuznetsov}, {Balanutza}, {Tatarnikov},
  {Metlov}, {Shatsky}, {Nadzhip}, {Burlak}, {Mikailov}, {Salmanov}, \&
  {Gaskell}}]{oknyanky2016}
{Oknyansky} V.~L. {et~al.}, 2016, The Astronomer's Telegram, 9015

\bibitem[{{Ozernoi} \& {Reinhardt}(1978)}]{ozernoi1978}
{Ozernoi} L.~M., {Reinhardt} M., 1978, \apss, 59, 171

\bibitem[{{Perets} {et~al}\mbox{.}(2016){Perets}, {Li}, {Lombardi}, \&
  {Milcarek}}]{perets2016_06}
{Perets} H.~B., {Li} Z., {Lombardi}, Jr. J.~C., {Milcarek}, Jr. S.~R., 2016,
  \apj, 823, 113

\bibitem[{{Piran} {et~al}\mbox{.}(2015){Piran}, {Svirski}, {Krolik}, {Cheng},
  \& {Shiokawa}}]{piran2015}
{Piran} T., {Svirski} G., {Krolik} J., {Cheng} R.~M., {Shiokawa} H., 2015,
  \apj, 806, 164

\bibitem[{{Pracy} {et~al}\mbox{.}(2009){Pracy}, {Kuntschner}, {Couch}, {Blake},
  {Bekki}, \& {Briggs}}]{pracy2009}
{Pracy} M.~B., {Kuntschner} H., {Couch} W.~J., {Blake} C., {Bekki} K., {Briggs}
  F., 2009, \mnras, 396, 1349

\bibitem[{{Pracy} {et~al}\mbox{.}(2014){Pracy}, {Owers}, {Zwaan}, {Couch},
  {Kuntschner}, {Croom}, \& {Sadler}}]{pracy2014}
{Pracy} M.~B., {Owers} M.~S., {Zwaan} M., {Couch} W., {Kuntschner} H., {Croom}
  S.~M., {Sadler} E.~M., 2014, \mnras, 443, 388

\bibitem[{{Quinlan}(1996)}]{quinlan1996}
{Quinlan} G.~D., 1996, \na, 1, 35

\bibitem[{{Quintero} {et~al}\mbox{.}(2004){Quintero}, {Hogg}, {Blanton},
  {Schlegel}, {Eisenstein}, {Gunn}, {Brinkmann}, {Fukugita}, {Glazebrook}, \&
  {Goto}}]{quintero2004}
{Quintero} A.~D. {et~al.}, 2004, \apj, 602, 190

\bibitem[{{Rees}(1988)}]{rees1988}
{Rees} M.~J., 1988, \nat, 333, 523

\bibitem[{{Renzini} {et~al}\mbox{.}(1995){Renzini}, {Greggio}, {di Serego
  Alighieri}, {Cappellari}, {Burstein}, \& {Bertola}}]{renzini1995}
{Renzini} A., {Greggio} L., {di Serego Alighieri} S., {Cappellari} M.,
  {Burstein} D., {Bertola} F., 1995, \nat, 378, 39

\bibitem[{{Roth} {et~al}\mbox{.}(2016){Roth}, {Kasen}, {Guillochon}, \&
  {Ramirez-Ruiz}}]{roth2016}
{Roth} N., {Kasen} D., {Guillochon} J., {Ramirez-Ruiz} E., 2016, \apj, 827, 3

\bibitem[{{Ryu} {et~al}\mbox{.}(2013){Ryu}, {Nobukawa}, {Nakashima}, {Tsuru},
  {Koyama}, \& {Uchiyama}}]{ryu2013}
{Ryu} S.~G., {Nobukawa} M., {Nakashima} S., {Tsuru} T.~G., {Koyama} K.,
  {Uchiyama} H., 2013, \pasj, 65, 33

\bibitem[{{Sandrinelli}, {Covino} \& {Treves}(2014){Sandrinelli}, {Covino}, \&
  {Treves}}]{sandrinelli2014}
{Sandrinelli} A., {Covino} S., {Treves} A., 2014, \aap, 562, A79

\bibitem[{{Saxton} {et~al}\mbox{.}(2012){Saxton}, {Soria}, {Wu}, \&
  {Kuin}}]{saxton2012}
{Saxton} C.~J., {Soria} R., {Wu} K., {Kuin} N.~P.~M., 2012, \mnras, 422, 1625

\bibitem[{{Saxton}, {Younsi} \& {Wu}(2016){Saxton}, {Younsi}, \&
  {Wu}}]{saxton2016a}
{Saxton} C.~J., {Younsi} Z., {Wu} K., 2016, \mnras, 461, 4295

\bibitem[{{Saxton} {et~al}\mbox{.}(2015){Saxton}, {Motta}, {Komossa}, \&
  {Read}}]{r_saxton2015}
{Saxton} R.~D., {Motta} S.~E., {Komossa} S., {Read} A.~M., 2015, \mnras, 454,
  2798

\bibitem[{{Schwarzschild}(1916)}]{schwarzschild1916}
{Schwarzschild} K., 1916, Abh.~Konigl.~Preuss.~Akad.~Wissenschaften Jahre,
  1916, 189

\bibitem[{{Severgnini} {et~al}\mbox{.}(2003){Severgnini}, {Caccianiga},
  {Braito}, {Della Ceca}, {Maccacaro}, {Wolter}, {Sekiguchi}, {Sasaki},
  {Yoshida}, {Akiyama}, {Watson}, {Barcons}, {Carrera}, {Pietsch}, \&
  {Webb}}]{severgnini2003}
{Severgnini} P. {et~al.}, 2003, \aap, 406, 483

\bibitem[{{Shappee} {et~al}\mbox{.}(2014){Shappee}, {Prieto}, {Grupe},
  {Kochanek}, {Stanek}, {De Rosa}, {Mathur}, {Zu}, {Peterson}, {Pogge},
  {Komossa}, {Im}, {Jencson}, {Holoien}, {Basu}, {Beacom}, {Szczygie{\l}},
  {Brimacombe}, {Adams}, {Campillay}, {Choi}, {Contreras}, {Dietrich},
  {Dubberley}, {Elphick}, {Foale}, {Giustini}, {Gonzalez}, {Hawkins}, {Howell},
  {Hsiao}, {Koss}, {Leighly}, {Morrell}, {Mudd}, {Mullins}, {Nugent},
  {Parrent}, {Phillips}, {Pojmanski}, {Rosing}, {Ross}, {Sand}, {Terndrup},
  {Valenti}, {Walker}, \& {Yoon}}]{shappee2014}
{Shappee} B.~J. {et~al.}, 2014, \apj, 788, 48

\bibitem[{{Sirko} \& {Goodman}(2003)}]{sirko2003}
{Sirko} E., {Goodman} J., 2003, \mnras, 341, 501

\bibitem[{{Smith} {et~al}\mbox{.}(2015){Smith}, {Mauerhan}, {Cenko},
  {Kasliwal}, {Silverman}, {Filippenko}, {Gal-Yam}, {Clubb}, {Graham},
  {Leonard}, {Horst}, {Williams}, {Andrews}, {Kulkarni}, {Nugent}, {Sullivan},
  {Maguire}, {Xu}, \& {Ben-Ami}}]{smith2015}
{Smith} N. {et~al.}, 2015, \mnras, 449, 1876

\bibitem[{{Stone} \& {Metzger}(2016)}]{stone2016a}
{Stone} N.~C., {Metzger} B.~D., 2016, \mnras, 455, 859

\bibitem[{{Stone} \& {van~Velzen}(2016)}]{stone2016b}
{Stone} N.~C., {van~Velzen} S., 2016, \apjl, 825, L14

\bibitem[{{Storchi-Bergmann}, {Baldwin} \& {Wilson}(1993){Storchi-Bergmann},
  {Baldwin}, \& {Wilson}}]{storchi1993}
{Storchi-Bergmann} T., {Baldwin} J.~A., {Wilson} A.~S., 1993, \apjl, 410, L11

\bibitem[{{Strubbe} \& {Murray}(2015)}]{strubbe2015}
{Strubbe} L.~E., {Murray} N., 2015, \mnras, 454, 2321

\bibitem[{{Strubbe} \& {Quataert}(2009)}]{strubbe2009}
{Strubbe} L.~E., {Quataert} E., 2009, \mnras, 400, 2070

\bibitem[{{Svirski}, {Piran} \& {Krolik}(2017){Svirski}, {Piran}, \&
  {Krolik}}]{svirski2015}
{Svirski} G., {Piran} T., {Krolik} J., 2017, \mnras, 467, 1426

\bibitem[{{Tanaka}(2013)}]{tanaka2013}
{Tanaka} T.~L., 2013, \mnras, 434, 2275

\bibitem[{{Terlevich} {et~al}\mbox{.}(1992){Terlevich}, {Tenorio-Tagle},
  {Franco}, \& {Melnick}}]{terlevich1992}
{Terlevich} R., {Tenorio-Tagle} G., {Franco} J., {Melnick} J., 1992, \mnras,
  255, 713

\bibitem[{{Tran}(2001)}]{tran2001}
{Tran} H.~D., 2001, \apjl, 554, L19

\bibitem[{{Ulmer}(1999)}]{ulmer1999}
{Ulmer} A., 1999, \apj, 514, 180

\bibitem[{{Ulrich}, {Maraschi} \& {Urry}(1997){Ulrich}, {Maraschi}, \&
  {Urry}}]{ulrich1997}
{Ulrich} M.-H., {Maraschi} L., {Urry} C.~M., 1997, \araa, 35, 445

\bibitem[{{van~Velzen}(2017)}]{vanvelzen2017}
{van~Velzen} S., 2017, ArXiv e-prints, 1707.03458

\bibitem[{{van~Velzen} {et~al}\mbox{.}(2016{\natexlab{a}}){van~Velzen},
  {Anderson}, {Stone}, {Fraser}, {Wevers}, {Metzger}, {Jonker}, {van der
  Horst}, {Staley}, {Mendez}, {Miller-Jones}, {Hodgkin}, {Campbell}, \&
  {Fender}}]{vanvelzen2016a}
{van~Velzen} S. {et~al.}, 2016{\natexlab{a}}, Science, 351, 62

\bibitem[{{van~Velzen} {et~al}\mbox{.}(2011){van~Velzen}, {Farrar}, {Gezari},
  {Morrell}, {Zaritsky}, {{\"O}stman}, {Smith}, {Gelfand}, \&
  {Drake}}]{vanvelzen2011b}
{van~Velzen} S. {et~al.}, 2011, \apj, 741, 73

\bibitem[{{van~Velzen} {et~al}\mbox{.}(2013){van~Velzen}, {Frail},
  {K{\"o}rding}, \& {Falcke}}]{vanvelzen2013}
{van~Velzen} S., {Frail} D.~A., {K{\"o}rding} E., {Falcke} H., 2013, \aap, 552,
  A5

\bibitem[{{van~Velzen}, {K{\"o}rding} \& {Falcke}(2011){van~Velzen},
  {K{\"o}rding}, \& {Falcke}}]{vanvelzen2011a}
{van~Velzen} S., {K{\"o}rding} E., {Falcke} H., 2011, \mnras, 417, L51

\bibitem[{{van~Velzen} {et~al}\mbox{.}(2016{\natexlab{b}}){van~Velzen},
  {Mendez}, {Krolik}, \& {Gorjian}}]{vanvelzen2016b}
{van~Velzen} S., {Mendez} A.~J., {Krolik} J.~H., {Gorjian} V.,
  2016{\natexlab{b}}, \apj, 829, 19

\bibitem[{{Vasiliev}(2014)}]{vasiliev2014a}
{Vasiliev} E., 2014, Classical and Quantum Gravity, 31, 244002

\bibitem[{{Vergani} {et~al}\mbox{.}(2010){Vergani}, {Zamorani}, {Lilly},
  {Lamareille}, {Halliday}, {Scodeggio}, {Vignali}, {Ciliegi}, {Bolzonella},
  {Bondi}, {Kova{\v c}}, {Knobel}, {Zucca}, {Caputi}, {Pozzetti}, {Bardelli},
  {Mignoli}, {Iovino}, {Carollo}, {Contini}, {Kneib}, {Le F{\`e}vre},
  {Mainieri}, {Renzini}, {Bongiorno}, {Coppa}, {Cucciati}, {de la Torre}, {de
  Ravel}, {Franzetti}, {Garilli}, {Kampczyk}, {Le Borgne}, {Le Brun}, {Maier},
  {Pello}, {Peng}, {Perez Montero}, {Ricciardelli}, {Silverman}, {Tanaka},
  {Tasca}, {Tresse}, {Abbas}, {Bottini}, {Cappi}, {Cassata}, {Cimatti},
  {Guzzo}, {Koekemoer}, {Leauthaud}, {Maccagni}, {Marinoni}, {McCracken},
  {Memeo}, {Meneux}, {Oesch}, {Porciani}, {Scaramella}, {Capak}, {Sanders},
  {Scoville}, \& {Taniguchi}}]{vergani2010}
{Vergani} D. {et~al.}, 2010, \aap, 509, A42

\bibitem[{{Verner} \& {Ferland}(1996)}]{verner1996}
{Verner} D.~A., {Ferland} G.~J., 1996, \apjs, 103, 467

\bibitem[{{Vreeswijk} {et~al}\mbox{.}(2014){Vreeswijk}
  {et~al.}}]{vreeswijk2014}
{Vreeswijk} P.~M., {et~al.}, 2014, \apj, 797, 24

\bibitem[{{Wang} \& {Merritt}(2004)}]{wang2004}
{Wang} J., {Merritt} D., 2004, \apj, 600, 149

\bibitem[{{Wang} {et~al}\mbox{.}(2011){Wang}, {Zhou}, {Wang}, {Lu}, \&
  {Xu}}]{wang2011}
{Wang} T.-G., {Zhou} H.-Y., {Wang} L.-F., {Lu} H.-L., {Xu} D., 2011, \apj, 740,
  85

\bibitem[{{Wegg} \& {Nate Bode}(2011)}]{wegg2011}
{Wegg} C., {Nate Bode} J., 2011, \apjl, 738, L8

\bibitem[{{Wevers} {et~al}\mbox{.}(2017){Wevers}, {van Velzen}, {Jonker},
  {Stone}, {Hung}, {Onori}, {Gezari}, \& {Blagorodnova}}]{wevers2017}
{Wevers} T., {van Velzen} S., {Jonker} P.~G., {Stone} N.~C., {Hung} T., {Onori}
  F., {Gezari} S., {Blagorodnova} N., 2017, \mnras, 471, 1694

\bibitem[{{Wyrzykowski} {et~al}\mbox{.}(2017){Wyrzykowski}, {Zieli{\'n}ski},
  {Kostrzewa-Rutkowska}, {Hamanowicz}, {Jonker}, {Arcavi}, {Guillochon},
  {Brown}, {Koz{\l}owski}, {Udalski}, {Szyma{\'n}ski}, {Soszy{\'n}ski},
  {Poleski}, {Pietrukowicz}, {Skowron}, {Mr{\'o}z}, {Ulaczyk}, {Pawlak},
  {Rybicki}, {Greiner}, {Kr{\"u}hler}, {Bolmer}, {Smartt}, {Maguire}, \&
  {Smith}}]{wyrzykowski2016}
{Wyrzykowski} {\L}. {et~al.}, 2017, \mnras, 465, L114

\bibitem[{{Xie} {et~al}\mbox{.}(2002){Xie}, {Liang}, {Xie}, \& {Dai}}]{xie2002}
{Xie} G.~Z., {Liang} E.~W., {Xie} Z.~H., {Dai} B.~Z., 2002, \aj, 123, 2352

\bibitem[{{Yamauchi} \& {Goto}(2005)}]{yamauchi2005}
{Yamauchi} C., {Goto} T., 2005, \mnras, 359, 1557

\bibitem[{{Yang} {et~al}\mbox{.}(2006){Yang}, {Tremonti}, {Zabludoff}, \&
  {Zaritsky}}]{yang2006}
{Yang} Y., {Tremonti} C.~A., {Zabludoff} A.~I., {Zaritsky} D., 2006, \apjl,
  646, L33

\bibitem[{{Yang} {et~al}\mbox{.}(2004){Yang}, {Zabludoff}, {Zaritsky}, {Lauer},
  \& {Mihos}}]{yang2004}
{Yang} Y., {Zabludoff} A.~I., {Zaritsky} D., {Lauer} T.~R., {Mihos} J.~C.,
  2004, \apj, 607, 258

\bibitem[{{Young}, {Shields} \& {Wheeler}(1977){Young}, {Shields}, \&
  {Wheeler}}]{young1977}
{Young} P.~J., {Shields} G.~A., {Wheeler} J.~C., 1977, \apj, 212, 367

\bibitem[{{Zabludoff} {et~al}\mbox{.}(1996){Zabludoff}, {Zaritsky}, {Lin},
  {Tucker}, {Hashimoto}, {Shectman}, {Oemler}, \& {Kirshner}}]{zabludoff1996}
{Zabludoff} A.~I., {Zaritsky} D., {Lin} H., {Tucker} D., {Hashimoto} Y.,
  {Shectman} S.~A., {Oemler} A., {Kirshner} R.~P., 1996, \apj, 466, 104

\bibitem[{{Zel'dovich} \& {Raizer}(1967)}]{zeldovich1967}
{Zel'dovich} Y.~B., {Raizer} Y.~P., 1967, {Physics of shock waves and
  high-temperature hydrodynamic phenomena}

\bibitem[{{Zhang} \& {Wang}(2006)}]{zhang2006b}
{Zhang} E.-P., {Wang} J.-M., 2006, \apj, 653, 137

\end{thebibliography}

\appendix
%dummy comment inserted by tex2lyx to ensure that this paragraph is not empty%dummy comment inserted by tex2lyx to ensure that this paragraph is not empty

\section{Currently identified UV/O TDE candidates}
\label{s.candidates}

Table~\ref{table.scores}
   provides a listing of currently identified TDE candidate events
   with spectral lines,
   including the transient previously identified by
   \citet{drake2011}
   as a SN
   (but see \citealp{generozov2016}
   suggesting it is too luminous for a TDE,
   and \citealp{moriya2017} suggesting an AGN flare).
Spectral types are labelled as:
  `He' for \HeII-bright states;
  `He+H' for both \HeII\, and \Ha\, emission;
  `H' for \Ha-bright;
   and `abs' for the absorption line comparison case.
The clouds' emission line characteristic velocity ($v$)
   is defined as in \cite{arcavi2014}.
Where the observational papers give a bolometric luminosity ($L$),
   we use those values.
Otherwise,
   we interpolate and estimate values
   from the published photometric light-curves.
SMBH mass estimates are from the references.
(\citealp{wevers2017}
   infer some smaller $m_\bullet$
   from host galaxy velocity dispersions,
   but we tabulate the older published values for the sake of consistency.)

Through dimensional analyses,
   it can be shown that the composite scores $v^4/L$ and $Lv^4/m_\bullet$
   should be correlated with ionization parameter
   (if details of the system geometry were otherwise equal).
In these terms,
   there seems to be a tendency for the \Ha-bright events
   to have lower scores than the \HeII-overbright events.
It is however not a strong trend,
   and perhaps not statistically significant.
We were unable to use Table~\ref{table.scores}
   to declare any clear, quantitative predictor
   of the \cite{arcavi2014} spectral sequence.
In the Appendix~\ref{s.scaling},
   we search for patterns and discriminants in a different way,
   under particular model assumptions.

%%%%%%%%%%%%%%%%%%%%%%%%%%%%%%
%%%%%%%%%%%%%%%%%%%%%%%%%%%%%%
%% Table A1 

\begin{table*}
\caption{%
Column 1: name of the TDE event or host.
Column 2: type of broad line system:
   He{\sc}-dominated emission;
   both He{\sc ii} and H$\alpha$ emission;
   H$\alpha$-dominated emission;
   absorption.
Column 3: SMBH mass estimate.
Column 4: time relative to peak luminosity.
column 5: blackbody temperature (alternative versions for PS1-10jh).
Column 6: photometry used (`-' marks bolometric estimates given by reference).
Column 7: luminosity (derived from photometry if necessary).
Column 8: velocity width of broad line.
Column 9: ionization estimator.
Column 10: ionization estimator using estimated $m_\bullet$.
Column 11: references:
(a) \citealt{gezari2012};
(b) \citealt{arcavi2014};
(c) \citealt{wang2011};
(d) \citealt{holoien2014};
(e) \citealt{vanvelzen2011b};
(f) \citealt{drake2011};
(g) \citealt{chornock2014};
(h) \citealt{miller2015};
(i) \citealt{holoien2016b};
(j) \citealt{holoien2016a};
(k) \citealt{wyrzykowski2016};
(l) \citealt{blagorodnova2017};
(m) \citealt{hung2017}.
}
\label{table.scores}
\begin{center}
$\begin{array}{llr@{.}lrr@{.}lcr@{.}lr@{.}lrr@{.}llllllllllllll}
\hline
\mbox{object}
&\mbox{type}
&\multicolumn{2}{c}{m_\bullet}
&\multicolumn{1}{c}{t}
&\multicolumn{2}{c}{T_\mathrm{bb}}
&\multicolumn{1}{c}{\mbox{band}}
&\multicolumn{2}{c}{L}
&\multicolumn{2}{c}{v}
&\multirow{2}{*}{$\log\left({{v^4}\over{L}}\right)$}
&\multicolumn{2}{c}{\multirow{2}{*}{$\log\left({{Lv^4}\over{m_\bullet^2}}\right)$}}
&\mbox{ref.}
\\
&
&\multicolumn{2}{c}{(10^6m_\odot)}
&\multicolumn{1}{c}{\mbox{(day)}}
&\multicolumn{2}{c}{\mbox{($10^4$K)}}
&
&\multicolumn{2}{c}{(10^9L_\odot)}
&\multicolumn{2}{c}{\mbox{($10^3$km/s)}}
\\
\hline
\\
\mbox{PS1-10jh}&\mbox{He}
&4&^{+4}_{-2}
&-22
&\ga2&9\pm0.2
&g_{_\mathrm{P1}}r_{_\mathrm{P1}}i_{_\mathrm{P1}}
&\ga22&6\pm3
&5&43\pm1.46
&\la1.6\pm1.1
&\ga3&1^{+2.3}_{-1.5}
&\mbox{a,b}
\\
\\
%%%%%%%%%%%%%%%%%%%%%%%%%%%%%%%%%%%%%%%%%%%%%%%%%%%%%%%%%%%%%%%%%%%%%%%%%%%%%%
%%%%%%%%%%%%%%%%%%%%%%%%%%%%%%%%%%%%%%%%%%%%%%%%%%%%%%%%%%%%%%%%%%%%%%%%%%%%%%
\mbox{PTF09ge}&\mbox{He}
&5&65^{+3.02}_{-0.98}
&-18
&2&1\pm0.3
&\mathrm{-}
&13&0\pm4.6
&10&07\pm0.67
&2.9\pm0.4
&3&6^{+1.2}_{-0.6}
&\mbox{b}
\\
\\
%%%%%%%%%%%%%%%%%%%%%%%%%%%%%%%%%%%%%%%%%%%%%%%%%%%%%%%%%%%%%%%%%%%%%%%%%%%%%%
%%%%%%%%%%%%%%%%%%%%%%%%%%%%%%%%%%%%%%%%%%%%%%%%%%%%%%%%%%%%%%%%%%%%%%%%%%%%%%
\mbox{SDSSJ0748}&\mbox{He+H}
&11
&78^{+2.29}_{-3.56}
&>0
&1&3\pm0.4
&g
&>1&4\pm0.6
&9&95\pm0.51
&<3.8\pm0.5
&>2&0^{+0.6}_{-0.8}
&\mbox{b,c}
\\
\\
%%%%%%%%%%%%%%%%%%%%%%%%%%%%%%%%%%%%%%%%%%%%%%%%%%%%%%%%%%%%%%%%%%%%%%%%%%%%%%
%%%%%%%%%%%%%%%%%%%%%%%%%%%%%%%%%%%%%%%%%%%%%%%%%%%%%%%%%%%%%%%%%%%%%%%%%%%%%%
\mbox{ASASSN-14ae}&\mbox{H}
&2&45^{+1.55}_{-0.74}
&0
&2&3\pm0.1
&V
&38&3\pm3.6
&3&60\pm0.18
&0.6\pm0.2
&3&0^{+1.3}_{-0.6}
&\mbox{b,d}
\\
&\mbox{H}
&\multicolumn{2}{c}{\mbox{\aabboovvee}}
&4
&2&1\pm0.1
&-
&10&5\pm0.4
&7&2\pm0.2
&2.4\pm0.1
&3&7^{+1.3}_{-0.6}
\\
&\mbox{H}
&\multicolumn{2}{c}{\mbox{\aabboovvee}}
&30
&1&6\pm0.1
&-
&5&6\pm0.4
&6&0\pm0.2
&2.4\pm0.2
&3&1^{+1.3}_{-0.6}
\\
&\mbox{H}
&\multicolumn{2}{c}{\mbox{\aabboovvee}}
&51
&1&6\pm0.1
&-
&2&9\pm0.2
&5&1\pm0.2
&2.4\pm0.2
&2&5^{+1.3}_{-0.6}
\\
&\mbox{H}
&\multicolumn{2}{c}{\mbox{\aabboovvee}}
&73
&2&1\pm0.2
&-
&1&7\pm0.2
&4&2\pm0.2
&2.3\pm0.2
&2&0^{+1.3}_{-0.6}
\\
&\mbox{He+H}
&\multicolumn{2}{c}{\mbox{\aabboovvee}}
&94
&2&1\pm0.3
&-
&1&3\pm0.3
&3&4\pm0.2
&2.0\pm0.3
&1&5^{+1.3}_{-0.7}
\\
&\mbox{He+H}
&\multicolumn{2}{c}{\mbox{\aabboovvee}}
&132
&2&0\pm0.2
&-
&0&3\pm0.2
&3&4\pm0.2
&2.6\pm0.6
&0&8^{+1.4}_{-0.9}
\\
\\
%%%%%%%%%%%%%%%%%%%%%%%%%%%%%%%%%%%%%%%%%%%%%%%%%%%%%%%%%%%%%%%%%%%%%%%%%%%%%%
%%%%%%%%%%%%%%%%%%%%%%%%%%%%%%%%%%%%%%%%%%%%%%%%%%%%%%%%%%%%%%%%%%%%%%%%%%%%%%
\mbox{ASASSN-15oi}&\mbox{He}
&25&^{+25}_{-13}
&7
&2&0\pm0.4
&-
&52&\pm25
&8&4\pm0.4
&2.0\pm0.5
&2&6^{+2.1}_{-1.1}
&\mbox{j}
\\
&\mbox{He+H}
&\multicolumn{2}{c}{\mbox{\aabboovvee}}
&21
&7&5\pm0.4
&-
&35&\pm17
&3&9\pm0.4
&0.8\pm0.6
&1&1^{+2.1}_{-1.2}
\\
\\
%%%%%%%%%%%%%%%%%%%%%%%%%%%%%%%%%%%%%%%%%%%%%%%%%%%%%%%%%%%%%%%%%%%%%%%%%%%%%%
%%%%%%%%%%%%%%%%%%%%%%%%%%%%%%%%%%%%%%%%%%%%%%%%%%%%%%%%%%%%%%%%%%%%%%%%%%%%%%
\mbox{ASASSN-14li}&\mbox{He+H}
&3&\pm2
&35
&3&5\pm0.4
&-
&83&2\pm2.6
&1&3\pm0.4
&-1.5\pm1.3
&1&4\pm1.9
&\mbox{h,i}
\\
\\
%%%%%%%%%%%%%%%%%%%%%%%%%%%%%%%%%%%%%%%%%%%%%%%%%%%%%%%%%%%%%%%%%%%%%%%%%%%%%%
%%%%%%%%%%%%%%%%%%%%%%%%%%%%%%%%%%%%%%%%%%%%%%%%%%%%%%%%%%%%%%%%%%%%%%%%%%%%%%
\mbox{OGLE16aaa}&\mbox{He+H}
&3&8^{+3.8}_{-1.9}
&-3
&\ga2&2
&I
&86&\pm4
&7&6\pm0.4
&1.6\pm0.2
&4&3^{+2.0}_{-1.0}
&\mbox{k}
\\
\\
%%%%%%%%%%%%%%%%%%%%%%%%%%%%%%%%%%%%%%%%%%%%%%%%%%%%%%%%%%%%%%%%%%%%%%%%%%%%%%
%%%%%%%%%%%%%%%%%%%%%%%%%%%%%%%%%%%%%%%%%%%%%%%%%%%%%%%%%%%%%%%%%%%%%%%%%%%%%%
\mbox{iPTF16fnl}&\mbox{He+H}
&6&3\pm0.4
&0
&1&9\pm0.2
&-
&2&6\pm0.4
&5&9\pm0.4
&2.7\pm0.3
&2&9^{+1.5}_{-0.7}
&\mbox{l}
\\
\\
%%%%%%%%%%%%%%%%%%%%%%%%%%%%%%%%%%%%%%%%%%%%%%%%%%%%%%%%%%%%%%%%%%%%%%%%%%%%%%
%%%%%%%%%%%%%%%%%%%%%%%%%%%%%%%%%%%%%%%%%%%%%%%%%%%%%%%%%%%%%%%%%%%%%%%%%%%%%%
\mbox{iPTF16axa}&\mbox{He+H}
&5&0^{+7.0}_{-2.9}
&>0
&3&0\pm0.33
&-
&>29&\pm4
&4&0\pm0.2
&<1.0\pm0.2
&>2&5^{+2.8}_{-1.2}
&\mbox{m}
\\
\\
%%%%%%%%%%%%%%%%%%%%%%%%%%%%%%%%%%%%%%%%%%%%%%%%%%%%%%%%%%%%%%%%%%%%%%%%%%%%%%
%%%%%%%%%%%%%%%%%%%%%%%%%%%%%%%%%%%%%%%%%%%%%%%%%%%%%%%%%%%%%%%%%%%%%%%%%%%%%%
\mbox{PTF09axc}&\mbox{H}
&2&69^{+0.66}_{-0.64}
&\approx5
&1&2\pm0.2
&-
&4&7\pm2.4
&11&89\pm0.22
&3.6\pm0.5
&4&1\pm0.7
&\mbox{b}
\\
\\
%%%%%%%%%%%%%%%%%%%%%%%%%%%%%%%%%%%%%%%%%%%%%%%%%%%%%%%%%%%%%%%%%%%%%%%%%%%%%%
%%%%%%%%%%%%%%%%%%%%%%%%%%%%%%%%%%%%%%%%%%%%%%%%%%%%%%%%%%%%%%%%%%%%%%%%%%%%%%
\mbox{PTF09djl}&\mbox{H}
&3&57^{+9.97}_{-2.96}
&2
&2&6\pm0.4
&-
&30&\pm14
&6&53\pm0.35
&1.8\pm0.5
&3&6^{+5.6}_{-1.7}
&\mbox{b}
\\
&
&\multicolumn{2}{c}{\mbox{\aabboovvee}}
&31
&2&0\pm0.3
&-
&15&\pm9
&6&53\pm0.35
&2.1\pm0.6
&3&3^{+5.6}_{-1.7}
\\
&
&\multicolumn{2}{c}{\mbox{\aabboovvee}}
&62
&>3&2
&-
&<124&
&6&53\pm0.35
&>1.2\pm0.2
&<4&2^{+5.6}_{-1.7}
\\
\\
%%%%%%%%%%%%%%%%%%%%%%%%%%%%%%%%%%%%%%%%%%%%%%%%%%%%%%%%%%%%%%%%%%%%%%%%%%%%%%
%%%%%%%%%%%%%%%%%%%%%%%%%%%%%%%%%%%%%%%%%%%%%%%%%%%%%%%%%%%%%%%%%%%%%%%%%%%%%%
\mbox{TDE2}&\mbox{H}
&35&52^{+55.31}_{-25.80}
&>40
&1&82\pm0.07
&g
&42&0\pm2.9
%&>10&7\pm0.5
&3&44\pm0.11
&0.5\pm1.3
%&<1.1\pm1.3
&0&7^{+3.4}_{-1.9}
%&>0&1^{+3.4}_{-1.9}
&\mbox{b,e}
\\
\\
%%%%%%%%%%%%%%%%%%%%%%%%%%%%%%%%%%%%%%%%%%%%%%%%%%%%%%%%%%%%%%%%%%%%%%%%%%%%%%
%%%%%%%%%%%%%%%%%%%%%%%%%%%%%%%%%%%%%%%%%%%%%%%%%%%%%%%%%%%%%%%%%%%%%%%%%%%%%%
\mbox{CSS100217}&\mbox{H}
&14&7^{+2.4}_{-2.0}
&-5
&1&62\pm0.18
&V
&439&\pm104
&1&20\pm0.04
&-2.3\pm0.3
&0&6\pm0.3
&\mbox{f}
\\
&
&\multicolumn{2}{c}{\mbox{\aabboovvee}}
&20
&1&35\pm0.08
&V
&267&\pm36
&1&36\pm0.04
&-1.9\pm0.2
&0&6^{+0.4}_{-0.3}
\\
&
&\multicolumn{2}{c}{\mbox{\aabboovvee}}
&84
&0&78\pm0.20
&V
&67&\pm16
&2&07\pm0.04
&-0.6\pm0.3
&0&6\pm0.5
\\
&
&\multicolumn{2}{c}{\mbox{\aabboovvee}}
&164
&\multicolumn{2}{c}{~}
&V
&<4&6\pm1.1
&1&63\pm0.04
&>0.2\pm0.3
&<-1&0\pm0.6
\\
\\
%%%%%%%%%%%%%%%%%%%%%%%%%%%%%%%%%%%%%%%%%%%%%%%%%%%%%%%%%%%%%%%%%%%%%%%%%%%%%%
%%%%%%%%%%%%%%%%%%%%%%%%%%%%%%%%%%%%%%%%%%%%%%%%%%%%%%%%%%%%%%%%%%%%%%%%%%%%%%
\mbox{PS1-11af}&\mbox{abs}
&8&0\pm2.0
&24
&1&7\pm0.3
&-
&22&1\pm0.5
&4&31\pm0.34
&1.2\pm0.3
&1&9\pm0.3
&\mbox{g}
\\
\\
\hline
\end{array}$
\end{center}
\end{table*}

\section{Observed vs expected velocities of the illuminated cloud}
\label{s.scaling}

\begin{figure}
\includegraphics[width=8.0cm]{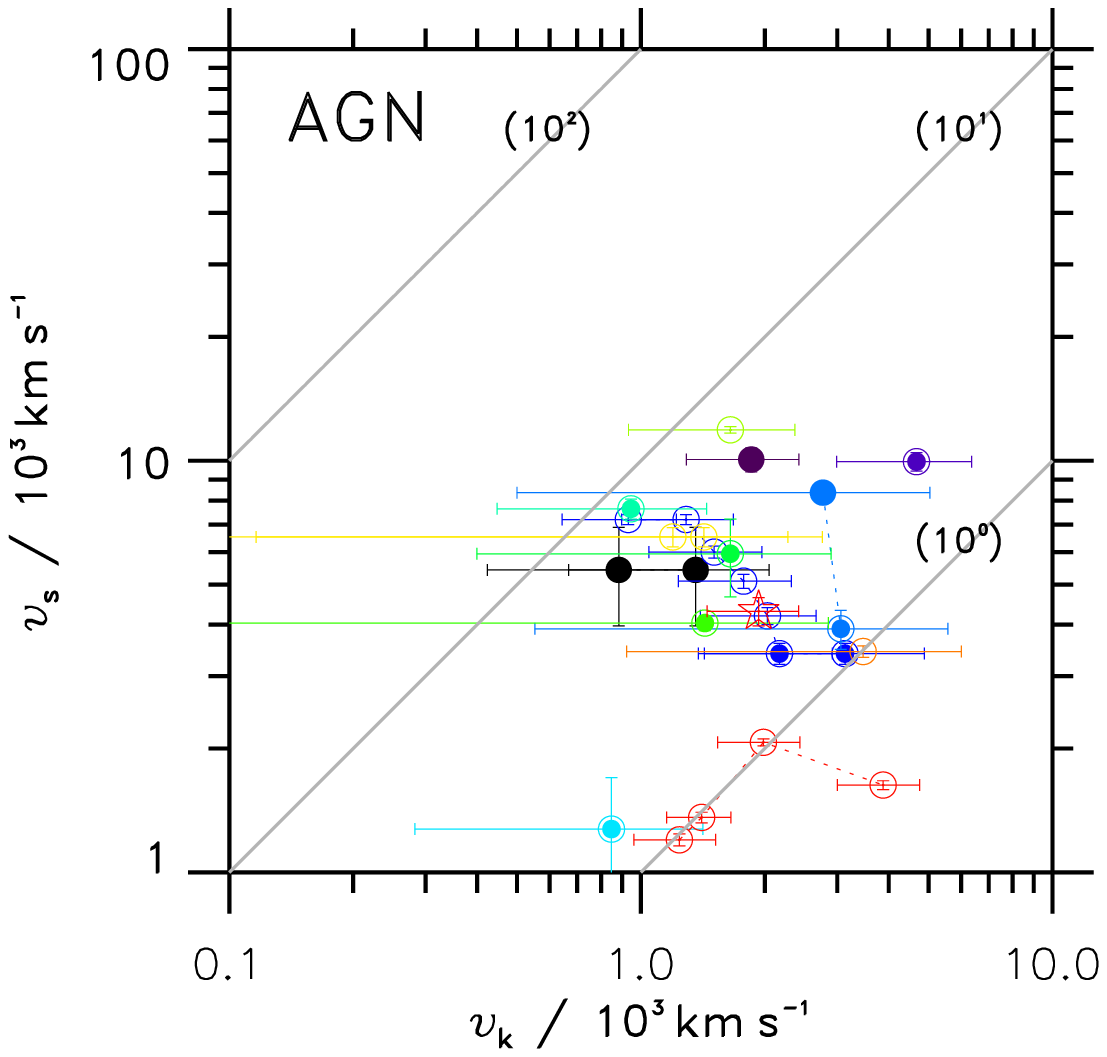}
\includegraphics[width=8.0cm]{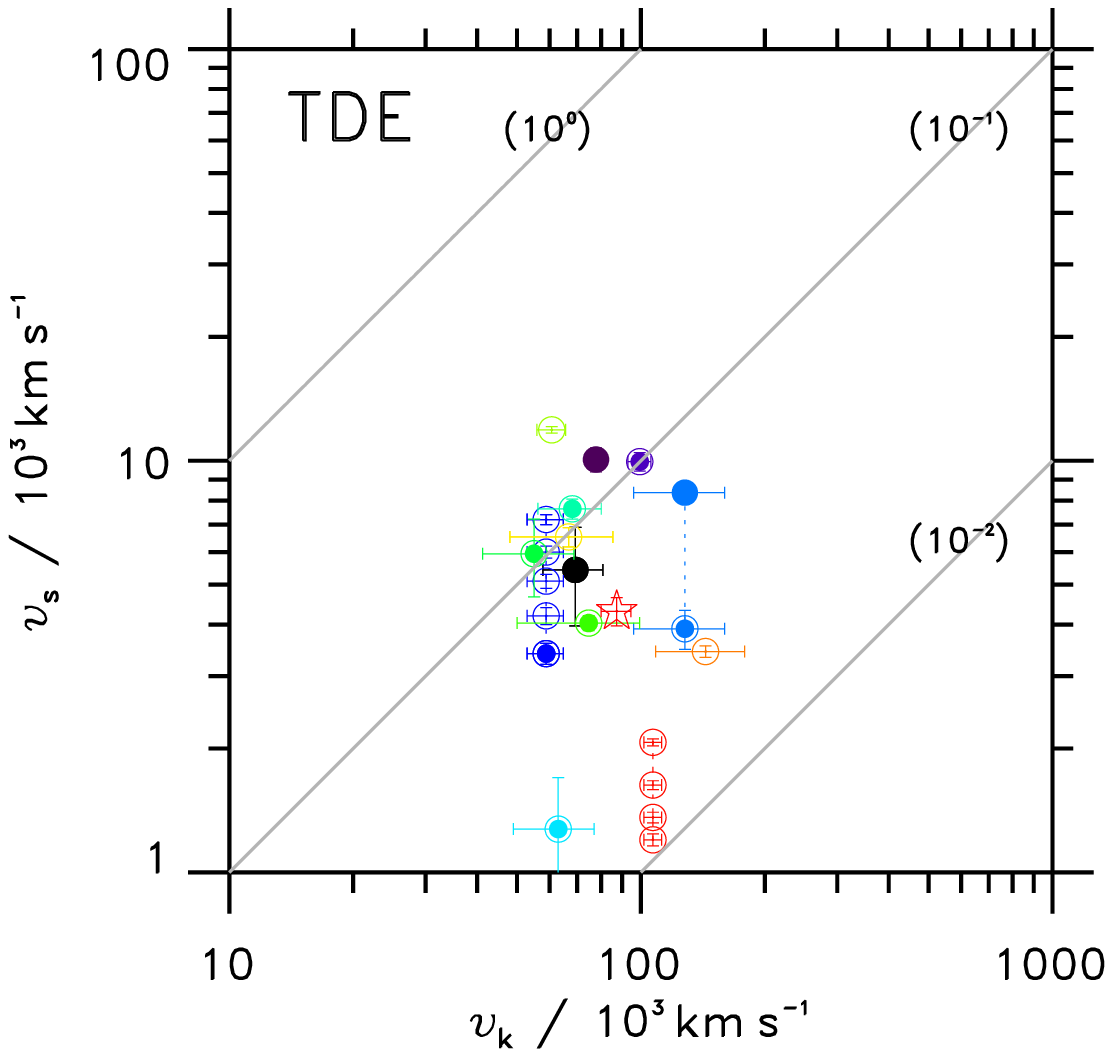}
\caption{%
Comparison of observed velocity widths ($v_\mathrm{s}$)
with expected Keplerian velocity ($v_{\mathrm{k}}$)
estimated using scaling assumptions from
AGN broad line regions (top) and TDE tidal radii (bottom).
Guidelines mark the ratio $v/v_{\mathrm{k}}$ in powers of ten.
Open circle symbols are events with H$\alpha$ emission;
   filled circles are events with \HeII\, emission;
   the starred event has broad absorption lines.
}
\label{fig.velocities} 
\end{figure}

The presently available observations ought to appear more orderly
   under assumption of an event type: TDE or AGN flare.
Empirically, the characteristic
radius of a steady active galaxy's BLR is
   $R_{\textsc{blr}}\approx R_{44}(L/10^{44}\,\mathrm{erg}\,\mathrm{s}^{-1})^{A}$
   where an index $A={\frac{1}{2}}$
   is the ideal if the BLR size were constrained by a threshold temperature
   (e.g. of dust sublimation).
From reverberation mapping in the 5100\AA\, monochromatic luminosity
   $\lambda L_{\lambda}$,
   \citet{kaspi2005} find $R_{44}=22.3\pm2.1$\,light\,day
   and $A=0.69\pm0.05$.
Given this sizing, and SMBH mass estimates \citep{arcavi2014},
   we obtain a characteristic orbital velocity of the BLR clouds
   ($v_{\mathrm{k}}$).
If however the event were a TDE,
   then the stellar tidal disruption radius provides the relevant
   $v_{\mathrm{k}}$ estimator.
Fig.~\ref{fig.velocities}
   compares observed spectroscopic line widths ($v_\mathrm{s}$)
   to the $v_{\mathrm{k}}$ estimates in both models.
Assuming stars of solar density,
   the TDE scenario works poorly:
   the measured velocities are $0.02$ to $0.2$ times predicted,
   and the values seem uncorrelated with $v_{\mathrm{k}}$;
   and neither does the AGN model show any significant correlation.
The measured velocities are a factor of a few higher than AGN expectations.
From these simple scaling arguments,
   neither the TDE nor AGN scenarios fits the UV/O events impressively well,
   but the AGN version may seem somewhat less discordant.
   %(at least having a wider spread of $v_\mathrm{k}$).
The weakness of the correlation between actual and expected velocities
   implies that the line-emitters are not causally related to the flaring event.
Rather, we suppose that they were pre-existing, dormant circumnuclear clouds
   that revived and reactivated as a BLR
   when irradiated by the explosive event.

\section{Cloudy calculations and the line ratios
in the photoionization cube parameter phase space}
\label{appendix.cubes}

For each SED, we run several cubes of calculations: exploring a rectangle
   of
   $(N_{\textsc{h}},n_{\textsc{h}},\Phi_{\textsc{h}})$
   in steps of $0.1$\,dex,
   and saving depth profiles along the $N_{\textsc{h}}$ or spatial axis.
We rerun extra cubes where the maximum column was set to
   $N_{\textsc{h}}=10^{23},10^{24},10^{25}~\mathrm{cm}^{-2}$.
To allay concerns about the validity of
   {\sc Cloudy}'s approximate treatment of electron scattering,
   we conservatively exclude opaque models in our final results.
The maximum $N_\textsc{h}$ for every $(n_\textsc{h},\Phi_\textsc{h})$
   is manually adjusted to give a maximum optical depth of $\tau_\mathrm{es}\le0.5$
   (often near $N_\textsc{h}\approx10^{23.8}~\mathrm{cm}^{-2}$).
From runs generated with different stopping columns,
   the coinciding parts of the line intensity vs depth profiles
   agree within an order of unity scale factor.

{\sc Cloudy} version 13.03 enables output of
   intensities from the illuminated face of the cloud,
   or the total of inward and outward faces.
Unless otherwise stated,
   we use the inward face,
   which is most relevant for light-echo effects.
   
In every numerical run, 
   \textsc{Cloudy} reports
   the mean energy of a photoionizing photon (\texttt{Average nu}),
   the intensity of ionizing photons (\texttt{I(nu\textgreater{}1ryd)}),
   and the total intensity (\texttt{Total inten}).
Using these, or numerical integration of the known SED function,
   provides a characteristic photon energy ($\overline{\epsilon}$,
   enabling conversion between the ionizing flux
   ($\Phi_{\textsc{h}}$, user selected),
   an assumed cloud radial position ($R_{\mathrm{c}}$),
   and the bolometric luminosity of the source ($L$).
%\begin{equation}
%L=4\upi R_{\mathrm{c}}^{2}\Phi_{\textsc{h}}\overline{\epsilon}\ .
%\end{equation}
If $L$ is known,
   then we infer the distance between the source
   and the cloud,
\begin{equation}
R_{\mathrm{c}}=\left[\frac{{L}}{{4\upi\overline{\epsilon}\Phi_{\textsc{h}}}}\right]^{1/2}%=
%{L_{\mathrm{o}}(T_{\mathrm{f}}/T_{\mathrm{o}})^{b/a}}
%\frac{}{{}}
%4\upi\overline{\epsilon}_{\mathrm{f}}\Phi_{\textsc{h}}
\ .\label{eq.cloud.radius}
\end{equation}
The emission line luminosity ($L_{\lambda}$) and the cloud mass ($M_{\mathrm{c}}$)
   depend on the cloud covering factor ($0$$<$$\Omega$$<$$1$)
   and the line's specific intensity
   ($I_\mathrm{line}$, calculated by \textsc{Cloudy}).
Generally, 
\begin{equation}
	L_{\lambda}
	=4\upi R_{\mathrm{c}}^{2}I_\mathrm{line}\Omega
	=\frac{{L}}{{\overline{\epsilon}\Phi_{\textsc{h}}}}I_\mathrm{line}\Omega
\label{eq.cloud.luminosity-1}
\end{equation}
\begin{equation}
	M_{\mathrm{c}}
	=4\upi R_{\mathrm{c}}^{2}\mu m_{\textsc{p}}N_{\textsc{h}}\Omega
	=\frac{{L}}{{\overline{\epsilon}\Phi_{\textsc{h}}}}\mu m_{\textsc{p}}N_{\textsc{h}}\Omega
	=\mu m_{\textsc{p}}\frac{{L_{\lambda}}}{{I_\mathrm{line}}}N_{\textsc{h}}
	\ ,
\label{eq.cloud.mass-1}
\end{equation}
   where $\mu$ is the mean molecular weight,
   and $m_{\mathrm{p}}$ is the proton mass.
We can either make a fiducial assumption about the covering factor
   (e.g. $\Omega=0.1$),
   or we can calculate a self-consistent value
   if both the line luminosity and continuum luminosity are known,
\begin{equation}
	\Omega
	=\frac{{L_{\lambda}}}{{L}}\frac{{\overline{\epsilon}\Phi_{\textsc{h}}}}{{I_\mathrm{line}}}
%=\frac{{L_{\lambda}}}{{L_{\mathrm{o}}(T_{\mathrm{f}}/T_{\mathrm{o}})^{b/a}}}
%\frac{{\overline{\epsilon}_{\mathrm{f}}\Phi_{\textsc{h}}}}{{I_\mathrm{line}}}
\ .\label{eq.Omega-1}
\end{equation}
In practice, $\Phi_{\textsc{h}}$ is a user-chosen runtime parameter;
   $\overline{\epsilon}$ is a constant for each SED;
   $I_\mathrm{line}$ is output from \textsc{Cloudy}
   for given
   $(N_{\textsc{h}},n_{\textsc{h}},\Phi_{\textsc{h}})$;
   $L_{\lambda}$ is the observed line luminosity;
   and the total luminosity $L$ is observationally inferred.
Thus we assemble cubes of results in the
   $(N_{\textsc{h}},n_{\textsc{h}},\Phi_{\textsc{h}})$
   parameter cube,
   and give them quantitative interpretation in the empirical context
   of a specific TDE candidate.
Imposing some commonsense conditions
   can whittle the possibilities,
   to leave a set of physically reasonable results. 
\begin{itemize}
\item
For self-consistency,
   the clouds must be outside
   the opaque photosphere of the actively luminous source
   ($R_{\mathrm{c}}>R_{\mathrm{o}}$,
   where $L_{\mathrm{o}}=4\upi R_{\mathrm{o}}^{2}\sigma T_{\mathrm{o}}^{4}$
   for a blackbody sphere). 
\item
Clouds can't cover more than the whole source ($\Omega<1$). 
\item
The \HeII/H$\alpha$ line ratio and the respective equivalent
widths can be checked for consistency with observations. 
\end{itemize}
Before interpreting the scale-independent \textsc{Cloudy} results
   to model a particular TDE candidate event,
   a minority of the numerical output requires filtering for artifacts.
For the hotter blackbody scenarios,
   a numerical instability sometimes occurs near the far end
   of the spatial grid.
%   regardless of the total column density.
In these incidents, \textsc{Cloudy} reports global non-convergence
   (which might have been the numercial problem reported by
   \citealt{strubbe2015}).
On close inspection,
   the majority of the profile is however stable.
We can retain these good sections, and excise the final $0.2$\,dex in $N_{\textsc{h}}$,
   and any unrealistically steep jumps in \HeII\, intensity.
   %rises with a logarithmic gradient greater than $2$.
After this initial filtering,
   a smaller minority of numerical spikes remain;
   but their sensitivity to resolution proves their spuriousness.
These glitches occur at uninterestingly low $N_{\textsc{h}}$,
   in configurations where the ionization parameter is low ($U<0.03$),
   so from some cubes
%From the $10^{6}$K and $10^{7}$K blackbody SEDs,
   we exclude models with %$U<0.03$
   $\Phi_{\textsc{h}}/n_{\textsc{h}}<10^{9}~\mathrm{cm}\,\mathrm{s}^{-1}$.
%Since we accumulated three cubes of results at different final $N_{\textsc{h}}$
%   for every hypothetical SED,
%   we have enough redundant and overlapping
%calculations to leave a finely sampled combined cube after filtering.

\begin{figure*}
\begin{tabular}{cc}
\includegraphics[width=82mm]{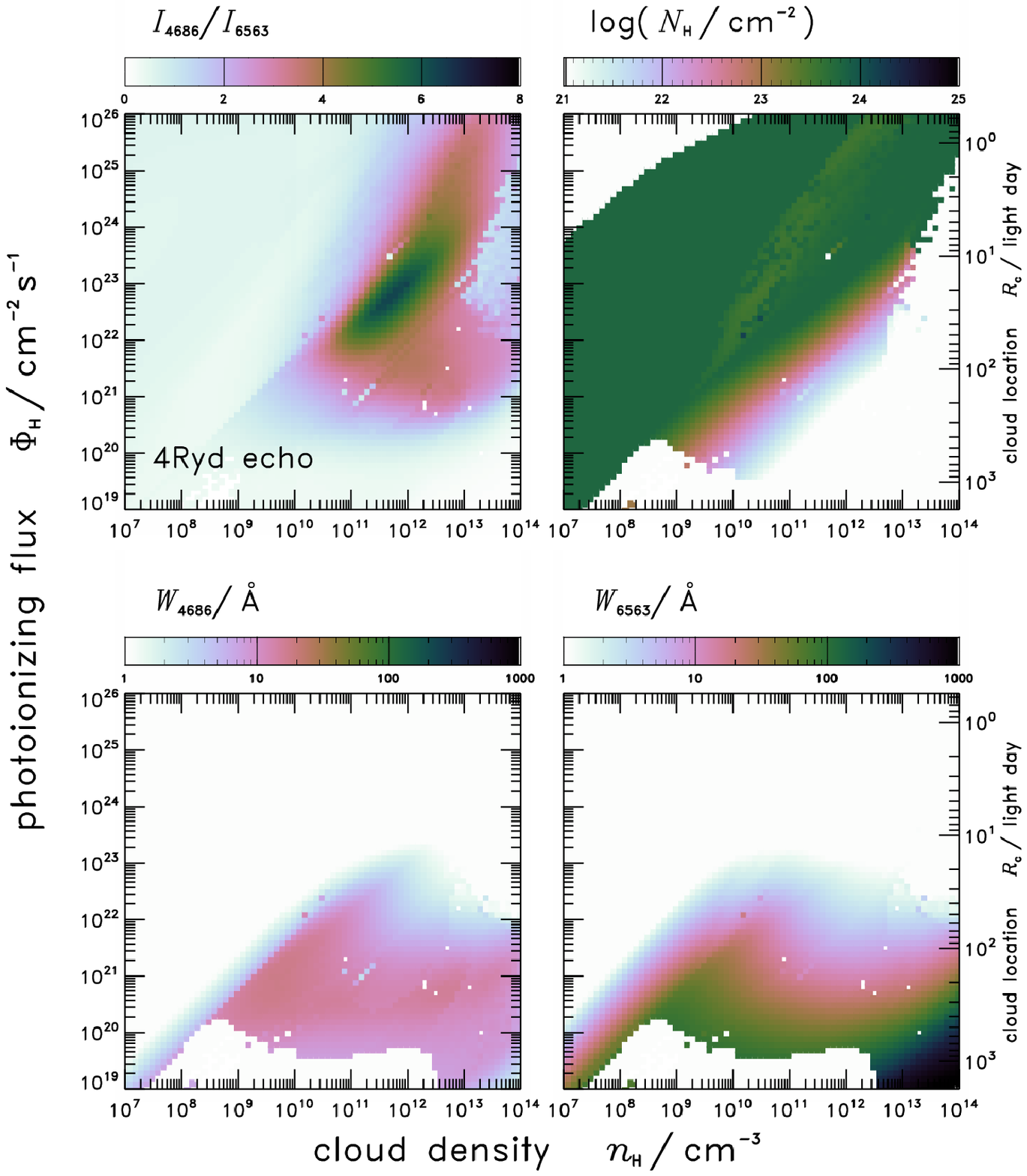}
&
\includegraphics[width=82mm]{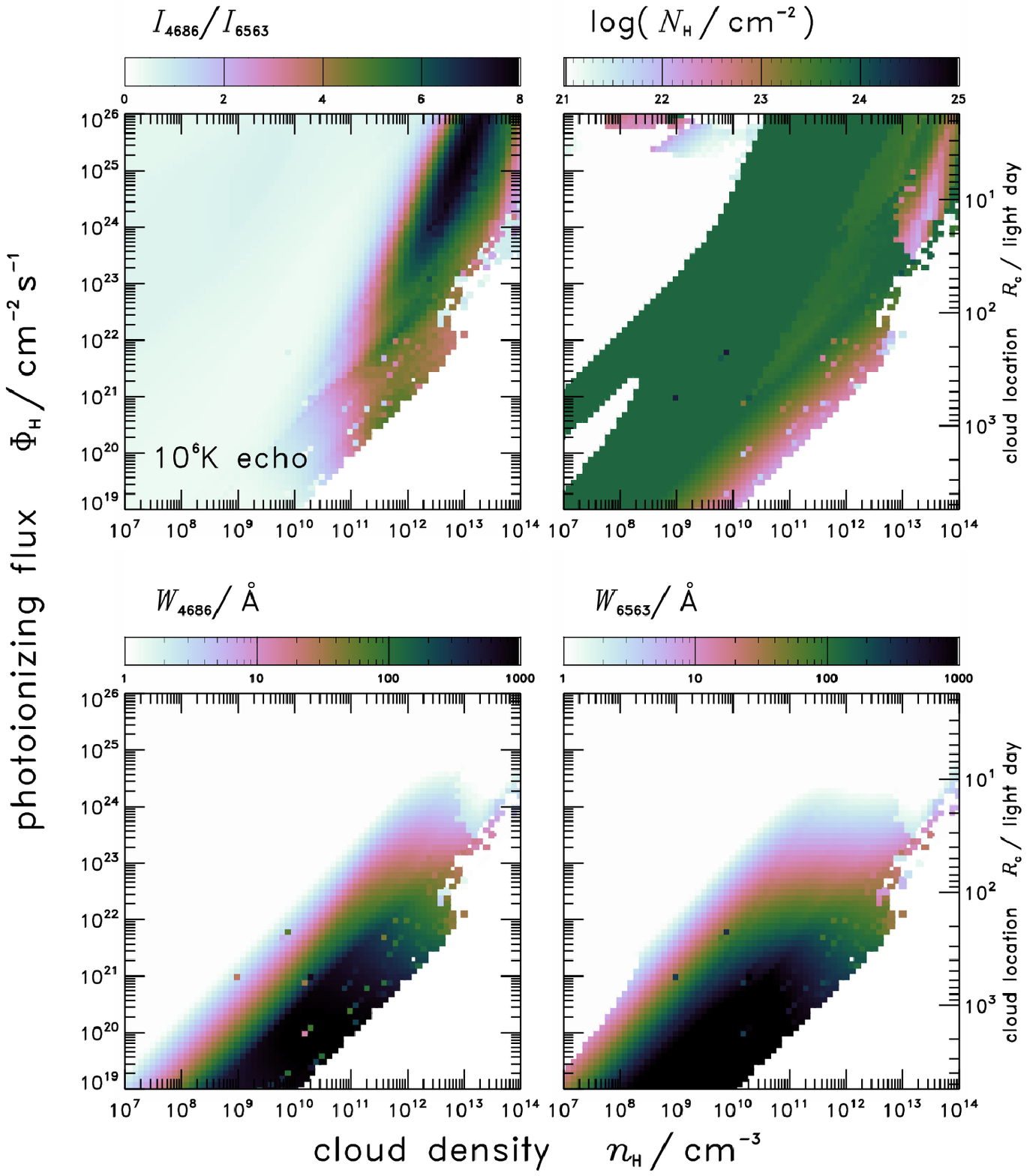}\\
\\
\\
\includegraphics[width=82mm]{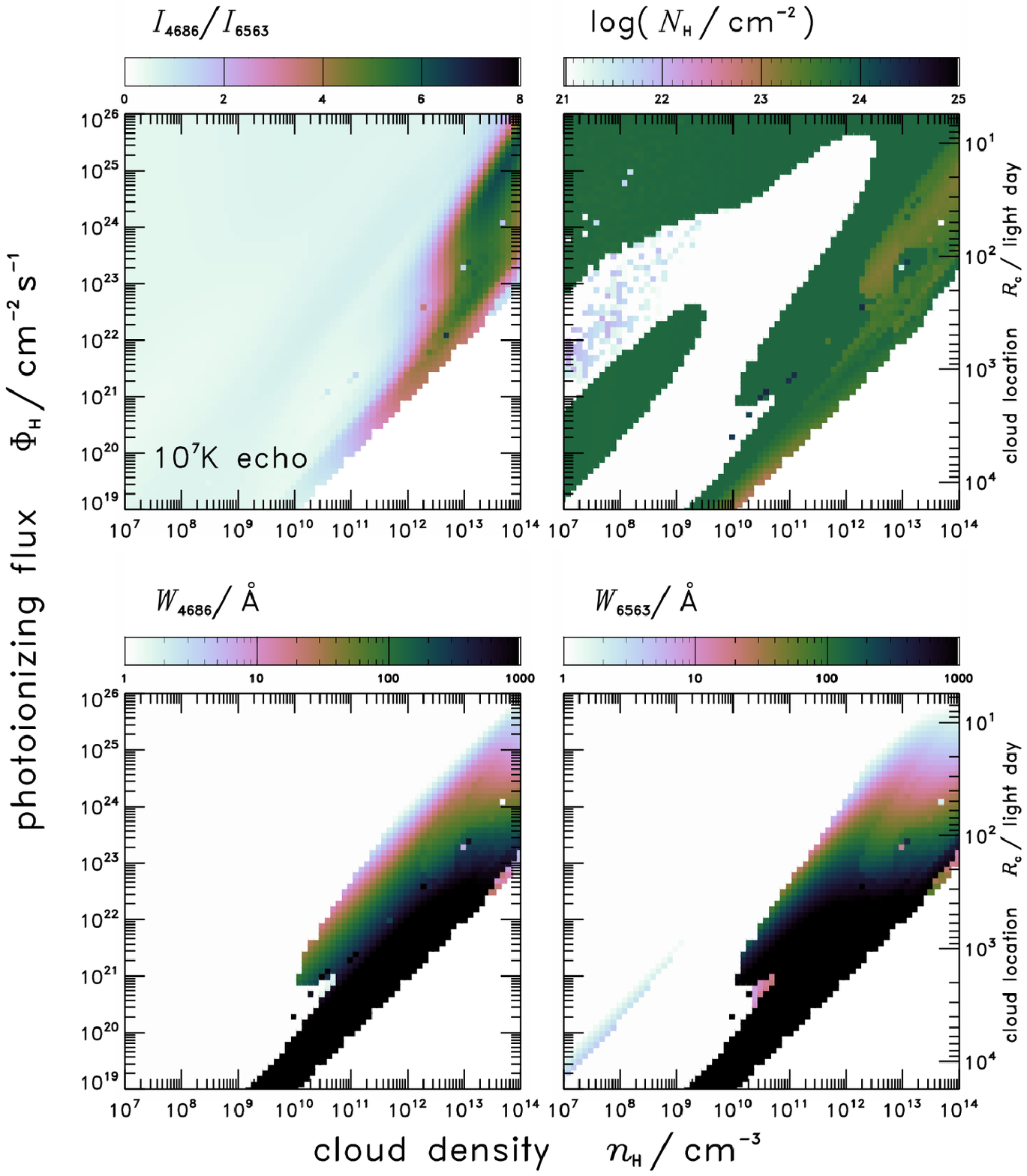}
\end{tabular}
\caption{%
Conditions where $N_\textsc{h}$ maximises
   the \HeII(4686\AA\,) to H$\alpha$(6563\AA\,) intensity ratio,
   at given $(n_{\textsc{h}},\Phi_{\textsc{h}})$.
%The rows show the effect of blackbody SEDs
%   except that the equivalent widths derive from
The equivalent widths derive from
   continuum estimates in a `light echo' model,
   assuming that the currently visible continuum is a blackbody of
   $3\times10^{4}$K,
   %$T_{2}=3\times10^{4}$K,
   and the earlier flash illuminating the clouds is a 4Ryd, $10^6$K or $10^{7}$K blackbody
   %(left and right blocks respectively). }
   (first, second and third blocks respectively).
These plots pertain to radation from the illuminated face of each cloud.
}
\label{fig.ratio.echo} 
\end{figure*}

The three-parameter cube of photoionization states in
   $(N_{\textsc{h}},n_{\textsc{h}},\Phi_{\textsc{h}})$
   is conventionally visualized in terms of cross-sections:
   the constant-$N_{\textsc{h}}$ flux-density planes
   \citep[e.g.][]{korista2004}.
For our purposes,
   in evaluating the possible conditions surrounding UV/O nuclear transients,
   it is more useful to project the non-linear surfaces
   to select $N_\textsc{h}$ values that maximise
   the \HeII/\Ha\, line ratio at each $(n_\textsc{h},\Phi_\textsc{h})$,
   to prioritise \HeII-bright events.
Fig.~\ref{fig.ratio.echo} shows the results of coarse mapping
   of the local peak ratio for given $(n_{\textsc{h}},\Phi_{\textsc{h}})$
   pairs on a grid with $0.1$dex steps on each axis,
   showing the effect of blackbody source SEDs.
%   on the light-echo of the illuminated face of a cloud.
Top-left panels of each block show maximum line ratios
   at each $(n_\textsc{h},\Phi_\textsc{h})$.
%Line ratios (top-left panels)
%   are a projection down along the $N_{\textsc{h}}$ axis
%   of the $(N_{\textsc{h}},n_{\textsc{h}},\Phi_{\textsc{h}})$ cube.
\HeII-overbright states occupy the minority region
   appearing as a dark (green-blue) inclined `sausage.'
If these conditions are unmet,
   a cloud can only be a \HeII+\Ha\, or \Ha-dominated event.
The lower two panels of each block
   show the equivalent widths $W_{4686}$ and $W_{6563}$
   (assuming $\Omega=0.1$)
   in the peak line ratio conditions.
   %$N_{\textsc{h}}$ conditions that gave the peak line ratio.
   %from AGN-like illumination produced small EW values.
The top-right panels show the $N_{\textsc{h}}$ values
    that gave each peak line ratio.
\HeII-bright states require column densities
   near the highest we investigated.
Echo models assume that the clouds were photoionized by an early,
   hard flash from a compact,
   hot blackbody.
After the radiation-dominated ball expands and cools for some time,
   the observed continuum settles near $3\times10^4$K.
The effect of the light-echo correction \textemdash{}
   early hard photoionization source
   superimposed on a later cooled continuum, according to the light crossing time
   \textemdash{} is to lower the equivalent widths.

%%%%%%%%%%%%%%%%%%%%%%%%%%%%%%%%%%%%%%%%%%%%%%%%
%%%%%%%%%%%%%%%%%%%%%%%%%%%%%%%%%%%%%%%%%%%%%%%%

\bsp	% typesetting comment
\label{lastpage}
\end{document}